\documentclass[12pt]{article}
\usepackage{amsmath,amssymb}
\usepackage{graphicx}
\usepackage{amscd}

\newcommand{\eqdef}{\stackrel{\text{def}}{=}}
\newcommand{\n}{\nonumber\\}
\newcommand{\bm}{\boldsymbol}
\renewcommand{\theequation}{\arabic{section}.\arabic{equation}}
\newcommand{\Romannumeral}[1]{\uppercase\expandafter{\romannumeral#1}}
\newcommand{\cF}{c_{\text{\tiny$\mathcal{F}$}}}

\allowdisplaybreaks[4]

\begin{document}

\baselineskip=20pt

%%%%%%%%%%%%%%%%%%%%%%%%%%%%%%%%%%%%%%%%%%%%%%%%%%%%%%%%%%%%
%                                                          %
%  Title page                                              %
%                                                          %
%%%%%%%%%%%%%%%%%%%%%%%%%%%%%%%%%%%%%%%%%%%%%%%%%%%%%%%%%%%%
\newfont{\elevenmib}{cmmib10 scaled\magstep1}
\newcommand{\preprint}{
% \vspace*{-20mm}
   \begin{flushleft}
     \elevenmib Yukawa\, Institute\, Kyoto\\
   \end{flushleft}\vspace{-1.3cm}
   \begin{flushright}\normalsize \sf
     DPSU-11-3\\
     YITP-11-35\\
     ~\\
%     {\tt arXiv:1104.0473[math-ph]}\\
%     April 2011
   \end{flushright}}
\newcommand{\Title}[1]{{\baselineskip=26pt
   \begin{center} \Large \bf #1 \\ \ \\ \end{center}}}
\newcommand{\Author}{\begin{center}
   \large \bf Satoru Odake${}^a$ and Ryu Sasaki${}^b$ \end{center}}
\newcommand{\Address}{\begin{center}
     $^a$ Department of Physics, Shinshu University,\\
     Matsumoto 390-8621, Japan\\
     ${}^b$ Yukawa Institute for Theoretical Physics,\\
     Kyoto University, Kyoto 606-8502, Japan
   \end{center}}
\newcommand{\Accepted}[1]{\begin{center}
   {\large \sf #1}\\ \vspace{1mm}{\small \sf Accepted for Publication}
   \end{center}}

\preprint
\thispagestyle{empty}

\Title{Discrete Quantum Mechanics}

\Author

\Address
\vspace{1cm}

\begin{abstract}
A comprehensive review of the discrete quantum mechanics with the pure
imaginary shifts and the real shifts is presented in parallel with the
corresponding results in the ordinary quantum mechanics. The main subjects
to be covered are the factorised Hamiltonians, the general structure of
the solution spaces of the Schr\"{o}dinger equation (Crum's theorem and its
modification), the shape invariance, the exact solvability in the
Schr\"{o}dinger picture as well as in the Heisenberg picture, the
creation/annihilation operators and the dynamical symmetry algebras,
the unified theory of exact and quasi-exact solvability based on the
sinusoidal coordinates, the infinite families of new orthogonal (the
exceptional) polynomials. Two new infinite families of orthogonal polynomials,
the $X_\ell$ Meixner-Pollaczek and the $X_\ell$ Meixner polynomials are
reported.
\end{abstract}


{\bf keywords}:
{discrete quantum mechanics, difference Schr\"{o}dinger equation,
 exact solvability, quasi-exact solvability, intertwining relations,
 shape invariance, Heisenberg operator solutions, closure relations,
 new orthogonal polynomials}

%%%%%%%%%%%%%%%%%%%%%%%%%%%%%%%%%%%%%%%%%%%%%%%%%%%%%%%%%%%%%%%
%                                                             %
%  1. Introduction                                            %
%                                                             %
%%%%%%%%%%%%%%%%%%%%%%%%%%%%%%%%%%%%%%%%%%%%%%%%%%%%%%%%%%%%%%%
\section{Introduction}
\label{intro}
\setcounter{equation}{0}

In the discrete quantum mechanics (dQM), developed by the present authors,
the time-independent Schr\"{o}dinger equation is a second order {\em difference}
equation instead of a second order {\em differential} equation in the ordinary
quantum mechanics (oQM). The basic framework of quantum mechanics,
such as the probability interpretation,
the Hilbert space structure, the continuous real time, etc. is unchanged and
the only difference is the  discrete forms of  various Hamiltonians.
The discrete quantum mechanics can be considered as a generalisation
(deformation) of the ordinary QM in the sense that a differential equation
can be obtained from a difference equation in an appropriate limit.
The generalisation is also consistent with Heisenberg's idea that a certain
fundamental length would appear in the theory at some stage of descending
the microscopic ladder.
This generalisation has turned out to be extremely fruitful. Most of the
concepts and methods of oQM can be easily transplanted to dQM, offering
a unified platform for the systematic understanding of various orthogonal
polynomials satisfying second order difference equations with the pure
{\em imaginary shifts} (idQM) as well as the {\em real shifts} (rdQM).
Sometimes these polynomials are said to have the {\em bi-spectral property}
\cite{gh}. They satisfy two types of spectral conditions, the difference
eigenvalue (Schr\"{o}dinger) equation \eqref{polyeq} and the three term
recurrence relations \eqref{threeterm}.
Those polynomials satisfying the difference equations with the real shifts
are also called the orthogonal polynomials of a {\em discrete variable\/}
\cite{nikiforov}.
They contain most of the ($q$)-hypergeometric orthogonal polynomials
belonging to the Askey scheme \cite{nikiforov,askey,gasper,ismail,koekswart}.
The power of the discrete QM is demonstrated most eloquently in the
discovery of various infinite families of {\em new orthogonal}
({\em exceptional\/}) {\em  polynomials}.
In less than two years after the discovery of the infinite family of the
exceptional Laguerre and Jacobi polynomials \cite{os16,os19,hos}, the
generic counterparts in dQM, the exceptional Wilson, Askey-Wilson, Racah
and $q$-Racah polynomials are constructed by the present authors
\cite{os17,os23}. These new orthogonal polynomials are expected to play
a significant role in mathematics, mathematical physics and related
disciplines.
The exceptional Jacobi polynomials
provide infinitely many global solutions of Fuchsian differential
equations with more than three and arbitrarily many regular singularities
\cite{hos}, although the locations of the extra singularities are rigidly
specified.

This topical review of dQM provides a comprehensive overview of the subjects
related to the {\em exactly and quasi-exactly solvable} quantum particle
dynamics \cite{os4}--\cite{os22}.
Due to the length constraints, we have to concentrate on the systems of
single degree of freedom, which is the most basic and best established
part of the theory.
The subjects to be covered in this article are the adaptation to dQM of
the methods and concepts of oQM accumulated over 80 years after the birth
of quantum mechanics.
After the general setting of the discrete quantum mechanics with the pure
imaginary shifts (idQM) as well as the real shifts (rdQM), we start with
the {\em factorised} Hamiltonians and the Schr\"{o}dinger equations
\cite{os4,os12,os13}.
The general structure of the solution spaces is explored by the
intertwining relations and Crum's theorem together with its modification
\cite{crum}--\cite{adler}, \cite{os15}--\cite{os22}.
Exact solvability in the Schr\"{o}dinger picture is explained by the
{\em shape invariance} \cite{genden,dks,os4,os12,os13}.
The {\em generic eigenvalue formula}, {\em unified Rodrigues formulas}
and the {\em forward/backward shift operators} are deduced.
The solvability in the Heisenberg picture is derived based on the
{\em closure relation} between the {\em sinusoidal coordinate} and the
Hamiltonian \cite{os7,os8}.
The {\em creation/annihilation} operators are introduced and their connection
with the {\em three term recurrence relations} of the orthogonal polynomials
is emphasised.
The {\em dynamical symmetry algebra} generated by the Hamiltonian and the
creation/annihilation operators are also established for all the solvable
systems in the Heisenberg picture \cite{os13,os14}. This includes the
dynamical realisation of the $q$-oscillator algebras \cite{os11}.
For the rdQM, the {\em dual pair} of orthogonal polynomials (the Leonard
pair \cite{leonard,terw}) and the related Askey-Wilson algebras
\cite{zhedanov} are explored.
The unified theory of exactly and quasi-exactly solvable dQM is another
major theme of this review. Based on the fundamental properties of various
sinusoidal coordinates, a simple recipe to construct a solvable dQM
Hamiltonian is provided \cite{os14}.
It generates all known exactly solvable theories and many new ones.
As a byproduct it also gives a recipe of constructing quasi-exactly
solvable dQM Hamiltonians.
The final major subject of the review is the new (exceptional)
orthogonal 
 polynomials \cite{os16}--\cite{os23},\cite{os18}--\cite{os21}.
 
Throughout the review we stress the similarity and differences among oQM,
idQM and rdQM. As far as possible we adopt the same notation and present
one all-inclusive formula for the three categories, oQM, idQM and rdQM.
The differences are exhibited by explicit examples.
We usually adopt three examples from each group denoted by the name of
the corresponding eigenpolynomials with an abbreviation.
For the oQM, they are the Hermite (H), the Laguerre (L) and the Jacobi (J)
polynomials.
For the idQM, we choose the special case of the Meixner-Pollaczek (MP)
(the parameter $\phi$ is fixed to $\phi=\frac{\pi}{2}$),
the Wilson (W) and the Askey-Wilson (AW) polynomials.
For the rdQM, the Meixner (M), the Racah (R) and the $q$-Racah ($q$R)
polynomials are selected.
In some cases, we present only one representative from each group.
They are usually the most generic ones, J, AW and $q$R or the simplest
ones H, MP and M.
For the full line-ups, the original references should be consulted
\cite{os12,os13,os14}. We will provide cross references to the original
articles with the section or the equation numbers.
 
This paper is organised as follows.
In section two various concepts and methods in dQM are explained.
These are: (a) the factorised Hamiltonians \eqref{factHam}--\eqref{rdQMHam}
\& the `Schr\"{o}dinger equations' for the eigenpolynomials
\eqref{polyeq}--\eqref{HtrdQM} together with the groundstate eigenfunctions
\eqref{phi0idQM}, \eqref{dicsrete_phi0} as the orthogonality weight
functions \eqref{ortho1}--\eqref{ortho2}.
(b) the intertwining relations connecting various Hamiltonians \eqref{intw1},
\eqref{intw12}--\eqref{intw21}, and eigenfunctions \eqref{phi01},
\eqref{phi12}, \eqref{phiss1}. This is a brief summary of Crum's theory
in our language. Its deformation \`a la Krein-Adler is also mentioned.
(c) shape invariance \eqref{shapeinv} and the full energy spectrum
\eqref{shapeenery} and the unified Rodrigues type formula
\eqref{phin=A..Aphi0} for all the eigenfunctions.
(d) solvability in the Heisenberg picture \eqref{quantsol} and the closure
relation \eqref{closurerel}. The formulas for the creation/annihilation
operators \eqref{apm}--\eqref{apmdefs} are given.
(e) dual closure relation \eqref{dualclosurerel} together with its
connection with the Askey-Wilson algebra.
In section three we provide the essence of the unified theory of exact
and quasi-exact solvability in dQM.
In section four the new (exceptional) orthogonal polynomials satisfying
second order differential (difference) equations are explored.
The exceptional Meixner-Pollaczek and the exceptional Meixner polynomials are
new results.
The final section is for a summary and comments.
Basic symbols and definitions are listed in Appendix.
A substantially shorter version of the present review is published recently
\cite{rsbull}.
We apologise to all the authors whose good  works could not be referred to
in the review due to the lack of space.

%%%%%%%%%%%%%%%%%%%%%%%%%%%%%%%%%%%%%%%%%%%%%%%%%%%%%%%%%%%%%%%
%                                                             %
%  2. Discrete Quantum Mechanics                              %
%                                                             %
%%%%%%%%%%%%%%%%%%%%%%%%%%%%%%%%%%%%%%%%%%%%%%%%%%%%%%%%%%%%%%%
\section{Discrete Quantum Mechanics}
\label{dQM}
\setcounter{equation}{0}

The dynamical variables of one-dimensional QM are the coordinate $x$ and
its conjugate momentum $p$, which is realised as a differential operator
$p=-i\hbar\frac{d}{dx}\equiv -i\hbar\partial_x$.
Hereafter we adopt the convention $\hbar=1$.
The dQM is a generalisation of oQM in which the Schr\"{o}dinger equation is
a difference equation instead of differential in the oQM
\cite{os4}--\cite{os7}, \cite{os12,os13}.
In other words, the Hamiltonian contains the momentum operator in
exponentiated forms $e^{\pm\beta p}$ which work as shift operators on
the wavefunction
\begin{equation}
  e^{\pm\beta p}\psi(x)=\psi(x\mp i\beta).
\end{equation}
For the two choices of the parameter $\beta$, either {\em real\/} or
pure {\em imaginary\/}, we have two types of dQM; with (\romannumeral1)
pure imaginary shifts $\beta=\gamma\in\mathbb{R}_{\neq0}$ (idQM),
or (\romannumeral2) real shifts $\beta=i$ (rdQM), respectively.
In the case of idQM, $\psi(x\mp i\gamma)$, we require the wavefunction
and potential functions etc to be {\em analytic\/} functions of $x$ with
their  domains including the real axis or a part of it on which the
dynamical variable $x$ is defined. In contrast, in the rdQM, the
difference equation gives constraints on wavefunctions only on equally
spaced lattice points. Then we choose, after proper rescaling, the
variable $x$ to take value of non-negative {\em integers\/}, with the
total number
either finite ($N+1$) or infinite ($x_{\text{max}}=N$ or $\infty$).
To sum up, the dynamical variable $x$ of the one dimensional dQM takes
continuous or discrete values:
\begin{equation}
  \text{idQM}:\ x\in\mathbb{R},\quad x\in(x_1,x_2)\ ;\qquad
% \label{xconti}
  \text{rdQM}:\  x\in\mathbb{Z}_{\ge0},\quad
  x\in[0,x_{\text{max}}].
  \label{xdiscr}
\end{equation}
Here $x_1$, $x_2$ may be finite, $-\infty$ or $+\infty$. Correspondingly,
the inner product of the wavefunctions has the following form:
\begin{equation}
  \text{idQM}:\ (f,g)=\int_{x_1}^{x_2}f(x)^*g(x)dx\ ;\qquad
  \text{rdQM}:\ (f,g)=\sum_{x=0}^{x_{\text{max}}}f(x)^*g(x),
  \label{inn_pro}
\end{equation}
and the norm of $f(x)$ is $|\!|f|\!|=\sqrt{(f,f)}$.

\bigskip
We will consider the Hamiltonians having a finite (rdQM with finite $N$)
or semi-infinite number of {\em discrete energy levels only}:
\begin{equation}
  0=\mathcal{E}(0)<\mathcal{E}(1)<\mathcal{E}(2)<\cdots.
  \label{positivesemi}
\end{equation}
The additive constant of the Hamiltonian is so chosen that the groundstate
energy vanishes. That is, the Hamiltonian is {\em positive semi-definite}.
It is a well known theorem in linear algebra that any positive
semi-definite hermitian matrix can be factorised as a product of a certain
matrix, say $\mathcal{A}$, and its hermitian conjugate $\mathcal{A^\dagger}$.
As we will see shortly, the Hamiltonians we consider  always have
factorised forms in one-dimension as well as in higher dimensions.

%%%%%%%%%%%%%%%%%%%%%%%%%%%%%%%%%%%%%%%%%%%
%                                         %
% 2.1 Factorised Hamiltonian              %
%                                         %
%%%%%%%%%%%%%%%%%%%%%%%%%%%%%%%%%%%%%%%%%%%
%\subsection{Factorised Hamiltonian and Strategy}
\subsection{Factorised Hamiltonian}
\label{sec:H}

The Hamiltonian we consider has a simple factorised form
\cite{infhull}
\begin{equation}
  \mathcal{H}\eqdef\mathcal{A}^\dagger \mathcal{A}\quad \text{or}\quad
  \mathcal{H}\eqdef\sum_{j=1}^D\mathcal{A}_j^\dagger\mathcal{A}_j\quad
  \text{in }D\ \text{dimensions}.
  \label{factHam}
\end{equation}
The operators $\mathcal{A}$ and $\mathcal{A}^\dagger$ in one dimensional
QM are:
\begin{align}
  \text{oQM}:\quad&
  \mathcal{A}\eqdef\frac{d}{dx}-\frac{dw(x)}{dx},\quad 
  \mathcal{A}^\dagger=-\frac{d}{dx}-\frac{dw(x)}{dx},\quad
  w(x)\in\mathbb{R},\quad\phi_0(x)=e^{w(x)},
  \label{oAdef}\\
  &\mathcal{H}=p^2+U(x),\quad
  U(x)\eqdef\bigl(\partial_x w(x)\bigr)^2+\partial_x^2w(x),
  \label{oUdef}\\
  \text{idQM}:\quad&
  \mathcal{A}\eqdef i\bigl(e^{\frac{\gamma}{2}p}\sqrt{V^*(x)}
  -e^{-\frac{\gamma}{2}p}\sqrt{V(x)}\,\bigr),\qquad
  \gamma\in\mathbb{R}_{\neq 0},\n
  &\mathcal{A}^{\dagger}=-i\bigl(\sqrt{V(x)}\,e^{\frac{\gamma}{2}p}
  -\sqrt{V^*(x)}\,e^{-\frac{\gamma}{2}p}\bigr),\quad
  V(x),V^*(x)\in \mathbb{C},
  \label{VVsdef}\\
  &\mathcal{H}=\sqrt{V(x)}e^{\gamma p}\sqrt{V^*(x)}+\sqrt{V^*(x)}
  e^{-\gamma p}\sqrt{V(x)}-V(x)-V^*(x),\\
  \text{rdQM}:\quad&
  \mathcal{A}\eqdef\sqrt{B(x)}-e^{\partial}\sqrt{D(x)},\quad
  \mathcal{A}^{\dagger}=\sqrt{B(x)}-\sqrt{D(x)}\,e^{-\partial},
  \label{ArdQM}\\
  &D(x)>0\ (\text{for $x>0$}),\ \ D(0)=0,\n
  &B(x)>0\ (\text{for $x\geq 0$}),\ \ B(N)=0\ (\text{for the finite case}),
  \label{BDbound}\\
  &\mathcal{H}=-\sqrt{B(x)}e^\partial\sqrt{D(x)}
  -\sqrt{D(x)}e^{-\partial}\sqrt{B(x)}+B(x)+D(x).
  \label{rdQMHam}
\end{align}
Here $w(x)$ is called {\em prepotential}. The function $V^*(x)$ in idQM
is an analytic function of $x$ obtained from $V(x)$ by the $*$-operation,
which is defined as follows.
If $f(x)=\sum\limits_{n}a_nx^n$, $a_n\in\mathbb{C}$, then
$f^*(x)\eqdef\sum\limits_{n}a_n^*x^n$, in which $a_n^*$ is the complex
conjugation of $a_n$. Obviously $f^{**}(x)=f(x)$ and $f(x)^*=f^*(x^*)$.
If a function satisfies $f^*=f$, then it takes real values on the real line.
To the best of our knowledge, the factorised Hamiltonian approach to dQM
is our invention.
The existing factorisation methods for dQM are not for the Hamiltonian
$\mathcal{H}$, but for Hamiltonian minus an eigenvalue
$\mathcal{H}-\lambda(n)$ \cite{discrete,spivinzhed}.
The condition $D(0)=0$ in rdQM \eqref{BDbound} is necessary for the term
$\psi(-1)$ not to appear in $\mathcal{H}\psi(0)$. Likewise $B(N)=0$ is
necessary for the finite case. Roughly speaking these correspond to the
regular singularities in oQM.
The Hamiltonians of idQM and rdQM can be written in a unified notation:
\begin{align}
  \text{dQM}:\quad&
  \mathcal{H}=\varepsilon\Bigl(
  \sqrt{V_+(x)}\,e^{\beta p}\sqrt{V_-(x)}
  +\!\sqrt{V_-(x)}\,e^{-\beta p}\sqrt{V_+(x)}
  -V_+(x)-V_-(x)\Bigr),\\
  &\hspace*{-5mm}\text{idQM}:\quad
  \beta=\gamma,\quad\,\varepsilon=1,\quad\ \,V_+(x)=V(x),
  \quad V_-(x)=V^*(x),\n
  &\hspace*{-5mm}\text{rdQM}:\quad
  \beta=i,\quad\varepsilon=-1,\quad V_+(x)=B(x),\quad V_-(x)=D(x).
  \label{Vpmdefs}
\end{align}

Multi-particle exactly solvable systems can be constructed in a similar way.
The prepotential approach is also useful in Calogero-Moser systems
\cite{Cal-Sut} in oQM \cite{bms,kps}:
\begin{equation}
  \text{oQM}:\quad
  \mathcal{A}_j\eqdef\frac{\partial}{\partial x_j}
  -\frac{\partial w(x)}{\partial x_j},
  \ \ \mathcal{A}_j^\dagger=-\frac{\partial}{\partial x_j}
  -\frac{\partial w(x)}{\partial x_j}\ \ (j=1,\ldots,D),\quad
  \phi_0(x)=e^{w(x)}.
  \label{oAdefj}
\end{equation}
We also adopt the factorised Hamiltonian for the discrete analogues of
the Calogero-Moser systems:
\begin{align}
  \text{idQM}:\quad&
  \mathcal{A}_j\eqdef i\bigl(e^{\frac{\gamma}{2}p_j}\sqrt{V_j^*(x)}
  -e^{-\frac{\gamma}{2}p_j}\sqrt{V_j(x)}\,\bigr),\quad
   V_j(x),V_j^*(x)\in\mathbb{C},\n
  &\mathcal{A}_j^{\dagger}=-i\bigl(\sqrt{V_j(x)}\,e^{\frac{\gamma}{2}p_j}
  -\sqrt{V_j^*(x)}\,e^{-\frac{\gamma}{2}p_j}\bigr)\ \ (j=1,\ldots, D),
  \label{VVsdefj}
\end{align}
which is markedly different from the approach of Ruijsenaars et al
\cite{RSvD}. The multiparticle version of rdQM is now 
being developed.

The Schr\"{o}dinger equation
\begin{equation}
  \mathcal{H}\phi_n(x)=\mathcal{E}(n)\phi_n(x)
  \ \ (n=0,1,2,\ldots),\quad
  0=\mathcal{E}(0)<\mathcal{E}(1)<\mathcal{E}(2)<\cdots,\quad
  \label{Sch_eq}
\end{equation}
is a second order differential (oQM) or difference equation (dQM) and
the {\em groundstate wavefunction} $\phi_0(x)$ is determined as a zero
mode of the operator $\mathcal{A}$ ($\mathcal{A}_j$) which is a first
order equation:
\begin{equation}
  \mathcal{A}\phi_0(x)=0\quad(\mathcal{A}_j\phi_0(x)=0,\ j=1,\ldots,D)\quad
  \Rightarrow\ \mathcal{H}\phi_0(x)=0.
  \label{Aphi0=0}
\end{equation}
By definition all the eigenfunctions are required to have a finite positive
norm:
\begin{equation}
  0<(\phi_n,\phi_n)<\infty.
  \label{finnorm}
\end{equation}
Throughout this paper we look for the eigenfunctions in a factorised form:
\begin{equation}
  \phi_n(x)=\phi_0(x)P_n(\eta(x)),
  \label{facteig}
\end{equation}
in which $P_n(\eta(x))$ is a degree $n$ (except for those discussed in
\S\ref{sec:Exce}) polynomial in the {\em sinusoidal coordinate} $\eta(x)$.
However, some basic results, such as  the (modified) Crum's theorem in
\S\,\ref{sec:Inter}--\ref{sec:ModCrum} and the shape invariance in
\S\,\ref{sec:Sha}  hold without this assumption \eqref{facteig}.
The explicit forms of the eigenfunctions will be given for various examples
\eqref{ex1}--\eqref{ex9}.
Here we will briefly introduce their general property for dQM.
The sinusoidal coordinate $\eta(x)$ in dQM is a {\em monotone increasing
function} of $x$ with the initial (boundary) condition
\begin{equation}
  \text{dQM}:\quad \eta(0)=0.
  \label{etabound}
\end{equation}
With this, we also normalise the eigenpolynomials $\{P_n(\eta)\}$ in rdQM as:
\begin{equation}
  \text{rdQM}:\quad P_n(0)=1\ \ (n=0,1,\ldots).
  \label{univnorm}
\end{equation}
In the multi-particle case $n$ will be a multi-index.
Here we require $\phi_0(x)$ to be chosen real and positive for the
physical values of $x$, which is always guaranteed in oQM.
One important distinction between a differential and a difference equation
is the {\em uniqueness} of the solution. For a linear differential equation,
the solution is unique when the initial conditions are specified.
In contrast, a solution of a difference Schr\"{o}dinger equation
\eqref{Sch_eq} multiplied by any periodic function of a period $i\beta$
($i\gamma$ for idQM, $1$ for rdQM) is another solution.
In the case of idQM, this non-uniqueness is removed by requiring the
finite norm condition for the groundstate wavefunction $\phi_0$,
$(\phi_0,\phi_0)<\infty$ and the hermiticity (self-adjointness)
condition of the Hamiltonian. For details, see \cite{os13} Appendix A.
For the rdQM, the periodic ambiguity of period 1 is harmless since the
values of eigenfunctions on the integer lattice points only count for
the inner products, etc.
It is also important to realise that the periodic ambiguity does not appear
in the polynomial solutions of \eqref{polyeq}.

It is important to stress that the second order equation for $P_n(\eta(x))$
is {\em square root free}:
\begin{equation}
 \widetilde{\mathcal{H}}P_n(\eta(x))=\mathcal{E}(n)P_n(\eta(x)).
 \label{polyeq}
\end{equation}
In other words, the similarity transformed Hamiltonian
$\widetilde{\mathcal{H}}$ in terms of the groundstate wavefunction
$\phi_0(x)$ has a much simpler form than the original Hamiltonian
$\mathcal{H}$:
\begin{align}
  &\widetilde{\mathcal{H}}\eqdef
  \phi_0(x)^{-1}\circ\mathcal{H}\circ\phi_0(x),
  \label{Htdef}\\
  &\text{oQM}:\quad
  \widetilde{\mathcal{H}}=-\frac{d^2}{dx^2}-2\frac{dw(x)}{dx}\frac{d}{dx},\\
  &\text{dQM}:\quad
  \widetilde{\mathcal{H}}
  =\varepsilon\Bigl(V_+(x)(e^{\beta p}-1)+V_-(x)(e^{-\beta p}-1)\Bigr)
  \label{HtdQM}\\
  &\phantom{\text{dQM}:\quad\widetilde{\mathcal{H}}}
  =\left\{\begin{array}{ll}
  V(x)(e^{\gamma p}-1)+V^*(x)(e^{-\gamma p}-1)&: \text{idQM}\\[2pt]
  B(x)(1-e^{\partial})+D(x)(1-e^{-\partial})&: \text{rdQM}
  \end{array}\right..
  \label{HtrdQM}
\end{align}
For all the examples discussed in this section, $\widetilde{\mathcal{H}}$
is {\em lower triangular\/}
\begin{equation}
  \widetilde{\mathcal{H}}\eta(x)^n
  =\mathcal{E}(n)\eta(x)^n+\text{lower orders in}\ \eta(x),
  \label{lowtri}
\end{equation}
in the special basis
\[
  1,\ \eta(x),\ \eta(x)^2,\ \ldots, \eta(x)^n,\ \ldots,
\]
spanned by the sinusoidal coordinate $\eta(x)$.
This situation is expressed as
\begin{equation}
  \widetilde{\mathcal{H}}\mathcal{V}_n\subseteq\mathcal{V}_n,\quad
  \mathcal{V}_n\eqdef
  \text{Span}\bigl[1,\eta(x),\ldots,\eta(x)^n\bigr].
  \label{Vndef}
\end{equation}

Here are some explicit examples.
For the oQM the {\em prepotential} $w(x)$ determines the potential $U(x)$
of the Hamiltonian $\mathcal{H}=p^2+U(x)$,
$U(x)\eqdef\bigl(\partial_x w(x)\bigr)^2+\partial_x^2w(x)$:
\begin{align}
  \text{H}:\quad&w(x)=-\tfrac12x^2,\hspace{26mm}-\infty<x<\infty,\n
  &U(x)=x^2-1,\hspace{25mm}\eta(x)=x,
  \label{ex1}\\
  \text{L}:\quad&w(x)=-\tfrac12x^2+g\log x,
  \hspace{24mm}g>0,\ \ 0<x<\infty,\n
  &U(x)=x^2+\frac{g(g-1)}{x^2}-1-2g,\qquad \eta(x)=x^2,
  \label{ex2}\\
  \text{J}:\quad&w(x)=g\log\sin x+h\log\cos x,
  \hspace{21mm}g>0,\ h>0,\ \ 0<x<\tfrac{\pi}{2},\n
  &U(x)=\frac{g(g-1)}{\sin^2x}+\frac{h(h-1)}{\cos^2x}-(g+h)^2,
  \quad\eta(x)=\cos2x.
  \label{ex3}
\end{align}
Let us note that $x=0$ for L and $x=0,\frac{\pi}{2}$ for J are the
{\em regular singular points} of the Fuchsian differential equations.
The monodromy at the regular singular point is determined by the
{\em characteristic exponent} $\rho$:
\begin{equation}
  M_\rho=e^{2\pi i\rho}.
  \label{monodromy}
\end{equation}
The corresponding exponents are $\rho=g$, $1-g$ for L and $\rho=g,1-g$
and $\rho=h,1-h$ for J.\\
For the idQM ($0<q<1$):
\begin{align}
  \text{MP}:\quad&V(x)=a+ix,\qquad a>0,\quad-\infty<x<\infty,\quad
  \gamma=1,\quad\eta(x)=x,
  \label{ex4}\\
  \text{W}:\quad&V(x)=\frac{\prod_{j=1}^4(a_j+ix)}{2ix(2ix+1)},\qquad
  \text{Re}(a_j)>0,\quad 0<x<\infty,\quad\gamma=1,\n
  &\eta(x)=x^2,\qquad
  \{a_1^*,a_2^*,a_3^*,a_4^*\}=\{a_1,a_2,a_3,a_4\}\ \ (\text{as a set}),\\
  \text{AW}:\quad&V(x)=\frac{\prod_{j=1}^4(1-a_je^{ix})}
  {(1-e^{2ix})(1-q\,e^{2ix})},\qquad |a_j|<1,\quad 0<x<\pi,\quad
  \gamma=\log q,\n
  &\eta(x)=1-\cos x,\qquad
  \{a_1^*,a_2^*,a_3^*,a_4^*\}=\{a_1,a_2,a_3,a_4\}\ \ (\text{as a set}).
  \label{ex6}
\end{align}
For the rdQM ($0<q<1$):
\begin{align}
  \text{M}:\quad&B(x)=\frac{c}{1-c}(x+\beta),\quad
  D(x)=\frac{1}{1-c}x,\quad\beta>0,\quad 0<c<1,\n
  &\eta(x)=x,\quad x_{\text{max}}=\infty,
  \label{ex7}\\
  \text{R}:\quad&B(x)=-\frac{(x+a)(x+b)(x+c)(x+d)}{(2x+d)(2x+1+d)},\n
  &D(x)=-\frac{(x+d-a)(x+d-b)(x+d-c)x}{(2x-1+d)(2x+d)},\quad
  \tilde{d}\eqdef a+b+c-d-1,\n
  &a=-N,\quad a+b>d>0,\quad 0<c<1+d,\n
  &\eta(x)=x(x+d),\quad x_{\text{max}}=N,
  \label{ex8}\\
  \text{$q$R}:\quad&B(x)=-\frac{(1-aq^x)(1-bq^x)(1-cq^x)(1-dq^x)}
  {(1-dq^{2x})(1-dq^{2x+1})},\n
  &D(x)=-\tilde{d}\,\frac{(1-a^{-1}dq^x)(1-b^{-1}dq^x)(1-c^{-1}dq^x)(1-q^x)}
  {(1-dq^{2x-1})(1-dq^{2x})},\quad
  \tilde{d}\eqdef abcd^{-1}q^{-1},\n
  &a=q^{-N},\quad 0<ab<d<1,\quad qd<c<1,\n
  &\eta(x)=(q^{-x}-1)(1-dq^x),\quad x_{\text{max}}=N.
  \label{ex9}
\end{align}
The R and $q$R cases admit four different parametrisations specified by
$\epsilon,\epsilon'=\pm 1$, see \cite{os12}. Here we present
$\epsilon=\epsilon'=1$ case only.
It should be emphasised that the sinusoidal coordinates $\eta(x)$ in rdQM
in general depends on the parameters in contradistinction to those of the
oQM or idQM.

The eigenfunctions of these examples have the form \eqref{facteig}.
The eigenpolynomials $P_n(\eta(x))$ are the following
(for notation, see \cite{koekswart}):\\
oQM :
\begin{align}
  \text{H}:\quad&
  P_n(\eta(x))=H_n(x)\eqdef(2x)^n
  {}_2F_0\Bigl(\genfrac{}{}{0pt}{}{-\frac{n}{2},\,-\frac{n-1}{2}}
  {-}\Bigm|-\frac{1}{x^2}\Bigr),
  \label{defH}\\
  \text{L}:\quad&
  P_n(\eta(x))=L_n^{(g-\frac12)}(x^2)
  \eqdef\frac{(g+\frac12)_n}{n!}
  {}_1F_1\Bigl(\genfrac{}{}{0pt}{}{-n}{g+\frac12}\Bigm|x^2\Bigr),
  \label{defL}\\
  \text{J}:\quad&
   P_n(\eta(x))=P_n^{(g-\frac12,h-\frac12)}(\cos 2x)
  \eqdef\frac{(g+\frac12)_n}{n!}
  {}_2F_1\Bigl(\genfrac{}{}{0pt}{}{-n,n+g+h}{g+\frac12}\Bigm|
  \frac{1-\cos2x}{2}\Bigr),
  \label{defJ}
\end{align}
idQM :
\begin{align}
  \text{MP}:\ \ &
  P_n(\eta(x))=P_n^{(a)}(x;\tfrac{\pi}{2})
  \eqdef\frac{(2a)_n}{n!}\,i^n
  {}_2F_1\Bigl(\genfrac{}{}{0pt}{}{-n,\,a+ix}{2a}\Bigm|2\Bigr),
  \label{defMP}\\
  \text{W}:\ \ &
  P_n(\eta(x))=W_n(x^2\,;a_1,a_2,a_3,a_4)
  \qquad(b_1\eqdef a_1+a_2+a_3+a_4)\n
  &\ \ \eqdef(a_1+a_2)_n(a_1+a_3)_n(a_1+a_4)_n\,
  {}_4F_3\Bigl(
  \genfrac{}{}{0pt}{}{-n,\,n+b_1-1,\,a_1+ix,\,a_1-ix}
  {a_1+a_2,\,a_1+a_3,\,a_1+a_4}\Bigm|1\Bigr),\!\!
  \label{defW}\\
  \text{AW}:\ \ &
  P_n(\eta(x))=p_n(\cos x\,;a_1,a_2,a_3,a_4|q)
  \qquad(b_4\eqdef a_1a_2a_3a_4)\n
  &\ \ \eqdef a_1^{-n}(a_1a_2,a_1a_3,a_1a_4\,;q)_n\,
  {}_4\phi_3\Bigl(\genfrac{}{}{0pt}{}{q^{-n},\,b_4q^{n-1},\,
  a_1e^{ix},\,a_1e^{-ix}}{a_1a_2,\,a_1a_3,\,a_1a_4}\!\!\Bigm|\!q\,;q\Bigr),
  \label{defAW}
\end{align}
rdQM :
\begin{align}
  \text{M}:\ \ &
  P_n(\eta(x))=M_n(x\,;\beta,c)
  \eqdef{}_2F_1\Bigl(\genfrac{}{}{0pt}{}{-n,\,-x}{\beta}\Bigm|1-c^{-1}\Bigr),
  \label{defM}\\
  \text{R}:\ \ &
  P_n(\eta(x))=R_n(x(x+d)\,;a-1,\tilde{d}-a,c-1,d-c)
  \eqdef{}_4F_3\Bigl(\genfrac{}{}{0pt}{}{-n,\,n+\tilde{d},\,-x,\,x+d}
  {a,\,b,\,c}\Bigm|1\Bigr),
  \label{defR}\\
  \text{$q$R}:\ \ &
  P_n(\eta(x))
  =R_n(q^{-x}+dq^x\,;aq^{-1},\tilde{d}a^{-1},cq^{-1},dc^{-1}|q)
  \eqdef{}_4\phi_3\Bigl(
  \genfrac{}{}{0pt}{}{q^{-n},\,\tilde{d}q^n,\,q^{-x},\,dq^x}
  {a,\,b,\,c}\Bigm|q\,;q\Bigr).
  \label{defqR}
 \end{align}

Obviously, the square of the groundstate wavefunction $\phi_0(x)^2$ provides
the positive definite orthogonality weight function for the polynomials:
\begin{align}
  \text{oQM, idQM}:\quad&
  \int_{x_1}^{x_2}\!\!\phi_0(x)^2P_n(\eta(x))P_n(\eta(x))dx=h_n\delta_{nm},
  \label{ortho1}\\
  \text{rdQM}:\qquad\quad\ \ &
  \sum_{x=0}^{x_{\text{max}}}\phi_0(x)^2P_n(\eta(x))P_n(\eta(x))
  =\frac{1}{d_n^2} \delta_{n\,m}.
  \label{ortho2}
\end{align}
Let us emphasise that the weight function, or $\phi_0(x)$ is determined as a
solution of a first order differential (difference) equation \eqref{Aphi0=0},
without recourse to a {\em moment problem}.
This situation becomes crucially important when various deformations of
orthogonal polynomials are considered.
For the oQM, the weight function is simply given by the prepotential
$\phi_0(x)^2=e^{2w(x)}$ and the normalisation constants are
\begin{equation}
  h_n=\left\{\begin{array}{ll}
  2^nn!\sqrt{\pi}&:\text{H}\\
  \frac{1}{2\,n!}\,\Gamma(n+g+\tfrac12)&:\text{L}\\[2pt]
  {\displaystyle
  \frac{\Gamma(n+g+\frac12)\Gamma(n+h+\frac12)}
  {2\,n!\,(2n+g+h)\Gamma(n+g+h)}}&:\text{J}
  \end{array}\right..
  \label{hnoQM}
\end{equation}
The explicit forms of the squared groundstate wavefunction (weight
function) $\phi_0(x)^2$ and the normalisation constants $h_n$ for the
above examples in pure imaginary shifts dQM are:
\begin{align}
  \phi_0(x)^2&=\left\{
  \begin{array}{ll}
  \displaystyle{\Gamma(a+ix)\Gamma(a-ix)}
  &:\text{MP}\\
  \displaystyle{
  \bigl(\Gamma(2ix)\Gamma(-2ix)\bigr)^{-1}
  \prod_{j=1}^4\Gamma(a_j+ix)\Gamma(a_j-ix)}
  &:\text{W}\\
  \displaystyle{
  (e^{2ix}\,;q)_{\infty}(e^{-2ix}\,;q)_{\infty}\prod_{j=1}^4
  \bigl((a_je^{ix}\,;q)_{\infty}(a_je^{-ix}\,;q)_{\infty}\bigr)^{-1}}
  &:\text{AW}
  \end{array}\right.,
  \label{phi0idQM}\\
  h_n&=\left\{
  \begin{array}{ll}
  \displaystyle{
  2\pi(2^{2a}n!)^{-1}\Gamma(n+2a)}
  &:\text{MP}\\
  \displaystyle{
  2\pi n!\,(n+b_1-1)_n\prod_{1\leq i<j\leq 4}\Gamma(n+a_i+a_j)\cdot
  \Gamma(2n+b_1)^{-1}}
  &:\text{W}\\
  \displaystyle{
  2\pi(b_4q^{n-1};q)_n(b_4q^{2n};q)_{\infty}(q^{n+1};q)_{\infty}^{-1}
  \prod_{1\leq i<j\leq 4}(a_ia_jq^n;q)_{\infty}^{-1}}
  &:\text{AW}
  \end{array}\right..
  \label{hnidQM}
\end{align}
For the rdQM, the zero-mode equation $\mathcal{A}\phi_0(x)=0$ \eqref{Aphi0=0}
is a two term recurrence relation, which can be solved elementarily by using
the boundary condition \eqref{BDbound}:
\begin{equation}
  \phi_0(x)^2=\prod_{y=0}^{x-1}\frac{B(y)}{D(y+1)},\quad\phi_0(0)=1.
  \label{genphi0rd}
\end{equation}
The explicit forms of $\phi_0(x)^2$ and $d_n^2$, which are related by
duality, are:
\begin{align}
  \phi_0(x)^2&=\left\{
  \begin{array}{ll}
  \displaystyle{\frac{(\beta)_x\, c^x}{x!}}
  &:\text{M}\\
  \displaystyle{
  \frac{(a,b,c,d)_x}{(1+d-a,1+d-b,1+d-c,1)_x}\,\frac{2x+d}{d}}
  &:\text{R}\\[8pt]
  \displaystyle{
  \frac{(a,b,c,d\,;q)_x}{(a^{-1}dq,b^{-1}dq,c^{-1}dq,q\,;q)_x\,\tilde{d}^x}\,
  \frac{1-dq^{2x}}{1-d}}
  &:\text{$q$R}
  \end{array}\right.,
  \label{dicsrete_phi0}\\
  d_n^2&=\left\{
  \begin{array}{ll}
  \displaystyle{\frac{(\beta)_n\,c^n}{n!}\times(1-c)^{\beta}}
  &:\text{M}\\
  {\displaystyle
  \frac{(a,b,c,\tilde{d})_n}
  {(1+\tilde{d}-a,1+\tilde{d}-b,1+\tilde{d}-c,1)_n}\,
  \frac{2n+\tilde{d}}{\tilde{d}}}&\\[8pt]
  {\displaystyle
  \quad\times
  \frac{(-1)^N(1+d-a,1+d-b,1+d-c)_N}{(\tilde{d}+1)_N(d+1)_{2N}}}
  &:\text{R}\\[8pt]
  {\displaystyle
  \frac{(a,b,c,\tilde{d}\,;q)_n}
  {(a^{-1}\tilde{d}q,b^{-1}\tilde{d}q,c^{-1}\tilde{d}q,q\,;q)_n\,d^n}\,
  \frac{1-\tilde{d}q^{2n}}{1-\tilde{d}}}&\\[8pt]
  {\displaystyle
  \quad\times
  \frac{(-1)^N(a^{-1}dq,b^{-1}dq,c^{-1}dq\,;q)_N\,\tilde{d}^Nq^{\frac12N(N+1)}}
  {(\tilde{d}q\,;q)_N(dq\,;q)_{2N}}}
  &:\text{$q$R}
  \end{array}\right..
  \label{dnrdQM}
\end{align}
It should be stressed that $\phi_0(x)^2$ for the rdQM \eqref{genphi0rd},
when continued analytically in $x$, vanish on the integer points outside
the defining domain, i.e., $x\in\mathbb{Z}_{<0}$ and
$x\in\mathbb{Z}_{>x_{\text{max}}}$.

To the best of our knowledge, the first example of a factorised idQM
Hamiltonian was mentioned in 2001 eq.(4.28) of \cite{degru} for the
Meixner-Pollaczek polynomial.
But it was not considered as a general principle. 

%%%%%%%%%%%%%%%%%%%%%%%%%%%%%%%%%%%%%%%%%%%%%%
%                                            %
% 2.2 Intertwining Relations: Crum's Theorem %
%                                            %
%%%%%%%%%%%%%%%%%%%%%%%%%%%%%%%%%%%%%%%%%%%%%%
\subsection{Intertwining Relations: Crum's Theorem}
\label{sec:Inter}

The general structure of the intertwining relations works equally well
for the oQM as well as the dQM. Let us denote by $\mathcal{H}^{[0]}$ the
original factorised Hamiltonian and by $\mathcal{H}^{[1]}$ its partner
(associated) Hamiltonian obtained by changing the order of
$\mathcal{A}^\dagger$ and $\mathcal{A}$:
\begin{equation}
  \mathcal{H}^{[0]}\eqdef\mathcal{A}^\dagger\mathcal{A},\qquad
  \mathcal{H}^{[1]}\eqdef\mathcal{A}\mathcal{A}^\dagger.
\end{equation}
One simple and most important consequence of the factorised Hamiltonian
\eqref{factHam} is the {\em intertwining relations\/}:
\begin{align}
  \mathcal{A}\mathcal{H}^{[0]}=\mathcal{A}\mathcal{A}^\dagger\mathcal{A}
  =\mathcal{H}^{[1]}\mathcal{A},\qquad
  \mathcal{A}^\dagger\mathcal{H}^{[1]}
  =\mathcal{A}^\dagger\mathcal{A}\mathcal{A}^\dagger
  =\mathcal{H}^{[0]}\mathcal{A}^\dagger,
  \label{intw1}
\end{align}
which are equally valid in the oQM and the dQM.
The pair of Hamiltonians $\mathcal{H}^{[0]}$ and $\mathcal{H}^{[1]}$ are
essentially {\em iso-spectral\/} and their eigenfunctions
$\{\phi_n^{[0]}(x)\}$ and $\{\phi_n^{[1]}(x)\}$ are related by the
Darboux-Crum transformations \cite{darboux,crum}:
\begin{align}
  &\mathcal{H}^{[0]}\phi_n^{[0]}(x)=\mathcal{E}(n)\phi_n^{[0]}(x)
  \quad(n=0,1,\ldots),\quad\mathcal{A}\phi_0^{[0]}(x)=0,\\
  &\mathcal{H}^{[1]}\phi_n^{[1]}(x)=\mathcal{E}(n)\phi_n^{[1]}(x)
  \quad(n=1,2,\ldots),\\
  &\phi_n^{[1]}(x)=\mathcal{A}\phi_n^{[0]}(x),\quad
  \phi_n^{[0]}(x)=\frac{\mathcal{A}^\dagger}{\mathcal{E}(n)}\phi_n^{[1]}(x)
  \quad(n=1,2,\ldots),
  \label{phi01}\\
  &(\phi_n^{[1]},\phi_m^{[1]})=
%  (\mathcal{A}\phi_n^{[0]},\mathcal{A}\phi_m^{[0]})=
%  (\phi_n^{[0]},\mathcal{A}^\dagger\mathcal{A}\phi_m^{[0]})=
  \mathcal{E}(n)(\phi_n,\phi_m)
  \quad(n,m=1,2,\ldots).
\end{align}
The partner Hamiltonian $\mathcal{H}^{[1]}$ has the lowest eigenvalue
$\mathcal{E}(1)$. If the groundstate energy $\mathcal{E}(1)$ is subtracted
from the partner Hamiltonian $\mathcal{H}^{[1]}$, it is again positive
semi-definite and can be factorised in terms of new operators
$\mathcal{A}^{[1]}$ and $\mathcal{A}^{[1]\dagger}$:
\begin{equation}
  \mathcal{H}^{[1]}=\mathcal{A}^{[1]\dagger}\mathcal{A}^{[1]}
  +\mathcal{E}(1),\quad\mathcal{A}^{[1]}\phi_1^{[1]}(x)=0.
\end{equation}
It should be stressed that in rdQM with a finite $N$, the size of the
Hamiltonian decreases by one, since the lowest eigenstate is removed,
see \cite{os22}.

By changing the orders of $\mathcal{A}^{[1]\dagger}$ and
$\mathcal{A}^{[1]}$, a new Hamiltonian $\mathcal{H}^{[2]}$ is defined:
\begin{equation}
  \mathcal{H}^{[2]}\eqdef\mathcal{A}^{[1]}\mathcal{A}^{[1]\dagger}
  +\mathcal{E}(1).
\end{equation}
These two Hamiltonians are intertwined by $\mathcal{A}^{[1]}$ and
$\mathcal{A}^{[1]\dagger}$:
\begin{align}
  &\mathcal{A}^{[1]}\bigl(\mathcal{H}^{[1]}-\mathcal{E}(1)\bigr)
  =\mathcal{A}^{[1]}\mathcal{A}^{[1]\dagger}\mathcal{A}^{[1]}
  =\bigl(\mathcal{H}^{[2]}-\mathcal{E}(1)\bigr)\mathcal{A}^{[1]},
  \label{intw12}\\
  &\mathcal{A}^{[1]\dagger}\bigl(\mathcal{H}^{[2]}-\mathcal{E}(1)\bigr)
  =\mathcal{A}^{[1]\dagger}\mathcal{A}^{[1]}\mathcal{A}^{[1]\dagger}
  =\bigl(\mathcal{H}^{[1]}-\mathcal{E}(1)\bigr)\mathcal{A}^{[1]\dagger}.
  \label{intw21}
\end{align}
The iso-spectrality of the two Hamiltonians $\mathcal{H}^{[1]}$ and
$\mathcal{H}^{[2]}$ and the relationship among their eigenfunctions
follow as before:
\begin{align}
  &\mathcal{H}^{[2]}\phi_n^{[2]}(x)=\mathcal{E}(n)\phi_n^{[2]}(x)
  \quad(n=2,3,\ldots),\\
  &\phi_n^{[2]}(x)=\mathcal{A}^{[1]}\phi_n^{[1]}(x),\quad
  \phi_n^{[1]}(x)=\frac{\mathcal{A}^{[1]\dagger}}
  {\mathcal{E}(n)-\mathcal{E}(1)}\phi_n^{[2]}(x)
  \quad(n=2,3,\ldots),
  \label{phi12}\\
  &(\phi_n^{[2]},\phi_m^{[2]})=
  \bigl(\mathcal{E}(n)-\mathcal{E}(1)\bigr)(\phi_n^{[1]},\phi_m^{[1]})
  \quad(n,m=2,3,\ldots),\\
  &\mathcal{H}^{[2]}=\mathcal{A}^{[2]\dagger}\mathcal{A}^{[2]}
  +\mathcal{E}(2),\quad\mathcal{A}^{[2]}\phi_2^{[2]}(x)=0.
\end{align}
This process can go on indefinitely by successively deleting the lowest
lying energy level:
\begin{align}
  &\mathcal{H}^{[s]}
  \eqdef\mathcal{A}^{[s-1]}\mathcal{A}^{[s-1]\dagger}+\mathcal{E}(s-1)
  =\mathcal{A}^{[s]\dagger}\mathcal{A}^{[s]}+\mathcal{E}(s),\\
  &\mathcal{H}^{[s]}\phi_n^{[s]}(x)=\mathcal{E}(n)\phi_n^{[s]}(x)
  \quad(n=s,s+1,\ldots),\quad\mathcal{A}^{[s]}\phi_s^{[s]}(x)=0,\\
  &\phi_n^{[s]}(x)=\mathcal{A}^{[s-1]}\phi_n^{[s-1]}(x),\quad
  \phi_n^{[s-1]}(x)=\frac{\mathcal{A}^{[s-1]\dagger}}
  {\mathcal{E}(n)-\mathcal{E}(s-1)}\phi_n^{[s]}(x)
  \quad(n=s,s+1,\ldots),
  \label{phiss1}\\
  &(\phi_n^{[s]},\phi_m^{[s]})=
  \bigl(\mathcal{E}(n)-\mathcal{E}(s-1)\bigr)(\phi_n^{[s-1]},\phi_m^{[s-1]})
  \quad(n,m=s,s+1,\ldots).
\end{align}
The quantities in the $s$-th step are defined by those in the ($s-1$)-st
step: ($s\geq 1$)
\begin{align}
  \text{oQM}:\quad&w^{[s]}(x)\eqdef\log|\phi^{[s]}_s(x)|,
  \label{w[s]}\\
  &\mathcal{A}^{[s]}\eqdef\partial_x-\partial_xw^{[s]}(x),\quad
  \mathcal{A}^{[s]\,\dagger}=-\partial_x-\partial_xw^{[s]}(x),\\
  \text{idQM}:\quad&V^{[s]}(x)\eqdef\sqrt{V^{[s-1]}(x-i\tfrac{\gamma}{2})
  V^{[s-1]*}(x-i\tfrac{\gamma}{2})}
  \ \frac{\phi^{[s]}_s(x-i\gamma)}{\phi^{[s]}_s(x)},
  \label{Vsformsq}\\
  &\mathcal{A}^{[s]}\eqdef i\bigl(e^{\frac{\gamma}{2}p}\sqrt{V^{[s]*}(x)}
  -e^{-\frac{\gamma}{2}p}\sqrt{V^{[s]}(x)}\bigr),\n
  &\mathcal{A}^{[s]\,\dagger}
  =-i\bigl(\sqrt{V^{[s]}(x)}\,e^{\frac{\gamma}{2}p}
  -\sqrt{V^{[s]*}(x)}\,e^{-\frac{\gamma}{2}p}\bigr),\\
  \text{rdQM}:\quad&B^{[s]}(x)\eqdef\sqrt{B^{[s-1]}(x+1)D^{[s-1]}(x+1)}\,
  \frac{\phi_s^{[s]}(x+1)}{\phi_s^{[s]}(x)},\\
  &D^{[s]}(x)\eqdef\sqrt{B^{[s-1]}(x)D^{[s-1]}(x)}\,
  \frac{\phi_s^{[s]}(x-1)}{\phi_s^{[s]}(x)},\\
  &\mathcal{A}^{[s]}\eqdef\sqrt{B^{[s]}(x)}-e^{\partial}\sqrt{D^{[s]}(x)},
  \quad
  \mathcal{A}^{[s]\dagger}=\sqrt{B^{[s]}(x)}-\sqrt{D^{[s]}(x)}\,e^{-\partial}.
\end{align}
The eigenfunctions at the $s$-th step have succinct {\em determinant forms}
in terms of the Wronskian (oQM) and the Casoratian (dQM)
\cite{crum,matveev,os15,os22}: ($n\geq s\geq 0$)
\begin{align}
  \text{oQM}:\quad&\text{W}\,[f_1,\ldots,f_m](x)
  \eqdef\det\Bigl(\frac{d^{j-1}f_k(x)}{dx^{j-1}}\Bigr)_{1\leq j,k\leq m}
  \quad(\text{Wronskian}),
  \label{wron}\\
  &\phi^{[s]}_n(x)=
  \frac{\text{W}\,[\phi_0,\phi_1,\ldots,\phi_{s-1},\phi_n](x)}
  {\text{W}\,[\phi_0,\phi_1,\ldots,\phi_{s-1}](x)},
  \label{crumwronsfin}\\
  \text{idQM}:\quad&\text{W}_{\gamma}[f_1,\ldots,f_m](x)
  \eqdef i^{\frac12m(m-1)}
  \det\Bigl(f_k(x+i\tfrac{m+1-2j}{2}\gamma)\Bigr)_{1\leq j,k\leq m}
  \ (\text{Casoratian}),\!\!\\
  &\phi^{[s]}_n(x)=\prod_{j=0}^{s-1}\sqrt{V^{[j]}(x+i\tfrac{s-j}{2}\gamma)}
  \cdot\frac{\text{W}_{\gamma}[\phi_0,\phi_1,\ldots,\phi_{s-1},\phi_n](x)}
  {\text{W}_{\gamma}[\phi_0,\phi_1,\ldots,\phi_{s-1}](x-i\frac{\gamma}{2})},\\
  \text{rdQM}:\quad&\text{W}_C[f_1,\ldots,f_m](x)
  \eqdef\det\Bigl(f_k(x+j-1)\Bigr)_{1\leq j,k\leq m}\quad
  (\text{Casoratian}),\\
  &\phi_n^{[s]}(x)=(-1)^s\prod_{k=0}^{s-1}\sqrt{B^{[k]}(x)}\cdot
  \frac{\text{W}_C[\phi_0,\phi_1,\ldots,\phi_{s-1},\phi_n](x)}
  {\text{W}_C[\phi_0,\phi_1,\ldots,\phi_{s-1}](x+1)}\\
  &\phantom{\phi_n^{[s]}(x)}
  =(-1)^s\prod_{k=0}^{s-1}\sqrt{D^{[k]}(x+s-k)}\cdot
  \frac{\text{W}_C[\phi_0,\phi_1,\ldots,\phi_{s-1},\phi_n](x)}
  {\text{W}_C[\phi_0,\phi_1,\ldots,\phi_{s-1}](x)}.
  \label{crumrdQM}
\end{align}
The norm of the $s$-th step eigenfunctions have a simple uniform expression:
\begin{equation}
  (\phi^{[s]}_n,\phi^{[s]}_m)
  =\prod_{j=0}^{s-1}\bigl(\mathcal{E}(n)-\mathcal{E}(j)\bigr)
  \cdot(\phi_n,\phi_m).
\end{equation}
This situation of the Crum's theorem is illustrated in Fig.\,1.

\bigskip

A quantum mechanical system with a factorised Hamiltonian
$\mathcal{H}=\mathcal{A}^\dagger\mathcal{A}$ together with the associated
one $\mathcal{H}^{[1]}=\mathcal{A}\mathcal{A}^\dagger$ is sometimes called
a `supersymmetric' QM \cite{witten,susyqm}. This appears rather a misnomer,
since as we have shown the factorised form is generic and it implies no
extra symmetry. The iso-spectrality is shared by all the associated
Hamiltonians, not merely by the first two. In this connection, the
transformation of mapping the $s$-th to the ($s+1$)-st associated Hamiltonian
is sometimes called susy transformation. It is also known as the Darboux
or Darboux-Crum transformation. Those covering multi-steps are sometimes
called `higher derivative' or `nonlinear' or `$\mathcal{N}$-fold' susy
transformations \cite{ais,plyu,aoy}.

% place of figure is moved temporarily.
% 
%%%%%%%%%%%%%%%%%%%%%%%%%%%%%%%%%%%%%%%%%%%%%%%%%%%%%%%%%%%%%%%%%%%%%%%%
\begin{center}
  \includegraphics{crumscheme2.epsi}
\end{center}
\begin{center}
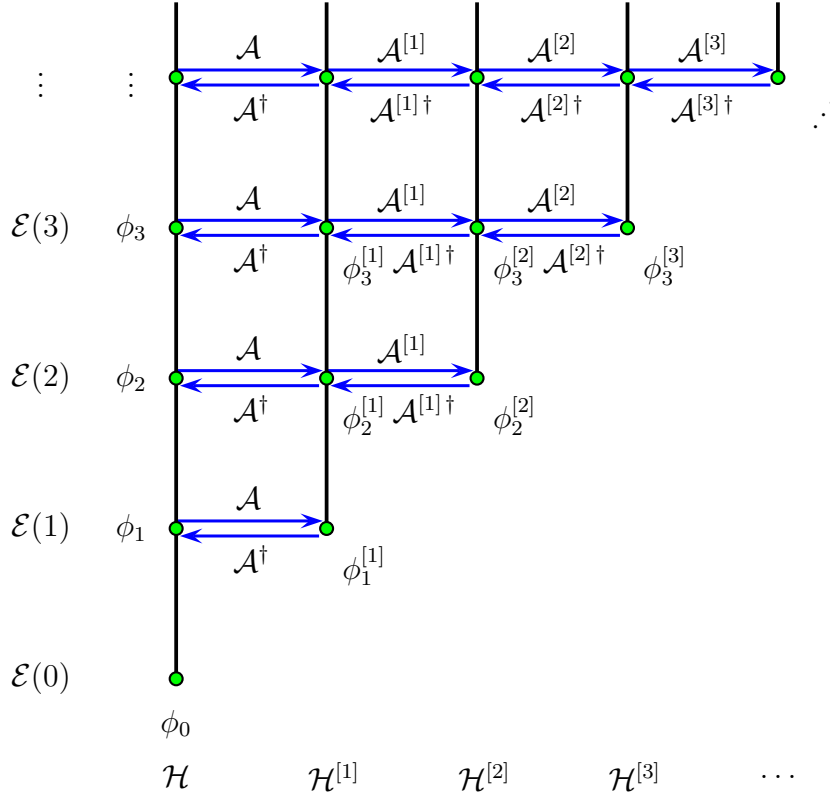
  Figure\,1: Schematic picture of the Crum's theorem
\end{center}
%%%%%%%%%%%%%%%%%%%%%%%%%%%%%%%%%%%%%%%%%%%%%%%%%%%%%%%%%%%%%%%%%%%%%%%%

%%%%%%%%%%%%%%%%%%%%%%%%%%%%%%%%%%%%%%%%%%%
%                                         %
% 2.3 Modified Crum's Theorem             %
%                                         %
%%%%%%%%%%%%%%%%%%%%%%%%%%%%%%%%%%%%%%%%%%%
\subsection{Modified Crum's Theorem}
\label{sec:ModCrum}

Crum's theorem provides a new  iso-spectral Hamiltonian system by deleting
successively the  lowest lying levels from the original Hamiltonian systems
$\mathcal{H}$ and $\{\phi_n(x)\}$.
The modification of Crum's theorem by Krein-Adler \cite{krein,adler} is
achieved by deleting a finite number of eigenstates indexed by a set of
non-negative distinct integers\footnote{
Although this notation $d_j$ conflicts with the notation of the normalisation
constant $d_n$ in \eqref{ortho2}, we think this does not cause any confusion
because the latter appears as $\frac{1}{d_n^2}\,\delta_{nm}$.
}
$\mathcal{D}\eqdef\{d_1,d_2,\ldots,d_{\ell}\}\subset\mathbb{Z}_{\ge0}^{\ell}$,
satisfying certain conditions to be specified later.
After the deletion, the new groundstate has the label $\mu$ given by
\begin{equation}
  \mu\eqdef\min\{n\,|\,n\in\mathbb{Z}_{\geq 0}\backslash\mathcal{D}\}.
\end{equation}
Corresponding to \eqref{w[s]}--\eqref{crumrdQM}, the new iso-spectral
Hamiltonian system is \cite{gos,os22}:
\begin{align} 
  &\bar{\mathcal{H}}\eqdef
  \bar{\mathcal{A}}^\dagger\bar{\mathcal{A}}+\mathcal{E}(\mu),
  \quad\bar{\mathcal{A}}\,\bar{\phi}_{\mu}(x)=0,
  \label{genmodhamdef}\\
  &\bar{\mathcal{H}}\bar{\phi}_n(x)=\mathcal{E}(n)\bar{\phi}_n(x)
   \quad(n\in\mathbb{Z}_{\geq 0}\backslash\mathcal{D}),\\
  \text{oQM}:\quad&\bar{\mathcal{H}}=p^2+\bar{U}(x),\quad
  \bar{U}(x)=U(x)-2\partial_x^2\Bigl(
  \log\text{W}\,[\phi_{d_1},\phi_{d_2},\ldots,\phi_{d_{\ell}}](x)\Bigr),
  \label{Ub1bs}\\
  &\bar{\mathcal{A}}\eqdef\partial_x-\partial_x\bar{w}(x),\quad
  \bar{\mathcal{A}}^{\dagger}=-\partial_x-\partial_x\bar{w}(x),\quad
  \bar{w}(x)\eqdef\log|\bar{\phi}_\mu(x)|,\\
  &\bar{\phi}_n(x)\eqdef
  \frac{\text{W}\,[\phi_{d_1},\phi_{d_2},\ldots,\phi_{d_{\ell}},\phi_n](x)}
  {\text{W}\,[\phi_{d_1},\phi_{d_2},\ldots,\phi_{d_{\ell}}](x)},\\
  \text{idQM}:\quad&\bar{V}(x)\eqdef
  \sqrt{V(x-i\tfrac{\ell}{2}\gamma)V^*(x-i\tfrac{\ell+2}{2}\gamma)}\n
  &\qquad\qquad\times
  \frac{\text{W}_{\gamma}\,[\phi_{d_1},\ldots,\phi_{d_{\ell}}]
  (x+i\tfrac{\gamma}{2})}
  {\text{W}_{\gamma}\,[\phi_{d_1},\ldots,\phi_{d_{\ell}}]
  (x-i\tfrac{\gamma}{2})}\,
  \frac{\text{W}_{\gamma}\,[\phi_{d_1},\ldots,\phi_{d_{\ell}},\phi_{\mu}]
  (x-i\gamma)}
  {\text{W}_{\gamma}\,[\phi_{d_1},\ldots,\phi_{d_{\ell}},\phi_{\mu}](x)},
  \label{Vb1bs+1}\\
  &\bar{\mathcal{A}}\eqdef i\Bigl(e^{\frac{\gamma}{2}p}\sqrt{\bar{V}^*(x)}
  -e^{-\frac{\gamma}{2}p}\sqrt{\bar{V}(x)}\,\Bigr),\n
  &\bar{\mathcal{A}}^{\dagger}
  =-i\Bigl(\sqrt{\bar{V}(x)}\,e^{\frac{\gamma}{2}p}
  -\sqrt{\bar{V}^*(x)}\,e^{-\frac{\gamma}{2}p}\Bigr),\\
  &F(x)\eqdef\frac{\sqrt{\prod_{j=0}^{\ell-1}V(x+i\tfrac{\ell-2j}{2}\gamma)
  V^*(x-i\tfrac{\ell-2j}{2}\gamma)}}
  {\text{W}_{\gamma}\,[\phi_{d_1},\ldots,\phi_{d_{\ell}}]
  (x-i\tfrac{\gamma}{2})\,
  \text{W}_{\gamma}\,[\phi_{d_1},\ldots,\phi_{d_{\ell}}]
  (x+i\tfrac{\gamma}{2})},\n
  &\bar{\phi}_n(x)\eqdef\sqrt{F(x)}\,
  \text{W}_{\gamma}\,[\phi_{d_1},\ldots,\phi_{d_{\ell}},\phi_n](x),
  \label{phib1bsn}\\
  \text{rdQM}:\quad&\bar{B}(x)=\sqrt{B(x+\ell)D(x+\ell+1)}\,
  \frac{\text{W}_C[\phi_{d_1},\ldots,\phi_{d_{\ell}}](x)}
  {\text{W}_C[\phi_{d_1},\ldots,\phi_{d_{\ell}}](x+1)}\n
  &\qquad\qquad\times
  \frac{\text{W}_C[\phi_{d_1},\ldots,\phi_{d_{\ell}},\phi_{\mu}](x+1)}
  {\text{W}_C[\phi_{d_1},\ldots,\phi_{d_{\ell}},\phi_{\mu}](x)},
  \label{Bbar}\\
  &\bar{D}(x)=\sqrt{B(x-1)D(x)}\,
  \frac{\text{W}_C[\phi_{d_1},\ldots,\phi_{d_{\ell}}](x+1)}
  {\text{W}_C[\phi_{d_1},\ldots,\phi_{d_{\ell}}](x)}\n
  &\qquad\qquad\times
  \frac{\text{W}_C[\phi_{d_1},\ldots,\phi_{d_{\ell}},\phi_{\mu}](x-1)}
  {\text{W}_C[\phi_{d_1},\ldots,\phi_{d_{\ell}},\phi_{\mu}](x)},
  \label{Dbar}\\
  &\bar{\mathcal{A}}\eqdef
  \sqrt{\bar{B}(x)}-e^{\partial}\sqrt{\bar{D}(x)},\quad
  \bar{\mathcal{A}}^{\dagger}=
  \sqrt{\bar{B}(x)}-\sqrt{\bar{D}(x)}\,e^{-\partial},
  \label{dbfQMAAd}\\
  &F(x)\eqdef\frac{\sqrt{\prod_{k=1}^{\ell}B(x+k-1)D(x+k)}}
  {\text{W}_C[\phi_{d_1},\ldots,\phi_{d_{\ell}}](x)\,
  \text{W}_C[\phi_{d_1},\ldots,\phi_{d_{\ell}}](x+1)},\n[2pt]
  &\bar{\phi}_n(x)=
  (-1)^{\ell}\sqrt{F(x)}\,
  \text{W}_C[\phi_{d_1},\ldots,\phi_{d_{\ell}},\phi_n](x),
  \label{phibarnD}
\end{align}
and 
\begin{equation}
  \text{oQM, dQM}:\quad(\bar{\phi}_n,\bar{\phi}_m)
  =\prod_{j=1}^{\ell}\bigl(\mathcal{E}(n)-\mathcal{E}(d_j)\bigr)
  \cdot(\phi_n,\phi_m)\quad(n,m\in\mathbb{Z}_{\geq 0}\backslash\mathcal{D}).
  \label{modnorm}
\end{equation}
It should be emphasised that the Hamiltonian $\bar{\mathcal{H}}$ as well as
the eigenfunction $\{\bar{\phi}_n(x)\}$ are symmetric with respect to
$d_1,\ldots,d_{\ell}$, and thus they are independent of the order of $\{d_j\}$.
In order to guarantee the positivity of the norm \eqref{modnorm} of all
the eigenfunctions $\{\bar{\phi}_n(x)\}$ of the  modified Hamiltonian,
the set of deleted energy levels $\mathcal{D}=\{d_1,\ldots,d_{\ell}\}$ must
satisfy the necessary and sufficient conditions \cite{krein,adler}
\begin{equation}
  \prod_{j=1}^{\ell}(m-d_j)\ge0,\quad m\in\mathbb{Z}_{\geq 0}.
\end{equation}
And the Hamiltonian $\bar{\mathcal{H}}$ is non-singular under this
condition.
The Crum's theorem in \S\,\ref{sec:Inter} corresponds to the choice
$\{d_1,d_2,\ldots,d_{\ell}\}=\{0,1,\ldots,\ell-1\}$.

Starting from an exactly solvable Hamiltonian, one can construct infinitely
many variants of exactly solvable Hamiltonians and their eigenfunctions by
Adler's and Garc\'ia et al's methods \cite{gos}. The resulting systems are,
however, not shape invariant, even if the starting system is. For the dQM
with real shifts, see the recent work \cite{os22} and a related work \cite{yz}.

%%%%%%%%%%%%%%%%%%%%%%%%%%%%%%%%%%%%%%%%%%%
%                                         %
% 2.4 Shape Invariance                    %
%                                         %
%%%%%%%%%%%%%%%%%%%%%%%%%%%%%%%%%%%%%%%%%%%
\subsection{Shape Invariance} 
\label{sec:Sha}

Shape invariance \cite{genden} is a sufficient condition for the exact
solvability in the Schr\"{o}dinger picture. Combined with Crum's theorem
\cite{crum}, or the factorisation method \cite{infhull} or the so-called
supersymmetric quantum mechanics \cite{dks,susyqm}, the totality of the
discrete eigenvalues and the corresponding eigenfunctions can be easily
obtained. It was shown by the present authors that the concept of shape
invariance worked equally well in the dQM with the pure imaginary shifts
\cite{os4,os13} as well as the real shifts \cite{os12}, providing
quantum mechanical explanation of the solvability of the Askey scheme
of hypergeometric orthogonal polynomials in general.
Throughout this review, we concentrate on the simple situations that
the entire spectra are discrete. When continuous spectra exist, however,
shape invariance and Crum's theorem fail to provide the full set of
eigenvalues and eigenfunctions.

In many cases the Hamiltonian contains some parameter(s),
$\bm{\lambda}=(\lambda_1,\lambda_2,\ldots)$.
Here we write the parameter dependence explicitly, $\mathcal{H}(\bm{\lambda})$,
$\mathcal{A}(\bm{\lambda})$, $\mathcal{E}(n;\bm{\lambda})$,
$\phi_n(x;\bm{\lambda})$, $P_n(\eta(x;\bm{\lambda});\bm{\lambda})$ etc,
since it is the central issue.
The shape invariance condition with a suitable choice of parameters is
\begin{equation}
  \mathcal{A}(\bm{\lambda})\mathcal{A}(\bm{\lambda})^{\dagger}
  =\kappa\mathcal{A}(\bm{\lambda}+\bm{\delta})^{\dagger}
  \mathcal{A}(\bm{\lambda}+\bm{\delta})+\mathcal{E}(1;\bm{\lambda}),
  \label{shapeinv}
\end{equation}
where $\kappa$ is a real positive parameter and $\bm{\delta}$ is the
shift of the parameters. In other words $\mathcal{H}^{[0]}$ and
$\mathcal{H}^{[1]}$ have the same shape, only the parameters are
shifted by $\bm{\delta}$.
The $s$-th step Hamiltonian $\mathcal{H}^{[s]}$ in \S\,\ref{sec:Inter}
is $\mathcal{H}^{[s]}=\kappa^s\mathcal{H}(\bm{\lambda}+s\bm{\delta})
+\mathcal{E}(s;\bm{\lambda})$. The energy spectrum and the excited state
wavefunction are determined by the data of the groundstate wavefunction
$\phi_0(x;\bm{\lambda})$ and the energy of the first excited state
$\mathcal{E}(1;\bm{\lambda})$ as follows \cite{dks,os4,os12,os13}:
\begin{align}
  &\mathcal{E}(n;\bm{\lambda})=\sum_{s=0}^{n-1}
  \kappa^s\mathcal{E}(1;\bm{\lambda}^{[s]}),\qquad
  \bm{\lambda}^{[s]}\eqdef\bm{\lambda}+{s}\bm{\delta},
  \label{shapeenery}\\
  &\phi_n(x;\bm{\lambda})\propto
  \mathcal{A}(\bm{\lambda}^{[0]})^{\dagger}
  \mathcal{A}(\bm{\lambda}^{[1]})^{\dagger}
  \mathcal{A}(\bm{\lambda}^{[2]})^{\dagger}
  \cdots
  \mathcal{A}(\bm{\lambda}^{[n-1]})^{\dagger}
  \phi_0(x;\bm{\lambda}^{[n]}).
  \label{phin=A..Aphi0}
\end{align}
The above formula for the eigenfunctions $\phi_n(x;\bm{\lambda})$ can be
considered as the {\em universal Rodrigues formula\/} for the Askey scheme
of hypergeometric polynomials and their $q$-analogues.
For the explicit forms of the Rodrigues type formula for each polynomial,
one only has to substitute the explicit forms of the operator
$\mathcal{A}(\bm{\lambda})$ and the groundstate wavefunction
$\phi_0(x;\bm{\lambda})$. For the nine explicit examples given in
\eqref{ex1}--\eqref{ex9}, it is straightforward to verify the shape
invariance conditions \eqref{shapeinv} and the energy \eqref{shapeenery}
and the eigenfunction \eqref{phin=A..Aphi0} formulas.

In the case of a finite number of bound states, e.g. the Morse potential,
the eigenvalue has a maximum at a certain level $n$,
$\mathcal{E}(n;\bm{\lambda})$. Beyond that level the formula
\eqref{shapeenery} ceases to work and the Rodrigues formula
\eqref{phin=A..Aphi0} does not provide a square integrable eigenfunctions,
although $\phi_m$ ($m>n$) continues to satisfy the Schr\"{o}dinger equation
with $\mathcal{E}(m;\bm{\lambda})$.

The above shape invariance condition \eqref{shapeinv} is equivalent to
the following conditions:
\begin{align}
  \text{oQM}:\quad&\bigl(\partial_xw(x;\bm{\lambda})\bigr)^2
  -\partial_x^2w(x;\bm{\lambda})=
  \bigl(\partial_xw(x;\bm{\lambda}+\bm{\delta})\bigr)^2
  +\partial_x^2w(x;\bm{\lambda}+\bm{\delta})+\mathcal{E}(1;\bm{\lambda}),
  \ \ \kappa=1,
  \label{shapeinvoQM}\\
  \text{idQM}:\quad&V(x-i\tfrac{\gamma}{2};\bm{\lambda})
  V^*(x-i\tfrac{\gamma}{2};\bm{\lambda})
  =\kappa^2\,V(x;\bm{\lambda}+\bm{\delta})
  V^*(x-i\gamma;\bm{\lambda}+\bm{\delta}),
  \label{shapeinvidQM1}\\
  &V(x+i\tfrac{\gamma}{2};\bm{\lambda})
  +V^*(x-i\tfrac{\gamma}{2};\bm{\lambda})
  =\kappa\bigl(V(x;\bm{\lambda}+\bm{\delta})
  +V^*(x;\bm{\lambda}+\bm{\delta})\bigr)-\mathcal{E}(1;\bm{\lambda}),
  \label{shapeinvidQM2}\\
  \text{rdQM}:\quad&B(x+1;\bm{\lambda})D(x+1;\bm{\lambda})
  =\kappa^2\,B(x;\bm{\lambda}+\bm{\delta})D(x+1;\bm{\lambda}+\bm{\delta}),
  \label{shapeinvrdQM1}\\
  &B(x;\bm{\lambda})+D(x+1;\bm{\lambda})
  =\kappa\bigl(B(x;\bm{\lambda}+\bm{\delta})
  +D(x;\bm{\lambda}+\bm{\delta})\bigr)+\mathcal{E}(1;\bm{\lambda}).
  \label{shapeinvrdQM2}
\end{align}
For the idQM, the first condition \eqref{shapeinvidQM1} is multiplicative.
If $V_1$, $V_1^*$, $\kappa_1$ and $V_2$, $V_2^*$, $\kappa_2$ satisfy the
condition independently, then $V=V_1V_2$, $V^*=V_1^*V_2^*$,
$\kappa=\kappa_1\kappa_2$ also satisfies it. The second conditions
\eqref{shapeinvidQM2} provides a substantial constraint.
The situation is the same in rdQM.

It is straightforward to verify the shape invariance for the nine examples
\eqref{ex1}--\eqref{ex9} in \S\ref{sec:H} with the following data:
\begin{align}
  \text{oQM}:\ \ \text{H}:\quad&
  \bm{\lambda}=\phi\ (\text{null}),\quad\bm{\delta}=\phi,\quad
  \kappa=1,\quad\mathcal{E}(n;\bm{\lambda})=2n,\\
  \phantom{\text{oQM}:\ \ }\text{L}:\quad&
  \bm{\lambda}=g,\quad\bm{\delta}=1,\quad
  \kappa=1,\quad\mathcal{E}(n;\bm{\lambda})=4n,\\
  \phantom{\text{oQM}:\ \ }\text{J}:\quad&
  \bm{\lambda}=(g,h),\quad\bm{\delta}=(1,1),\quad
  \kappa=1,\quad\mathcal{E}(n;\bm{\lambda})=4n(n+g+h).
\end{align}
It should be stressed that the above shape invariant transformation
$\bm{\lambda}\to\bm{\lambda}+\bm{\delta}$,
$\mathcal{H}^{[s]}\to \mathcal{H}^{[s+1]}$ for L and J, that is,
$g\to g+1$, $h\to h+1$, {\em preserves the monodromy} \eqref{monodromy}
at the regular singularities.
\begin{align}
  \text{idQM}:\ \ \text{MP}:\quad&
  \bm{\lambda}=a,\quad\bm{\delta}=\tfrac12,\quad\kappa=1,\quad
  \mathcal{E}(n;\bm{\lambda})=2n,\\
  \phantom{\text{idQM}:\ \ }\text{W}:\quad&
  \bm{\lambda}=(a_1,a_2,a_3,a_4),\quad
  \bm{\delta}=(\tfrac12,\tfrac12,\tfrac12,\tfrac12),\quad\kappa=1,\n
  &\mathcal{E}(n;\bm{\lambda})=4n(n+b_1-1),\\
  \phantom{\text{idQM}:\ \ }\text{AW}:\quad&
  q^{\bm{\lambda}}=(a_1,a_2,a_3,a_4),\quad
  \bm{\delta}=(\tfrac12,\tfrac12,\tfrac12,\tfrac12),\quad\kappa=q^{-1},\n
  &\mathcal{E}(n;\bm{\lambda})=(q^{-n}-1)(1-b_4q^{n-1}),\\
  \text{rdQM}:\ \ \text{M}:\quad&
  \bm{\lambda}=(\beta,c),\quad\bm{\delta}=(1,0),\quad\kappa=1,\quad
  \mathcal{E}(n;\bm{\lambda})=n,\\
  \phantom{\text{rdQM}:\ \ }\text{R}:\quad&
  \bm{\lambda}=(a,b,c,d),\quad\bm{\delta}=(1,1,1,1),\quad\kappa=1,\quad
  \mathcal{E}(n;\bm{\lambda})=4n(n+\tilde{d}),\\
  \phantom{\text{rdQM}:\ \ }\text{$q$R}:\quad&
  q^{\bm{\lambda}}=(a,b,c,d),\quad\bm{\delta}=(1,1,1,1),\quad
  \kappa=q^{-1},\n
  &\mathcal{E}(n;\bm{\lambda})=(q^{-n}-1)(1-\tilde{d}q^{n-1}),
\end{align}
where $q^{(\lambda_1,\lambda_2,\ldots)}
=(q^{\lambda_1},q^{\lambda_2},\ldots)$.
For more complete lists of exactly solvable oQM see \cite{infhull,susyqm}
and for idQM and rdQM see \cite{os13} and \cite{os12}. It is well known
that the solutions of Schr\"{o}dinger equations provide those of the
corresponding Fokker-Planck equations \cite{risken}, which describe the
time evolution of probability density functions.
This connection can be readily generalised to discrete Schr\"{o}dinger
equations, which would correspond to, for example, discretised stochastic
processes, the Markov chains. The birth and death processes
\cite{ismail,bdp} are the best known examples.
It is interesting to point out that all the exactly solvable rdQM examples
in \cite{os12} and their modifications \cite{os22,os23} provide also
{\em exactly solvable birth and death processes} \cite{bdproc}.

The simplest example of rdQM, the Charlier polynomial, has no shiftable
parameter. Its shape invariance relation
$\mathcal{A}\mathcal{A}^\dagger-\mathcal{A}^\dagger\mathcal{A}=1$
gives another realisation of the oscillator algebra.
Likewise, the simplest example of idQM, the $q$-Hermite ($q$H) polynomial
has no parameter other than $q$. It is obtained as a special case of the
Askey-Wilson (AW) \eqref{ex6} by setting $a_j=0$, $j=1,\ldots,4$.
Like the harmonic oscillator (Hermite) case, it is shape invariant and
the shape invariance relation ($\kappa=q^{-1}$) \eqref{shapeinv} becomes
the {\em $q$-oscillator algebra}
$\mathcal{A}\mathcal{A}^\dagger -q^{-1}\mathcal{A}^\dagger \mathcal{A}
=q^{-1}-1$ itself \cite{os11}.

The shape invariance and the Crum's theorem imply that
$\phi_n(x;\bm{\lambda})$ and $\phi_{n-1}(x;\bm{\lambda}+\bm{\delta})$ are
mapped to each other by the operators $\mathcal{A}(\bm{\lambda})$ and
$\mathcal{A}(\bm{\lambda})^{\dagger}$:
\begin{align}
  \mathcal{A}(\bm{\lambda})\phi_n(x;\bm{\lambda})
  &=f_n(\bm{\lambda})
  \phi_{n-1}\bigl(x;\bm{\lambda}+\bm{\delta}\bigr)
  \times\left\{\begin{array}{ll}
  1&:\text{oQM, idQM}\\
  \frac{1}{\sqrt{B(0;\bm{\lambda})}}&:\text{rdQM}
  \end{array}\right.,
  \label{Aphi=fphi}\\
  \mathcal{A}(\bm{\lambda})^{\dagger}
  \phi_{n-1}\bigl(x;\bm{\lambda}+\bm{\delta}\bigr)
  &=b_{n-1}(\bm{\lambda})\phi_n(x;\bm{\lambda})
  \times\left\{\begin{array}{ll}
  1&:\text{oQM, idQM}\\
  \sqrt{B(0;\bm{\lambda})}&:\text{rdQM}
  \end{array}\right..
  \label{Adphi=bphi}
\end{align}
Here the constants $f_n(\bm{\lambda})$ and $b_{n-1}(\bm{\lambda})$ depend
on the normalisation of $\phi_n(x;\bm{\lambda})$ but their product does
not. It gives the energy eigenvalue,
\begin{equation}
  \mathcal{E}(n;\bm{\lambda})=f_n(\bm{\lambda})b_{n-1}(\bm{\lambda}).
\end{equation}
The factor $\sqrt{B(0;\bm{\lambda})}$ for rdQM is introduced for later
convenience. For our choice of $\phi_n(x;\bm{\lambda})$ and
$P_n(\eta(x;\bm{\lambda});\bm{\lambda})$, the data for $f_n(\bm{\lambda})$
and $b_{n-1}(\bm{\lambda})$ are:
\begin{align}
  \text{oQM}:\quad&f_n(\bm{\lambda})=\left\{
  \begin{array}{ll}
  2n&:\text{H}\\
  -2&:\text{L}\\
  -2(n+g+h)&:\text{J}
  \end{array}\right.\!,
  \quad
  b_{n-1}(\bm{\lambda})=\left\{
  \begin{array}{ll}
  1&:\text{H}\\
  -2n&:\text{L,J}
  \end{array}\right.\!,\\
  \text{idQM}:\quad&f_n(\bm{\lambda})=\left\{
  \begin{array}{ll}
  2&:\text{MP}\\
  -n(n+b_1-1)&:\text{W}\\
  q^{\frac{n}{2}}(q^{-n}-1)(1-b_4q^{n-1})&:\text{AW}
  \end{array}\right.\!,
  \quad
  b_{n-1}(\bm{\lambda})=\left\{
  \begin{array}{ll}
  n&:\text{MP}\\
  -1&:\text{W}\\
  q^{-\frac{n}{2}}&:\text{AW}
  \end{array}\right.\!,\\
  \text{rdQM}:\quad&f_n(\bm{\lambda})=\mathcal{E}(n;\bm{\lambda}),\quad
  b_{n-1}(\bm{\lambda})=1.
\end{align}
By removing the groundstate contributions, the forward and backward shift
operators acting on the polynomial eigenfunctions,
$\mathcal{F}(\bm{\lambda})$ and $\mathcal{B}(\bm{\lambda})$, are introduced:
\begin{align}
  \mathcal{F}(\bm{\lambda})&\eqdef
  \phi_0(x;\bm{\lambda}+\bm{\delta})^{-1}\circ
  \mathcal{A}(\bm{\lambda})\circ\phi_0(x;\bm{\lambda})
  \times\left\{\begin{array}{ll}
  1&:\text{oQM, idQM}\\
  \sqrt{B(0;\bm{\lambda})}&:\text{rdQM}
  \end{array}\right.
  \label{Fdef}\\
  &=\left\{\begin{array}{ll}
  \cF\frac{d}{d\eta}
  &:\text{oQM}\\
  i\varphi(x)^{-1}(e^{\frac{\gamma}{2}p}-e^{-\frac{\gamma}{2}p})
  &:\text{idQM}\\[2pt]
  B(0;\bm{\lambda})\varphi(x;\bm{\lambda})^{-1}(1-e^{\partial})
  &:\text{rdQM}
  \end{array}\right.,\\
  \mathcal{B}(\bm{\lambda})&\eqdef
  \phi_0(x;\bm{\lambda})^{-1}\circ
  \mathcal{A}(\bm{\lambda})^{\dagger}
  \circ\phi_0(x;\bm{\lambda}+\bm{\delta})
  \times\left\{\begin{array}{ll}
  1&:\text{oQM, idQM}\\
  \frac{1}{\sqrt{B(0;\bm{\lambda})}}&:\text{rdQM}
  \end{array}\right.
  \label{Bdef}\\
  &=\left\{\begin{array}{ll}
  -4\cF^{-1}c_2(\eta)\bigl(\frac{d}{d\eta}
  +\frac{c_1(\eta,\bm{\lambda})}{c_2(\eta)}\bigr)
  &:\text{oQM}\\[2pt]
  -i\bigl(V(x;\bm{\lambda})e^{\frac{\gamma}{2}p}
  -V^*(x;\bm{\lambda})e^{-\frac{\gamma}{2}p}\bigr)\varphi(x)
  &:\text{idQM}\\[2pt]
  \frac{1}{B(0;\bm{\lambda})}\bigl(B(x;\bm{\lambda})
  -D(x;\bm{\lambda})e^{-\partial}\bigr)\varphi(x;\bm{\lambda})
  &:\text{rdQM}
  \end{array}\right.,
\end{align}
where $\cF$, $c_1(\eta,\bm{\lambda})$ and $c_2(\eta)$ are
\begin{equation}
  \cF\eqdef\left\{
  \begin{array}{ll}
  1&:\text{H}\\
  2&\!:\text{L}\\
  -4&\!:\text{J}
  \end{array}\right.\!\!,
  \ \ c_1(\eta,\bm{\lambda})\eqdef\left\{
  \begin{array}{ll}
  -\frac12&\!:\text{H}\\
  g+\tfrac12-\eta&\!:\text{L}\\
  h-g-(g+h+1)\eta&\!:\text{J}
  \end{array}\right.\!\!,
  \ \ c_2(\eta)\eqdef\left\{
  \begin{array}{ll}
  \frac14&\!:\text{H}\\
  \eta&\!:\text{L}\\
  1-\eta^2&\!:\text{J}
  \end{array}\right.\!\!,\!\!
  \label{cF,c1,c2}
\end{equation}
and the auxiliary functions $\varphi(x)$ are
\begin{equation}
  \text{idQM}:\ \ \varphi(x)=\left\{
  \begin{array}{ll}
  1&:\text{MP}\\
  2x&:\text{W}\\
  2\sin x&:\text{AW}
  \end{array}\right.\!,
  \qquad
  \text{rdQM}:\ \ \varphi(x;\bm{\lambda})=\left\{
  \begin{array}{ll}
  1&:\text{M}\\
  {\displaystyle\frac{2x+d+1}{d+1}}&:\text{R}\\[6pt]
  {\displaystyle\frac{q^{-x}-dq^{x+1}}{1-dq}}&:\text{$q$R}
  \end{array}\right.\!.
\end{equation}
Then the above relations \eqref{Aphi=fphi}--\eqref{Adphi=bphi} become
\begin{align}
  \mathcal{F}(\bm{\lambda})\check{P}_n(x;\bm{\lambda})
  &=f_n(\bm{\lambda})\check{P}_{n-1}(x;\bm{\lambda}+\bm{\delta}),
  \label{forwardrel}\\
  \mathcal{B}(\bm{\lambda})\check{P}_{n-1}(x;\bm{\lambda}+\bm{\delta})
  &=b_{n-1}(\bm{\lambda})\check{P}_n(x;\bm{\lambda}),
  \label{backwardrel}
\end{align}
where we have used the notation
\begin{equation}
  \check{P}_n(x;\bm{\lambda})\eqdef P_n(\eta(x;\bm{\lambda});\bm{\lambda}).
  \label{cPn}
\end{equation}
Corresponding to \eqref{factHam}, the forward and backward shift operators
give a factorisation of the similarity transformed Hamiltonian \eqref{Htdef},
\begin{equation}
  \widetilde{\mathcal{H}}(\bm{\lambda})
  =\mathcal{B}(\bm{\lambda})\mathcal{F}(\bm{\lambda}).
  \label{htilbf}
\end{equation}

%%%%%%%%%%%%%%%%%%%%%%%%%%%%%%%%%%%%%%%%%%%%%
%                                           %
% 2.5 Solvability in the Heisenberg Picture %
%                                           %
%%%%%%%%%%%%%%%%%%%%%%%%%%%%%%%%%%%%%%%%%%%%%
\subsection{Solvability in the Heisenberg Picture}
\label{sec:Hei}

As is well known the Heisenberg operator formulation is central to
quantum field theory. The creation/annihilation operators of the
harmonic oscillators are the cornerstones of modern quantum physics.
However, until recently, it had been generally conceived that the
Heisenberg operator solutions are intractable. Here we show that most of
the shape invariant dQM Hamiltonian systems are exactly solvable in
the Heisenberg picture, too \cite{os7,os8}. To be more precise, the
Heisenberg operator of the sinusoidal coordinate $\eta(x)$
\begin{equation}
  e^{it\mathcal{H}}\eta(x)e^{-it\mathcal{H}}
  \label{etaHei}
\end{equation}
can be evaluated in a closed form. It is well known that any orthogonal
polynomials satisfy the {\em three term recurrence relations}
\cite{askey,ismail}
\begin{equation}
  \eta P_n(\eta)=A_nP_{n+1}(\eta)+B_nP_n(\eta)+C_nP_{n-1}(\eta)\ \ (n\geq 0),
  \label{threeterm}
\end{equation}
with $P_{-1}(\eta)=0$. Here the coefficients $A_n$, $B_n$ and $C_n$ are
real and $A_{n-1}C_n>0$ ($n\geq 1$). Conversely all the polynomials
starting with degree 0 and satisfy the above three term recurrence
relations are orthogonal (Favard's theorem \cite{Chihara}).
These relations can also be considered as an eigenvalue equation, in
which $\eta$ is the eigenvalue. For the polynomials in rdQM with the
universal normalisation $P_n(0)=1$ \eqref{univnorm}, the coefficients
of the three term recurrence relations are restricted by the condition
\begin{equation}
  \text{rdQM}:\quad B_n=-(A_n+C_n)\quad(n=0,1,\ldots).
  \label{ABCrel}
\end{equation}
For the factorised quantum mechanical eigenfunctions \eqref{facteig},
these relations mean
\begin{equation}
  \eta(x)\phi_n(x)=A_n\phi_{n+1}(x)+B_n\phi_n(x)+C_n\phi_{n-1}(x)
  \ \ (n\geq 0).
  \label{threetermphi}
\end{equation}
In other words, the operator $\eta(x)$ acts like a creation operator
which sends the eigenstate $n$ to $n+1$ as well as like an annihilation
operator, which maps an eigenstate $n$ to $n-1$.
This fact combined with the well known result that the annihilation/creation
operators of the harmonic oscillator are the positive/negative frequency
part of the Heisenberg operator solution for the coordinate $x$ is the
starting point of this subsection.
As will be shown below the {\em sinusoidal coordinate} $\eta(x)$ undergoes
sinusoidal motion \eqref{quantsol}, whose frequencies depend on the energy.
Thus it is not harmonic in general.
To the best of our knowledge, the sinusoidal coordinate was first introduced
in a rather broad sense for general (not necessarily solvable) potentials as
a useful means for coherent state research by Nieto and Simmons \cite{nieto}.

The sufficient condition for the closed form expression of the Heisenberg
operator \eqref{etaHei} is the {\em closure relation\/}
\begin{equation}
  [\mathcal{H},[\mathcal{H},\eta(x)]\,]
  =\eta(x)\,R_0(\mathcal{H})+[\mathcal{H},\eta(x)]\,R_1(\mathcal{H})
  +R_{-1}(\mathcal{H}).
  \label{closurerel}
\end{equation}
Here the coefficients $R_i(y)$ are polynomials in $y$. It is easy to see
that the cubic commutator $[\mathcal{H},[\mathcal{H},[\mathcal{H},\eta(x)]]]
\equiv(\text{ad}\,\mathcal{H})^3\eta(x)$ is reduced to $\eta(x)$ and
$[\mathcal{H},\eta(x)]$ with $\mathcal{H}$ depending coefficients:
\begin{align}
  (\text{ad}\,\mathcal{H})^3\eta(x)&=
  [\mathcal{H},\eta(x)]R_0(\mathcal{H})
  +[\mathcal{H},[\mathcal{H},\eta(x)]]\,R_1(\mathcal{H})\n
  &=\eta(x)\,R_0(\mathcal{H})R_1(\mathcal{H})
  +[\mathcal{H},\eta(x)]\,\bigl(R_1(\mathcal{H})^2+R_0(\mathcal{H})\bigr)
  +R_{-1}(\mathcal{H})R_{1}(\mathcal{H}),
\end{align}
in which the definition $(\text{ad}\,\mathcal{H})X\eqdef[\mathcal{H},X]$
is used. It is trivial to see that all the higher commutators
$(\text{ad}\,\mathcal{H})^n\eta(x)$ can also be reduced to $\eta(x)$ and
$[\mathcal{H},\eta(x)]$ with $\mathcal{H}$ depending coefficients.
The second order closure \eqref{closurerel} simply reflects the
Schr\"{o}dinger equation, which is a second order differential or difference
equation. Thus we arrive at
\begin{align}
  e^{it\mathcal{H}}\eta(x)e^{-it\mathcal{H}}
  &=\sum_{n=0}^\infty\frac{(it)^n}{n!}(\text{ad}\,\mathcal{H})^n\eta(x)\n
  &=[\mathcal{H},\eta(x)]
  \frac{e^{i\alpha_+(\mathcal{H})t}-e^{i\alpha_-(\mathcal{H})t}}
  {\alpha_+(\mathcal{H})-\alpha_-(\mathcal{H})}
  -R_{-1}(\mathcal{H})R_{0}(\mathcal{H})^{-1}\n
  &\quad
  +\bigl(\eta(x)+R_{-1}(\mathcal{H})R_0(\mathcal{H})^{-1}\bigr)
  \frac{-\alpha_-(\mathcal{H})e^{i\alpha_+(\mathcal{H})t}
  +\alpha_+(\mathcal{H})e^{i\alpha_-(\mathcal{H})t}}
  {\alpha_+(\mathcal{H})-\alpha_-(\mathcal{H})}.
  \label{quantsol}
\end{align}
This simply means that $\eta(x)$ oscillates sinusoidally with two
energy-dependent ``frequencies'' $\alpha_\pm(\mathcal{H})$ given by
\begin{gather}
  \alpha_\pm(\mathcal{H})=\tfrac12\bigl(R_1(\mathcal{H})\pm
  \sqrt{R_1(\mathcal{H})^2+4R_0(\mathcal{H})}\,\bigr),
  \label{alpmdef}\\
  \alpha_+(\mathcal{H})+\alpha_-(\mathcal{H})=R_1(\mathcal{H}),
  \quad
  \alpha_+(\mathcal{H})\alpha_-(\mathcal{H})=-R_0(\mathcal{H}).
  \label{freqpm}
\end{gather}
The energy spectrum is determined by the over-determined recursion relations
$\mathcal{E}(n+1)=\mathcal{E}(n)+\alpha_+(\mathcal{E}(n))$
and $\mathcal{E}(n-1)=\mathcal{E}(n)+\alpha_-(\mathcal{E}(n))$
with $\mathcal{E}(0)=0$.
It should be stressed that for the known spectra $\{\mathcal{E}(n)\}$
determined by the shape invariance, the quantity inside the square root
in the definition of $\alpha_\pm(\mathcal{H})$ \eqref{alpmdef} for each $n$:
\[
  R_1(\mathcal{E}(n))^2+4R_0(\mathcal{E}(n))
\]
becomes a complete square and the the above two conditions are consistent.
For oQM, the Hamiltonian and the sinusoidal coordinate satisfying the closure
relation \eqref{closurerel} are classified and then the eigenfunctions have
the factorised form \eqref{facteig} \cite{os7}.
For dQM we assume \eqref{facteig}.
The {\em annihilation} and {\em creation} operators $a^{(\pm)}$ are
extracted from this exact Heisenberg operator solution:
\begin{align}
  &e^{it\mathcal{H}}\eta(x)e^{-it\mathcal{H}}
  =a^{(+)}e^{i\alpha_+(\mathcal{H})t}+a^{(-)}e^{i\alpha_-(\mathcal{H})t}
  -R_{-1}(\mathcal{H})R_0(\mathcal{H})^{-1},
  \label{apm}\\
  &a^{(\pm)}\eqdef\pm\Bigl([\mathcal{H},\eta(x)]-\bigl(\eta(x)
  +R_{-1}(\mathcal{H})R_0(\mathcal{H})^{-1}\bigr)\alpha_{\mp}(\mathcal{H})
  \Bigr)
  \bigl(\alpha_+(\mathcal{H})-\alpha_-(\mathcal{H})\bigr)^{-1}\n
  &\phantom{a^{(\pm)}}=
  \pm\bigl(\alpha_+(\mathcal{H})-\alpha_-(\mathcal{H})\bigr)^{-1}
  \Bigl([\mathcal{H},\eta(x)]+\alpha_{\pm}(\mathcal{H})\bigl(\eta(x)
  +R_{-1}(\mathcal{H})R_0(\mathcal{H})^{-1}\bigr)\Bigr),
  \label{apmdefs}\\
  & a^{(+)\,\dagger}=a^{(-)},\quad
  a^{(+)}\phi_n(x)=A_n\phi_{n+1}(x),\quad
  a^{(-)}\phi_n(x)=C_n\phi_{n-1}(x).
  \label{apmphi}
\end{align}
It should be stressed that the annihilation operator and creation operators
are hermitian conjugate of each other and they act on the eigenstate
\eqref{apmphi}.
Simple commutation relations
\begin{equation}
  [\mathcal{H},a^{(\pm)}]=a^{(\pm)}\alpha_{\pm}(\mathcal{H}),
  \label{[H,apm]}
\end{equation}
follow from \eqref{apmdefs} and \eqref{closurerel}.
Commutation relations of $a^{(\pm)}$ are expressed in terms of the
coefficients of the three term recurrence relation by \eqref{apmphi}:
\begin{align}
  &a^{(-)}a^{(+)}\phi_n=A_nC_{n+1}\phi_n,\quad
  a^{(+)}a^{(-)}\phi_n=C_nA_{n-1}\phi_n,\n
  &\Rightarrow
  \ [a^{(-)},a^{(+)}]\phi_n=(A_nC_{n+1}-A_{n-1}C_n)\phi_n.
  \label{[a-,a+]}
\end{align}
These relations simply mean the operator relations
\begin{gather}
  a^{(-)}a^{(+)}=f(\mathcal{H}),\quad
  a^{(+)}a^{(-)}=g(\mathcal{H}),
  \label{apamg}
\end{gather}
in which $f$ and $g$ are analytic functions of $\mathcal{H}$ explicitly
given for each example. In other words, $\mathcal{H}$ and $a^{(\pm)}$ form
a so-called quasi-linear algebra \cite{vinzhed}.
Various {\em dynamical symmetry algebras} associated with exactly solvable
QM, including the $q$-oscillator algebra, were explicitly identified in
\cite{os12,os13}.
It should be stressed that the situation is quite different from those
of the wide variety of proposed annihilation/creation operators for various
quantum systems \cite{coherents}, most of which were introduced within
the framework of `algebraic theory of coherent states', without exact
solvability. In all such cases there is no guarantee for symmetry
relations like \eqref{apamg}.
The explicit form of the annihilation operator \eqref{apmdefs} allows us to
define the {\em coherent state} as its eigenvector,
$a^{(-)}\psi(\alpha,x)=\alpha\psi(\alpha,x)$, $\alpha\in\mathbb{C}$.
See \cite{os7,os13} for various coherent states.

The excited state wavefunctions $\{\phi_n(x)\}$ are obtained by the
successive action of the creation operator $a^{(+)}$ on the groundstate
wavefunction $\phi_0(x)$.
This is the exact solvability in the Heisenberg picture.

The data for the three examples in oQM \eqref{ex1}--\eqref{ex3} are:
\begin{align}
  \text{H}:\quad&R_1(y)=0,\quad R_0(y)=4,\quad R_{-1}(y)=0,\\
  \text{L}:\quad&R_1(y)=0,\quad R_0(y)=16,\quad R_{-1}(y)=-8(y+2g+1),\\
  \text{J}:\quad&R_1(y)=8,\quad R_0(y)=16\bigl(y+(g+h)^2-1\bigr),\quad
  R_{-1}(y)=16(g-h)(g+h-1).
\end{align}
The data for the three examples in idQM \eqref{ex4}--\eqref{ex6} are:
\begin{align}
  \text{MP}:\quad&R_1(y)=0,\quad R_0(y)=4,\quad R_{-1}(y)=0,\\
  \text{W}:\quad&R_1(y)=2,\quad R_0(y)=4y+b_1(b_1-2),\quad
  R_{-1}(y)=-2y^2+(b_1-2b_2)y+(2-b_1)b_3,\n
  &\hspace{15mm}b_2\eqdef\!\!\sum_{1\leq j<k\leq 4}a_ja_k,\quad
  b_3\eqdef\!\!\!\sum_{1\leq j<k<l\leq 4}a_ja_ka_l,\\
  \text{AW}:\quad&R_1(y)=(q^{-\frac12}-q^{\frac12})^2y',\quad
  R_0(y)=(q^{-\frac12}-q^{\frac12})^2
  \bigl(y^{\prime\,2}-(1+q^{-1})^2b_4\bigr),\n
  &R_{-1}(y)=\tfrac12(q^{-\frac12}-q^{\frac12})^2
  \bigl((b_1+q^{-1}b_3)y'-(1+q^{-1})(b_3+q^{-1}b_1b_4)\bigr)-R_0(y),\n
  &\hspace{15mm}y'\eqdef y+1+q^{-1}b_4,\quad
  b_1\eqdef\sum_{j=1}^4a_j,\quad
  b_3\eqdef\!\!\!\sum_{1\leq j<k<l\leq 4}a_ja_ka_l.
\end{align}
The closure relation and the creation/annihilation operators of the MP
polynomials were obtained in 2001 in \cite{degru}.
The data for the three examples in rdQM \eqref{ex7}--\eqref{ex9} are:
\begin{align}
  \text{M}:\quad&R_1(y)=0,\quad R_0(y)=1,\quad
  R_{-1}(y)=-\frac{1+c}{1-c}y-\frac{\beta c}{1-c},\\
  \text{R}:\quad&R_1(y)=2,\quad R_0(y)=4y+\tilde{d}^{\,2}-1,\n
  &R_{-1}(y)=2y^2+\bigl(2(ab+bc+ca)-(1+d)(1+\tilde{d})\bigr)y
  +abc(\tilde{d}-1),\\
  \text{$q$R}:\quad&R_1(y)=(q^{-\frac12}-q^{\frac12})^2y',
  \ \ R_0(y)=(q^{-\frac12}-q^{\frac12})^2
  \bigl(y^{\prime\,2}-(q^{-\frac12}+q^{\frac12})^2\tilde{d}\,\bigr),
  \ \ y'\eqdef y+1+\tilde{d},\n
  &R_{-1}(y)=(q^{-\frac12}-q^{\frac12})^2\Bigl((1+d)y^{\prime\,2}
  -\bigl(a+b+c+d+\tilde{d}+(ab+bc+ca)q^{-1}\bigr)y'\n
  &\phantom{R_{-1}(y)=(q^{-\frac12}-q^{\frac12})^2\Bigl(}
  +(1-a)(1-b)(1-c)(1-\tilde{d}q^{-1})\n
  &\phantom{R_{-1}(y)=(q^{-\frac12}-q^{\frac12})^2\Bigl(}
  +\bigl(a+b+c-1-d\tilde{d}+(ab+bc+ca)q^{-1}\bigr)(1+\tilde{d})\Bigr).
\end{align}

For oQM, the necessary and sufficient condition for the existence of the
sinusoidal coordinate satisfying the closure relation \eqref{closurerel}
is analysed in Appendix A of \cite{os7}. It was shown that such systems
constitute a sub-group of the shape invariant oQM. We also mention that
exact Heisenberg operator solutions for independent sinusoidal coordinates
as many as the degree of freedom were derived for the Calogero systems
based on any root system \cite{os9}. These are novel examples of
infinitely many multi-particle Heisenberg operator solutions.

There were attempts to relate $q$-oscillator algebras to the difference
equation of the $q$-Hermite polynomials \cite{qosci}. None of them is
based on a Hamiltonian, thus hermiticity is not manifest and the logic for
factorization is unclear. There are several derivations of
annihilation/creation operators together with the dynamical symmetry
algebras (including the $q$-oscillator) acting on the polynomials.

%%%%%%%%%%%%%%%%%%%%%%%%%%%%%%%%%%%%%%%%%%%
%                                         %
% 2.6  Dual Polynomials in rdQM           %
%                                         %
%%%%%%%%%%%%%%%%%%%%%%%%%%%%%%%%%%%%%%%%%%%
\subsection{Dual Polynomials in rdQM}
\label{sec:Dual}

The {\em dual\/} polynomials are an important concept in the theory of
orthogonal polynomials of a discrete variable \cite{nikiforov}, that is
the orthogonal polynomials appearing in rdQM.
We show that the dual polynomials arise naturally as the solutions of the
original eigenvalue problem, the Schr\"{o}dinger equation \eqref{Sch_eq}
or \eqref{polyeq}, obtained in an alternative way \cite{os12}.
We do believe this derivation of the duality is more intuitive than the
existing ones \cite{leonard,terw}.
The Hamiltonian $\mathcal{H}$ \eqref{rdQMHam} for rdQM is a real symmetric
{\em tri-diagonal\/} (Jacobi) matrix of a finite ($x_{\text{max}}=N$)
or infinite ($x_{\text{max}}=\infty$) dimension:
\begin{align}
  \mathcal{H}&=(\mathcal{H}_{x,y})_{0\leq x,y\leq x_{\text{max}}},\quad
  \mathcal{H}_{x,y}=\mathcal{H}_{y,x},\\
  \mathcal{H}_{x,y}&=
  -\sqrt{B(x)D(x+1)}\,\delta_{x+1,y}-\sqrt{B(x-1)D(x)}\,\delta_{x-1,y}
  +\bigl(B(x)+D(x)\bigr)\delta_{x,y}.
  \label{Jacobiform}
\end{align}
It is well-known that the spectrum of a Jacobi matrix is {\em simple}
\eqref{positivesemi}, that is no degeneracy.
The factor $\mathcal{A}^{\dagger}$ \eqref{ArdQM} is a lower triangular
matrix with the diagonal and sub-diagonal entries only, and $\mathcal{A}$
is upper triangular having the diagonal and super-diagonal entries only.
Throughout this paper we adopt the (standard) convention that the $(0,0)$
element of the matrix is at the upper left corner.
The similarity transformed Hamiltonian $\widetilde{\mathcal{H}}$
\eqref{HtrdQM} is again tri-diagonal:
\begin{equation}
  \widetilde{\mathcal{H}}
  =(\widetilde{\mathcal{H}}_{x,y})_{0\leq x,y\leq x_{\text{max}}},\quad
  \widetilde{\mathcal{H}}_{x,y}=B(x)(\delta_{x,y}-\delta_{x+1,y})
  +D(x)(\delta_{x,y}-\delta_{x-1,y}).
\end{equation}

The similarity transformed eigenvalue problem
$\widetilde{\mathcal{H}}\,v(x)=\mathcal{E}\,v(x)$ \eqref{polyeq}
can be rewritten into an explicit matrix form with the change of the
notation $v(x)\to{}^t(Q_0,Q_1,\ldots,Q_x,\ldots)$
\begin{equation}
  \sum_{y=0}^{x_{\text{max}}}
  \widetilde{\mathcal{H}}_{x,y}Q_y=\mathcal{E}Q_x\quad
  (x=0,1,\ldots,x_{\text{max}}).
\end{equation}
Because of the tri-diagonality of $\widetilde{\mathcal{H}}$, the above
eigenvalue equations are in fact the three term recurrence relations
for $\{Q_x\}$ as polynomials in $\mathcal{E}$:
\begin{equation}
  \mathcal{E}Q_x(\mathcal{E})
  =B(x)\bigl(Q_x(\mathcal{E})-Q_{x+1}(\mathcal{E})\bigr)
  +D(x)\bigl(Q_x(\mathcal{E})-Q_{x-1}(\mathcal{E})\bigr)\quad
  (x=0,1,\ldots,x_{\text{max}}).
  \label{dual3term}
\end{equation}
Starting with the boundary (initial) condition $Q_0=1$,
$Q_x(\mathcal{E})$ is determined as a degree $x$ polynomial in $\mathcal{E}$.
It is easy to see
\begin{equation}
  Q_x(0)=1\quad(x=0,1,\ldots,x_{\text{max}}).
  \label{Qxzero}
\end{equation}
When $\mathcal{E}$ is replaced by the actual value of the $n$-th eigenvalue
$\mathcal{E}(n)$ \eqref{lowtri} in $Q_x(\mathcal{E})$, we obtain the
explicit form of the eigenvector
\begin{equation}
  \sum_{y=0}^{x_{\text{max}}}\widetilde{\mathcal{H}}_{x,y}Q_y(\mathcal{E}(n))
  =\mathcal{E}(n)Q_x(\mathcal{E}(n))\quad
  (x=0,1,\ldots,x_{\text{max}}).
\end{equation}
In the finite dimensional case,
$\{Q_0(\mathcal{E}),\ldots,Q_N(\mathcal{E})\}$ are determined by
\eqref{dual3term} for $x=0,\ldots,N-1$. The last equation
\begin{align}
  \mathcal{E}Q_N(\mathcal{E})=
  D(N)\bigl(Q_N(\mathcal{E})-Q_{N-1}(\mathcal{E})\bigr),
\end{align}
is the degree $N+1$ algebraic equation (characteristic equation) for
the determination of all the eigenvalues $\{\mathcal{E}(n)\}$.

We now have two expressions (polynomials) for the eigenvectors of the
problem \eqref{polyeq} belonging to the eigenvalue $\mathcal{E}(n)$;
$P_n(\eta(x))$ and $Q_x(\mathcal{E}(n))$.
Due to the simplicity of the spectrum of the Jacobi matrix, they must
be equal up to a multiplicative factor $\alpha_n$,
\[
  P_n(\eta(x))=\alpha_n Q_x(\mathcal{E}(n))\quad
  (x=0,1,\ldots,x_{\text{max}}),
\]
which turns out to be unity because of the boundary (initial) condition
at $x=0$ \eqref{etabound}, \eqref{univnorm}, \eqref{Qxzero} (or at $n=0$
\eqref{Aphi0=0}, \eqref{univnorm}, \eqref{Qxzero});
($n_{\text{max}}\eqdef x_{\text{max}}$)
\begin{align}
  P_n(\eta(0))=P_n(0)&=1=Q_0(\mathcal{E}(n)),\qquad\qquad\ \ \,
  (n=0,1,\ldots,n_{\text{max}}),\\
  P_0(\eta(x))&=1=Q_x(0)=Q_x(\mathcal{E}(0)),\quad
  (x=0,1,\ldots,x_{\text{max}}).
\end{align}

We have established that two polynomials, $\{P_n(\eta)\}$ and its
{\em dual\/} polynomial $\{Q_x(\mathcal{E})\}$, coincide at the
integer lattice points:
\begin{equation}
  P_n(\eta(x))=Q_x(\mathcal{E}(n))\quad (n=0,1,\ldots,n_{\text{max}}\ ;
  \ x=0,1,\ldots,x_{\text{max}}).
  \label{Duality}
\end{equation}
The completeness relation, which is dual to the orthogonality relation
\eqref{ortho2},
\begin{align}
  \sum_{n=0}^{n_{\text{max}}}d_n^2P_n(\eta(x))P_n(\eta(y))
  =\sum_{n=0}^{n_{\text{max}}}d_n^2Q_x(\mathcal{E}(n))Q_y(\mathcal{E}(n))
  =\frac{1}{\phi_0(x)^2}\,\delta_{x,y},
  \label{Qnortho}
\end{align}
is now understood as the orthogonality relation of the dual polynomials
$Q_x(\mathcal{E})$, and the previous normalisation constant $d_n^2$ is
now the orthogonality measure.

The real symmetric (hermitian) matrix $\mathcal{H}_{x,y}$
\eqref{Jacobiform} can be expressed in terms of the complete set of the
eigenvalues and the corresponding normalised eigenvectors
\begin{align}
  \mathcal{H}_{x,y}&=\sum_{n=0}^{n_{\text{max}}}
  \mathcal{E}(n)\hat{\phi}_n(x)\hat{\phi}_n(y),\\
  \hat{\phi}_n(x)&=d_n\phi_0(x)P_n(\eta(x))=d_n\phi_0(x)Q_x(\mathcal{E}(n)).
\end{align}
The very fact that it is tri-diagonal can be easily verified by using
the difference equation for the polynomial $P_n(\eta(x))$ or the three
term recurrence relations for $Q_x(\mathcal{E}(n))$.

Here is a list of the dual correspondence:
\begin{align}
  &x\leftrightarrow n,\quad
  \eta(x)\leftrightarrow \mathcal{E}(n),\quad
  \eta(0)=0\leftrightarrow\mathcal{E}(0)=0,
  \label{duality1}\\
  &B(x)\leftrightarrow -A_n,\quad
  D(x)\leftrightarrow -C_n,\quad
  \frac{\phi_0(x)}{\phi_0(0)}\leftrightarrow\frac{d_n}{d_0}.
  \label{duality2}
\end{align}
The functions $B(x)$ and $D(x)$ govern the difference equation for the
polynomials $P_n(\eta)$, the solution of which requires the knowledge of
the sinusoidal coordinate $\eta(x)$.
The same quantities $B(x)$ and $D(x)$ specify the three term recurrence
of the dual polynomials $Q_x(\mathcal{E})$ without the knowledge of the
spectrum $\mathcal{E}(n)$. It is required for them to be the eigenvectors
of the eigenvalue problem \eqref{polyeq}.
Likewise, $A_n$ and $C_n$ in \eqref{threeterm} specify the polynomials
$P_n(\eta)$ without the knowledge of the sinusoidal coordinate.
As for the dual polynomial $Q_x(\mathcal{E}(n))$, $A_n$ and $C_n$
provide the difference equation (in $n$), the solution of which needs
the explicit form of $\mathcal{E}(n)$. Let us stress that it is the
eigenvalue problem \eqref{polyeq} with the specific Hamiltonian
\eqref{rdQMHam} that determines the polynomials $P_n(\eta)$ and their dual
$Q_x(\mathcal{E})$, the spectrum $\mathcal{E}(n)$, the sinusoidal
coordinate $\eta(x)$ and the orthogonality measures $\phi_0(x)^2$
and $d_n^2$.

By comparing the explicit Heisenberg operator solution for $\eta(x)$ with
the three term recurrence relations \eqref{threeterm}, the coefficients
$A_n$ and $C_n$ are determined:
\begin{align}
  A_0&=R_{-1}(0)R_0(0)^{-1},\quad C_0=0,\\
  A_n&=\frac{R_{-1}(\mathcal{E}(n))
  +\eta(1)\bigl(\mathcal{E}(n)-B(0)\bigr)\alpha_+(\mathcal{E}(n))}
  {\alpha_+(\mathcal{E}(n))
  \bigl(\alpha_+(\mathcal{E}(n))-\alpha_-(\mathcal{E}(n))\bigr)}
  \quad(n\geq 1),
  \label{Anformula2}\\
  C_n&=\frac{R_{-1}(\mathcal{E}(n))
  +\eta(1)\bigl(\mathcal{E}(n)-B(0)\bigr)\alpha_-(\mathcal{E}(n))}
  {\alpha_-(\mathcal{E}(n))
  \bigl(\alpha_-(\mathcal{E}(n))-\alpha_+(\mathcal{E}(n))\bigr)}
  \quad(n\geq 1).
  \label{Cnformula2}
\end{align}
There is an important relation among the four important quantities:
\begin{equation}
  A_0\mathcal{E}(1)+B(0)\eta(1)=0.
  \label{ABrel}
\end{equation}
For the details of the derivation, see (4.45)--(4.53) of \cite{os12}.

%%%%%%%%%%%%%%%%%%%%%%%%%%%%%%%%%%%%%%%%%%%
%                                         %
% 2.7  Dual Closure Relation in dQM       %
%                                         %
%%%%%%%%%%%%%%%%%%%%%%%%%%%%%%%%%%%%%%%%%%%
\subsection{Dual Closure Relation in dQM}
\label{sec:Dualclo}

The {\em dual closure relation\/} has the same form as the closure
relation \eqref{closurerel} with the roles of the Hamiltonian $\mathcal{H}$
and the sinusoidal coordinate $\eta(x)$ interchanged \cite{os12}:
\begin{align}
  [\eta,[\eta,\mathcal{H}]\,]&=\mathcal{H}\,R_0^{\text{dual}}(\eta)
  +[\eta,\mathcal{H}]\,R_1^{\text{dual}}(\eta)+R_{-1}^{\text{dual}}(\eta),
  \label{dualclosurerel}
\end{align}
in which
\begin{align}
  R_1^{\text{dual}}(\eta(x))&=
  \bigl(\eta(x-i\beta)-\eta(x)\bigr)+\bigl(\eta(x+i\beta)-\eta(x)\bigr),
  \label{dualclcon1}\\
  R_0^{\text{dual}}(\eta(x))&=
  -\bigl(\eta(x-i\beta)-\eta(x)\bigr)\bigl(\eta(x+i\beta)-\eta(x)\bigr),\\
  R_{-1}^{\text{dual}}(\eta(x))&
  =\varepsilon\bigl(V_+(x)+V_-(x)\bigr)R_0^{\text{dual}}(\eta(x)).
  \label{dualclcon3}
\end{align}

The dual closure relation is the characteristic feature shared by all
the `Hamiltonians' $\widetilde{\mathcal{H}}$ which map a polynomial
in $\eta(x)$ into another.
Therefore its dynamical contents are not so constraining as the closure
relation, {\em except for\/} the rdQM exactly solvable case, where the
closure relation and the dual closure relations are on the same footing
and they form a dynamical symmetry algebra which is sometimes called
the Askey-Wilson algebra \cite{zhedanov,glz,terw,os12}.

%%%%%%%%%%%%%%%%%%%%%%%%%%%%%%%%%%%%%%%%%%%
%                                         %
% 2.8  Bochner's Theorem                  %
%                                         %
%%%%%%%%%%%%%%%%%%%%%%%%%%%%%%%%%%%%%%%%%%%
\subsection{Bochner's Theorem}
\label{sec:Boch}

In 1884 \cite{routh},  Routh showed that polynomials satisfying the three
term recurrence relations and a second order differential equation were
one of the {\em classical} polynomials, the Hermite, Laguerre, Jacobi
and Bessel. 
Later in 1929 \cite{bochner}, Bochner classified all polynomial solutions to
second order Sturm-Liouville operators with polynomial coefficients and
arrived at the same conclusions.
See \S\,20.1 of \cite{ismail} for more details.
This was a kind of No-Go theorem in oQM, since it declared
that no essentially new exactly solvable oQM could be achieved as the
solutions of the ordinary Schr\"{o}dinger equation.
Thus avoiding Bochner's theorem was one of the strongest motivations
for the introduction of the discrete quantum mechanics, whose difference
Schr\"{o}dinger equations were not constrained by the theorem and provided
various exactly solvable examples \cite{os12,os13}.
As shown in \S\ref{sec:Exce}, the new orthogonal (exceptional ($X_{\ell}$)
Laguerre and Jacobi) polynomials \cite{gomez,quesne,os16,os19} were discovered
in an attempt to evade the restrictions of the theorem by allowing the
polynomials to start at degree $\ell\ge1$.
The new orthogonal polynomials in dQM, the exceptional ($X_{\ell}$) Wilson,
Askey-Wilson, Racah and $q$-Racah polynomials satisfying second order
difference equations, were constructed by the present authors in less
than two years.

Here we present Bochner's theorem for dQM.
As in the original Bochner's theorem, the conditions for the lowest
three degrees $n=0,1$ and 2 are essential to determine the constraints.
Reformulation of Bochner's theorem for second order difference equations
was pursued by several authors.
In \cite{gh2}, it was shown (in our language) that if orthogonal
polynomials in $\eta(x)=\cos x$ satisfy $q$-difference equations
\eqref{polyeq}, then the conditions characterise the Askey-Wilson
polynomials.
In \cite{vinzhed2}, it was shown that if the $q$-difference equations
\eqref{polyeq} have polynomial solutions, then the sinusoidal coordinate
$\eta(x)$ is at most $q$-quadratic and that the polynomials are
{\em at most} Askey-Wilson polynomials.
In these papers, the distinction between the pure imaginary and the real
shifts is blurred.

The starting point is the difference equation for the polynomials
\eqref{polyeq} with \eqref{HtdQM}:
\begin{align}
  &\varepsilon V_+(x)\bigl(P_n(\eta(x-i\beta))- P_n(\eta(x))\bigr)
  +\varepsilon V_-(x)\bigl(P_n(\eta(x+i\beta))- P_n(\eta(x))\bigr)\n
  &\quad=\mathcal{E}(n)P_n(\eta(x))\quad(n=0,1,\ldots),
  \label{ddifeq}
\end{align}
and the three term recurrence relations \eqref{threeterm}.
The above equation is trivially satisfied for $n=0$, since $\mathcal{E}(0)=0$.
Two relations for $n=1$ and $n=2$ are linear equations
$$
  \begin{pmatrix}
  P_1(\eta(x-i\beta))- P_1(\eta(x))&P_1(\eta(x+i\beta))- P_1(\eta(x))\\
  P_2(\eta(x-i\beta))- P_2(\eta(x))&P_2(\eta(x+i\beta))- P_2(\eta(x))
  \end{pmatrix}
  \begin{pmatrix}
  V_+(x)\\V_-(x)
  \end{pmatrix}
  =
  \begin{pmatrix}
  \varepsilon^{-1}\mathcal{E}(1)P_1(\eta(x))\\
  \varepsilon^{-1}\mathcal{E}(2)P_2(\eta(x))
  \end{pmatrix},
$$
which determine the potential functions $V_+(x)$ and $V_-(x)$ uniquely.
By using the explicit forms of $P_1$ and $P_2$,
\begin{equation}
  A_0P_1(\eta)=\eta-B_0,\quad
  A_0A_1P_2(\eta)=(\eta-B_0)(\eta-B_1)-A_0C_1,
\end{equation}
derived from the three term recurrence relations \eqref{threeterm},
the above linear equation gives
\begin{equation}
  V_{\pm}(x)=\frac{S_2+S_1\eta(x\pm i\beta)}
  {\bigl(\eta(x\mp i\beta)-\eta(x)\bigr)
  \bigl(\eta(x\mp i\beta)-\eta(x\pm i\beta)\bigr)},
  \label{Vpmsol}
\end{equation}
in which $S_1$ and $S_2$ are a linear and a quadratic polynomial in $\eta(x)$,
respectively:
\begin{align}
  S_1&=-\varepsilon^{-1}\mathcal{E}(1)A_0P_1(\eta(x)),\n
  S_2&=\varepsilon^{-1}\mathcal{E}(2)A_0A_1P_2(\eta(x))
  -\varepsilon^{-1}\mathcal{E}(1)A_0P_1(\eta(x))
  \bigl(A_0P_1(\eta(x))-B_1\bigr).
\end{align}
These $V_{\pm}$ have essentially the same forms as those which will be
presented in \S\,\ref{sec:Uni}.
If the symmetric shift-addition property of the sinusoidal coordinate
$\eta(x)$ \eqref{eta1+etam1} is satisfied, the above expressions for
$V_{\pm}(x)$ \eqref{Vpmsol} are equivalent to $V_\pm$ in
\eqref{V+-sol}-\eqref{V+-t} with $L=2$.
Moreover when the symmetric shift-multiplication property
\eqref{eta1*etam1} is satisfied, these potential functions give the
exactly solvable models.
For rdQM, the sinusoidal coordinates satisfying \eqref{eta1+etam1}
can be classified into five types \eqref{etaform1'}--\eqref{etaform5'}
\cite{os12} and they also satisfy \eqref{eta1*etam1}.

%%%%%%%%%%%%%%%%%%%%%%%%%%%%%%%%%%%%%%%%%%%%%%%%%%%%%%%%%%%%%%%
%                                                             %
% 3. Unified Theory Exact and Quasi-exact Solvability         %
%                                                             %
%%%%%%%%%%%%%%%%%%%%%%%%%%%%%%%%%%%%%%%%%%%%%%%%%%%%%%%%%%%%%%%
\section{Unified Theory of Exactly Solvable dQM}
\label{sec:Uni}
\setcounter{equation}{0}

In this section we present a simple theory of constructing exactly
solvable `Hamiltonians' in dQM based on the two required properties of
the sinusoidal coordinates \eqref{eta1+etam1}-\eqref{eta1*etam1}.
The general strategy is to construct the similarity transformed `Hamiltonian'
$\widetilde{\mathcal{H}}$ \eqref{HtdQM} in such a way that it maps a
polynomial in $\eta(x)$ into another:
\begin{equation}
  \widetilde{\mathcal{H}}\mathcal{V}_n \subseteq
  \mathcal{V}_{n+L-2}\subset\mathcal{V}_{\infty}\quad
  (n\in\mathbb{Z}_{\geq 0}),
  \label{htvn}
\end{equation}
where $L$ is a fixed positive integer.
Here $\mathcal{V}_n$ is defined by \eqref{Vndef} and
$\mathcal{V}_{\infty}\eqdef\lim_{n\to\infty} \mathcal{V}_n$.
When $L=2$, the above relation \eqref{htvn} is simply the
{\em lower triangularity} of the `Hamiltonian' $\widetilde{\mathcal{H}}$,
leading to exact solvability.

In the following we will take the similarity transformed Hamiltonian
$\widetilde{\mathcal{H}}$ \eqref{Htdef} instead of $\mathcal{H}$ as the
starting point. That is, we reverse the argument and construct directly
the `Hamiltonian' $\widetilde{\mathcal{H}}$ \eqref{HtdQM} based on the
{\em sinusoidal coordinate\/} $\eta(x)$.
This section is a brief review of \cite{os14}.

%%%%%%%%%%%%%%%%%%%%%%%%%%%%%%%%%%%%%%%%%%%
%                                         %
% 3.1 Potential Functions                 %
%                                         %
%%%%%%%%%%%%%%%%%%%%%%%%%%%%%%%%%%%%%%%%%%%
\subsection{Potential Functions}
\label{sec:pot_fn}

The general form of the `Hamiltonian' $\widetilde{\mathcal{H}}$
mapping a polynomial in $\eta(x)$ into another is achieved by the
following form of the potential functions $V_{\pm}(x)$:
\begin{align}
  V_{\pm}(x)&=\frac{\widetilde{V}_{\pm}(x)}
  {\bigl(\eta(x\mp i\beta)-\eta(x)\bigr)
  \bigl(\eta(x\mp i\beta)-\eta(x\pm i\beta)\bigr)},
  \label{V+-sol}\\[4pt]
  \widetilde{V}_{\pm}(x)&=\sum_{\genfrac{}{}{0pt}{}{k,l\geq 0}{k+l\leq L}}
  v_{k,l}\,\eta(x)^k\eta(x\mp i\beta)^l,
  \label{V+-t}
\end{align}
where $L$ is a natural number indicating the degree of $\eta(x)$ in
$\widetilde{V}_{\pm}(x)$ and $v_{k,l}$ are real constants, with the
constraint $\sum\limits_{k+l=L}v_{k,l}^2\neq0$. It is important that
the same $v_{k,l}$ appears in both $\widetilde{V}_{\pm}(x)$.
The `Hamiltonian' $\widetilde{\mathcal{H}}$ with the above $V_\pm(x)$
maps a degree $n$ polynomial in $\eta(x)$ to a degree $n+L-2$ polynomial.
This can be shown elementarily based on the two basic properties of the
sinusoidal coordinates called the symmetric shift-addition property:
\begin{equation}
  \eta(x-i\beta)+\eta(x+i\beta)=(2+r_1^{(1)})\eta(x)+r_{-1}^{(2)},
  \label{eta1+etam1}
\end{equation}
and the symmetric shift-multiplication property:
\begin{equation}
  \eta(x-i\beta)\eta(x+i\beta)
  =\bigl(\eta(x)-\eta(-i\beta)\bigr)\bigl(\eta(x)-\eta(i\beta)\bigr),
  \label{eta1*etam1}
\end{equation}
together with $\eta(x)\neq\eta(x-i\beta)\neq\eta(x+i\beta)\neq\eta(x)$.
Here $r_1^{(1)}$ and $r_{-1}^{(2)}$ are real parameters.
In fact these parameters also appear in the three functions $R_0$, $R_1$
and $R_{-1}$ in the closure relation \eqref{closurerel}:
\begin{equation}
  R_1(y)=r_1^{(1)}y+r_1^{(0)},\quad
  R_0(y)=r_0^{(2)}y^2+r_0^{(1)}y+r_0^{(0)},\quad
  R_{-1}(y)=r_{-1}^{(2)}y^2+r_{-1}^{(1)}y+r_{-1}^{(0)}.
  \label{Ricoeff}
\end{equation}

Here are the lists of the known sinusoidal coordinates satisfying the
above two conditions \eqref{eta1+etam1}-\eqref{eta1*etam1}.
There are eight sinusoidal coordinates for the idQM:
\begin{alignat}{4}
  \text{(\romannumeral1)}:&\quad&\eta(x)&=x,
  &\quad -\infty<\,&x<\infty,&\quad&\gamma=1,
  \label{etaform1}\\
  \text{(\romannumeral2)}:&\quad&\eta(x)&=x^2,
  &\quad 0<\,&x<\infty,&\quad&\gamma=1,
  \label{etaform2}\\
  \text{(\romannumeral3)}:&\quad&\eta(x)&=1-\cos x,
  &\quad 0<\,&x<\pi,&\quad&\gamma\in\mathbb{R}_{\neq 0},
  \label{etaform3}\\
  \text{(\romannumeral4)}:&\quad&\eta(x)&=\sin x,
  &\quad -\tfrac{\pi}{2}<\,&x<\tfrac{\pi}{2},
  &\quad&\gamma\in\mathbb{R}_{\neq 0},
  \label{etaform4}\\
  \text{(\romannumeral5)}:&\quad&\eta(x)&=1-e^{-x},
  &\quad -\infty<\,&x<\infty,&\quad&\gamma\in\mathbb{R}_{\neq 0},
  \label{etaform5}\\
  \text{(\romannumeral6)}:&\quad&\eta(x)&=e^x-1,
  &\quad -\infty<\,&x<\infty,&\quad&\gamma\in\mathbb{R}_{\neq 0},
  \label{etaform6}\\
  \text{(\romannumeral7)}:&\quad&\eta(x)&=\cosh x-1,
  &\quad 0<\,&x<\infty,&\quad&\gamma\in\mathbb{R}_{\neq 0},
  \label{etaform7}\\
  \text{(\romannumeral8)}:&\quad&\eta(x)&=\sinh x,
  &\quad -\infty<\,&x<\infty, &\quad&\gamma\in\mathbb{R}_{\neq 0},
  \label{etaform8}
\end{alignat}
and five sinusoidal coordinates for the rdQM: ($0<q<1$)
\begin{alignat}{2}
  \text{(\romannumeral1)}':&\quad&\eta(x)&=x,
  \label{etaform1'}\\
  \text{(\romannumeral2)}':&\quad&\eta(x)&=\epsilon'x(x+d),
  \qquad \qquad \quad \
  \epsilon'=\Bigl\{\begin{array}{ll}
  1&\text{for }\ d>-1\\
  -1&\text{for }\ d<-N
  \end{array},
  \label{etaform2'}\\
  \text{(\romannumeral3)}':&\quad&\eta(x)&=1-q^x,
  \label{etaform3'}\\
  \text{(\romannumeral4)}':&\quad&\eta(x)&=q^{-x}-1,
  \label{etaform4'}\\
  \text{(\romannumeral5)}':&\quad&\eta(x)&=\epsilon'(q^{-x}-1)(1-dq^x),\quad
  \epsilon'=\Bigl\{\begin{array}{ll}
  1&\text{for }\ d<q^{-1}\\
  -1&\text{for }\ d>q^{-N}
  \end{array}.
  \label{etaform5'}
\end{alignat}
As shown in detail in \S4C of \cite{os12}, the above five sinusoidal
coordinates for rdQM \eqref{etaform1'}--\eqref{etaform5'} exhaust all
the solutions of \eqref{eta1+etam1}--\eqref{eta1*etam1} up to a
multiplicative factor. On the other hand, those for idQM
(\romannumeral1)--(\romannumeral8) \eqref{etaform1}--\eqref{etaform8}
are merely typical examples satisfying all the postulates for the
sinusoidal coordinate \eqref{eta1+etam1}--\eqref{eta1*etam1} and the
extra one used for the shape invariance, (3.50) of \cite{os14}.

The essential part of the formula \eqref{V+-sol} is the denominators.
They have the same form as the generic formula for the coefficients of
the three term recurrence relations of the orthogonal polynomials,
\eqref{Anformula2} and \eqref{Cnformula2} (see (4.52) and (4.53) in
\cite{os12}).
The translation rules are the {\em duality\/} correspondence itself,
\eqref{duality1}-\eqref{duality2} (see also (3.14)--(3.18) in \cite{os12}):
\begin{gather}
  \mathcal{E}(n)\to\eta(x),\qquad-A_n\to V_+(x),
  \qquad-C_n\to V_-(x),\n
  \alpha_+\bigl(\mathcal{E}(n)\bigr)\to\eta(x-i\beta)-\eta(x),\quad
  \alpha_-\bigl(\mathcal{E}(n)\bigr)\to\eta(x+i\beta)-\eta(x).
\end{gather}

Some of the parameters $v_{k,l}$ in \eqref{V+-t} are redundant.
It is sufficient to keep $v_{k,l}$ with $l=0,1$.
The remaining $2L+1$ parameters $v_{k,l}$ ($k+l\leq L$, $l=0,1$) are
independent, with one of which corresponds to the overall normalisation
of the Hamiltonian. See \S\,II of \cite{os14} for the actual proof of the
property \eqref{htvn}.

The $L=2$ case is exactly solvable. This corresponds to the general
hypergeometric equations having at most degree two polynomial coefficients.
Since the Hamiltonian of the polynomial space $\widetilde{\mathcal{H}}$
is expressed as an upper triangular matrix, its eigenvalues and
eigenvectors are easily obtained explicitly.
For the solutions of a full quantum mechanical problem, however,
one needs the square-integrable groundstate wavefunction $\phi_0(x)$
\eqref{Aphi0=0}, which is essential for the existence of the Hamiltonian
$\mathcal{H}$ and the verification of its hermiticity. These conditions
would usually restrict the ranges of the parameters $v_{0,0},\ldots,v_{2,0}$.
It is easy to verify that the explicit examples of the potential functions
$V(x)$, $V^*(x)$ and $B(x)$, $D(x)$ in \eqref{ex4}--\eqref{ex9} are simply
reproduced by proper choices of the parameters $\{v_{k,l}\}$.
It should be stressed that the above form of the potential function
\eqref{V+-sol}--\eqref{V+-t} provides a {\em unified proof\/} of the
shape invariance relation \eqref{shapeinv}, the closure relation
\eqref{closurerel} and the dual closure relation \eqref{dualclosurerel}
in the $\widetilde{\mathcal{H}}$ scheme, see \cite{os14} for more details.
This is in good contrast with the various explicit examples presented
in the preceding section. The two solution methods, the shape invariance
and the closure relation are verified for each example.

%%%%%%%%%%%%%%%%%%%%%%%%%%%%%%%%%%%%%%%%%%%
%                                         %
% 3.2 Askey-Wilson algebra                %
%                                         %
%%%%%%%%%%%%%%%%%%%%%%%%%%%%%%%%%%%%%%%%%%%
\subsection{Askey-Wilson Algebra}
\label{AW3}

Here we briefly comment on the Askey-Wilson algebra, which are generated
by the closure plus the dual closure relations. By simply expanding the
double commutators in the closure \eqref{closurerel} and the dual closure
\eqref{dualclosurerel} relations, we obtain two cubic relations generated
by the two operators $\mathcal{H}$ and $\eta$:
\begin{align}
  &\mathcal{H}^2\eta-(2+r_1^{(1)})\mathcal{H}\eta\mathcal{H}
  +\eta\mathcal{H}^2-r_1^{(0)}(\mathcal{H}\eta+\eta\mathcal{H})
  -r_0^{(0)}\eta
  =r_{-1}^{(2)}\mathcal{H}^2+r_{-1}^{(1)}\mathcal{H}+r_{-1}^{(0)},
  \label{expandclos}\\
  &\eta^2\mathcal{H}-(2+r_1^{(1)})\eta\mathcal{H}\eta+\mathcal{H}\eta^2
  -r_{-1}^{(2)}(\eta\mathcal{H}+\mathcal{H}\eta)
  +\eta(-i\beta)\eta(i\beta)\mathcal{H}
  =r_1^{(0)}\eta^2+r_{-1}^{(1)}\eta+\varepsilon v_{0,0}.
  \label{expanddualclos}
\end{align}
{}From its structure, the closure relation is at most linear in $\eta$
and at most quadratic in $\mathcal{H}$.
So the l.h.s.\! of \eqref{expandclos} has terms containing one factor of
$\eta$ and the r.h.s, none. It is simply $R_{-1}(\mathcal{H})$.
Likewise, the l.h.s.\! of \eqref{expanddualclos} has terms containing
one factor of $\mathcal{H}$ and the r.h.s, none. It is simply
$R^{\text{dual}}_{-1}(\eta)$. In \eqref{expanddualclos},
$\eta(-i\beta)\eta(i\beta)$ is just a real number, not an operator.
These have the same form as the so-called Askey-Wilson algebra, which has
many different expressions \cite{zhedanov,glz,vinzhed,terw}.
While the Askey-Wilson algebra has no inherent structure, the closure
relation \eqref{closurerel} has the right structure to lead to the
Heisenberg operator solution for $\eta(x)$, whose positive and negative
frequency parts are the annihilation-creation operators \cite{os7,os12,os13}.
It is the Hamiltonian and the annihilation-creation operators that form
the {\em dynamical symmetry algebra\/} of the system \cite{os12,os13},
not the closure or dual-closure relations, nor the Askey-Wilson algebra
relations.

%%%%%%%%%%%%%%%%%%%%%%%%%%%%%%%%%%%%%%%%%%%
%                                         %
% 3.3 Quasi-exactly solvable              %
%                                         %
%%%%%%%%%%%%%%%%%%%%%%%%%%%%%%%%%%%%%%%%%%%
\subsection{Quasi-Exact Solvability in dQM}
\label{sec:QES}

The unified theory is general enough to generate
{\em quasi-exactly solvable\/} Hamiltonians in the same manner.
The quasi-exact solvability means, in contrast to the exact solvability,
that only a finite number of energy eigenvalues and the corresponding
eigenfunctions can be obtained exactly.
Many examples are known in the oQM \cite{morozov,turb,ush}, but only a
few are known in dQM in spite of the proposal that the $sl(2,R)$ algebra
characterisation of quasi-exact solvability could be extended to
difference Schr\"{o}dinger equations \cite{turb}.
The unified theory also incorporates the known examples of
quasi-exactly solvable Hamiltonians in dQM \cite{os10,deltaqes,newqes}.
A new type of quasi-exactly solvable Hamiltonians is constructed.
The present approach reveals the common structure underlying the exactly
and quasi-exactly solvable theories, in particular, the important roles
played by the sinusoidal coordinates.

The higher $L\ge3$ cases are obviously non-solvable. Among them, the
tame non-solvability of $L=3$ and $4$ can be made quasi-exactly solvable
(QES) by adding suitable compensation terms. This is a simple generalisation
of the method of Sasaki \& Takasaki \cite{st1} for multi-particle QES in
the oQM.
For a given positive integer $M$, let us try to find a QES `Hamiltonian'
$\widetilde{\mathcal{H}}$, or more precisely its modification
$\widetilde{\mathcal{H}}'$, having an invariant polynomial subspace
$\mathcal{V}_M$:
\begin{equation}
  \widetilde{\mathcal{H}}'\mathcal{V}_M\subseteq\mathcal{V}_M.
\end{equation}
For $L=3$, $\widetilde{\mathcal{H}}'$ is defined by adding one single
compensation term of degree one
\begin{equation}
  \widetilde{\mathcal{H}}'\eqdef\widetilde{\mathcal{H}}-e_0(M)\eta(x),
  \label{L3ham}
\end{equation}
and we have achieved the quasi-exact solvability
$\widetilde{\mathcal{H}}'\mathcal{V}_M\subseteq\mathcal{V}_M$.
Known discrete QES examples belong to this class \cite{deltaqes,newqes}.

For $L=4$ case, $\widetilde{\mathcal{H}}'$ is defined by adding a linear
and a quadratic in $\eta(x)$ compensation terms to the Hamiltonian
$\widetilde{\mathcal{H}}$:
\begin{equation}
  \widetilde{\mathcal{H}}'\eqdef
  \widetilde{\mathcal{H}}-e_0(M)\eta(x)^2-e_1(M)\eta(x),
  \label{L4ham}
\end{equation}
and one condition between $v_{3,1}$ and $v_{4,0}$ is imposed.
Then we have
$\widetilde{\mathcal{H}}'\mathcal{V}_M\subseteq\mathcal{V}_M$.
This type of QES theory is new.
For explicit forms of $e_0(M)$ and $e_1(M)$, see \cite{os14}.

%%%%%%%%%%%%%%%%%%%%%%%%%%%%%%%%%%%%%%%%%%%%%%%%%%%%%%%%%%%%%%%
%                                                             %
% 4. New Orthogonal Polynomials                               %
%                                                             %
%%%%%%%%%%%%%%%%%%%%%%%%%%%%%%%%%%%%%%%%%%%%%%%%%%%%%%%%%%%%%%%
\section{New Orthogonal Polynomials}
\label{sec:Exce}
\setcounter{equation}{0}

In this section we present a brief review of the recent hot topic,
the discovery of infinitely many new orthogonal polynomials satisfying
second order differential or difference equations. They are obtained as
the main parts of eigenfunctions of exactly solvable quantum mechanics
in all the three categories, oQM, idQM and rdQM.
They are the exceptional ($X_{\ell}$) Laguerre (XL) and Jacobi (XJ)
polynomials in oQM, the exceptional Meixner-Pollaczek (XMP), Wilson (XW) and
Askey-Wilson (XAW) polynomials in idQM and the exceptional Meixner (XM),
Racah (XR) and $q$-Racah (X$q$R) polynomials in rdQM.
The XMP and XM are {\em new results\/}.
For any positive integer $\ell$ these $X_{\ell}$ polynomials have degrees
$\ell+n$, $n=0,1,2,\ldots$ and are orthogonal with each other
\eqref{orthooQM}, \eqref{orthoidQM}, \eqref{orthordQM} with respect
to explicitly given positive definite weight functions.
Thus they do not satisfy the three term recurrence relations
\eqref{threeterm} and therefore the restrictions of Bochner's theorem
and its discrete counterparts in \S\ref{sec:Boch} do not apply.
As will be explained later, the quantum mechanical Hamiltonians for
these exceptional polynomials are deformations of those for the original
polynomials in terms of a degree $\ell$ eigenpolynomial with twisted
parameters. The deformation terms satisfy the same type of equations as
the original polynomials, but with the twisted parameters.
The parameters are so chosen that the deformed system retain
the shape invariance, too. The general scheme of the deformation is common
for various exceptional polynomials as shown in detail below.
The original system corresponds to $\ell=0$.

It should be stressed that the $X_{\ell}$ Jacobi polynomials
provide infinitely many global solutions of Fuchsian differential equations
having $3+\ell$ regular singularities \cite{hos}.

%%%%%%%%%%%%%%%%%%%%%%%%%%%%%%%%%%%%%%%%%%%
%                                         %
% 4.1 Deforming Polynomials               %
%                                         %
%%%%%%%%%%%%%%%%%%%%%%%%%%%%%%%%%%%%%%%%%%%
\subsection{Deforming Polynomials}
\label{subsec:xil}

In order to introduce new orthogonal polynomials, we prepare
the deforming polynomials. The deforming polynomials $\xi_{\ell}(\eta)$
are eigenpolynomials $P_{\ell}(\eta)$ with twisted parameters
(or twisted coordinate) and they are crucial for the construction of new
orthogonal polynomials.
Two identities relating $\xi_{\ell}$ with $\bm{\lambda}$ and
$\bm{\lambda}+\bm{\delta}$, \eqref{xil(l+d)oQM}--\eqref{xil(l)oQM},
\eqref{xil(l+d)idQM}--\eqref{xil(l)idQM} and
\eqref{xil(l+d)rdQM}--\eqref{xil(l)rdQM},
will play important roles in the derivation of various results.

For oQM there are two types of the exceptional Jacobi polynomials,
called XJ1 and XJ2. Their confluent limits give the exceptional Laguerre
polynomials, XL1 and XL2.
The deforming polynomials are:
\begin{equation}
  \xi_{\ell}(\eta;\bm{\lambda})\eqdef\left\{
  \begin{array}{ll}
  P_{\ell}(-\eta;\bm{\lambda}+(\ell-1)\bm{\delta})
  &\hspace{46.2mm}:\text{XL1}\\[2pt]
  P_{\ell}(\eta;\mathfrak{t}(\bm{\lambda}+(\ell-1)\bm{\delta})),
  &\mathfrak{t}(\bm{\lambda})\eqdef\left\{
  \begin{array}{ll}
  -\lambda_1-1&:\text{XL2}\\
  (\lambda_1,-\lambda_2-1)&:\text{XJ1}\\
  (-\lambda_1-1,\lambda_2)&:\text{XJ2}
  \end{array}\right.
  \end{array}\right..
  \label{xiloQM}
\end{equation}
Explicitly they are
\begin{equation}
  \xi_{\ell}(\eta;\bm{\lambda})=\left\{
  \begin{array}{ll}
  L_{\ell}^{(g+\ell-\frac32)}(-\eta)&:\text{XL1}\\
  L_{\ell}^{(-g-\ell-\frac12)}(\eta)&:\text{XL2}
  \end{array}\right.,\quad
  \xi_{\ell}(\eta;\bm{\lambda})=\left\{
  \begin{array}{ll}
  P_{\ell}^{(g+\ell-\frac32,-h-\ell-\frac12)}(\eta)&:\text{XJ1}\\
  P_{\ell}^{(-g-\ell-\frac12,h+\ell-\frac32)}(\eta)&:\text{XJ2}
  \end{array}\right..
\end{equation}
The two identities relating $\xi_{\ell}(\eta;\bm{\lambda})$ and
$\xi_{\ell}(\eta;\bm{\lambda}+\bm{\delta})$ are:
\begin{align}
  &d_1(\bm{\lambda}+\ell\bm{\delta})\xi_{\ell}(\eta;\bm{\lambda})
  +d_2(\eta)\partial_{\eta}\xi_{\ell}(\eta;\bm{\lambda})
  =d_1(\bm{\lambda})\xi_{\ell}(\eta;\bm{\lambda}+\bm{\delta}),
  \label{xil(l+d)oQM}\\
  &d_3(\bm{\lambda},\ell)\xi_{\ell}(\eta;\bm{\lambda}+\bm{\delta})
  +\frac{c_2(\eta)}{d_2(\eta)}\,
  \partial_{\eta}\xi_{\ell}(\eta;\bm{\lambda}+\bm{\delta})
  =d_3(\bm{\lambda}+\ell\bm{\delta},\ell)\xi_{\ell}(\eta;\bm{\lambda}),
  \label{xil(l)oQM}
\end{align}
where $d_1(\bm{\lambda})$, $d_2(\eta)$ and $d_3(\bm{\lambda},\ell)$
are given by
\begin{align}
  &d_1(\bm{\lambda})\eqdef\left\{
  \begin{array}{ll}
  1&:\text{XL1}\\
  g+\frac12&:\text{XL2, XJ2}\\[2pt]
  h+\frac12&:\text{XJ1}
  \end{array}\right.\!\!,
  \quad
  d_2(\eta)\eqdef\left\{
  \begin{array}{ll}
  1&:\text{XL1}\\
  -\eta&:\text{XL2}\\
  \mp(1\pm \eta)&:\text{XJ1/XJ2}
  \end{array}\right.\!\!,
  \label{d1,d2}\\
  &d_3(\bm{\lambda},\ell)\eqdef\left\{
  \begin{array}{ll}
  g+\ell-\frac12&:\text{XL1, XJ1}\\
  1&:\text{XL2}\\
  h+\ell-\frac12&:\text{XJ2}
  \end{array}\right.\!\!.
\end{align}
Note that these two identities, \eqref{xil(l+d)oQM} and \eqref{xil(l)oQM},
imply the differential equation for the deforming polynomial,
\begin{align}
  &c_2(\eta)\partial_{\eta}^2\xi_{\ell}(\eta;\bm{\lambda})
  +\tilde{c}_1(\eta,\bm{\lambda},\ell)
  \partial_{\eta}\xi_{\ell}(\eta;\bm{\lambda})
  =-\tfrac14\widetilde{\mathcal{E}}(\ell;\bm{\lambda})
  \xi_{\ell}(\eta;\bm{\lambda}),
  \label{xildiffeqoQM}\\
  &\tilde{c}(\eta,\bm{\lambda},\ell)\eqdef\left\{
  \begin{array}{ll}
  c_1(-\eta,\bm{\lambda}+(\ell-1)\bm{\delta})&:\text{XL1}\\
  c_1(\eta,\mathfrak{t}(\bm{\lambda}+(\ell-1)\bm{\delta}))
  &:\text{XL2, XJ1, XJ2}
  \end{array}\right.,\\
  &\widetilde{\mathcal{E}}(\ell;\bm{\lambda})\eqdef\left\{
  \begin{array}{ll}
  -\mathcal{E}(\ell;\bm{\lambda})&:\text{XL1}\\
  \mathcal{E}(\ell;\mathfrak{t}(\bm{\lambda}))
  &:\text{XL2, XJ1, XJ2}
  \end{array}\right.,
\end{align}
which is to be compared with the differential equation for the eigenpolynomial
\eqref{polyeq},
\begin{equation}
  c_2(\eta)\partial_{\eta}^2P_n(\eta;\bm{\lambda})
  +c_1(\eta,\bm{\lambda})\partial_{\eta}P_n(\eta;\bm{\lambda})
  =-\tfrac14\mathcal{E}(n;\bm{\lambda})P_n(\eta;\bm{\lambda}).
  \label{PndiffeqoQM}
\end{equation}
Note that $\mathcal{E}(\ell;\mathfrak{t}(\bm{\lambda}))
=\mathcal{E}(\ell;\mathfrak{t}(\bm{\lambda}+(\ell-1)\bm{\delta}))$.
The deforming polynomial $\xi_\ell(\eta;\bm{\lambda})$ has the same
sign in the orthogonality domain, that is $(0,\infty)$ for L and
$(-1,1)$ for J, see (2.39), (2.40) of \cite{os18} and (3.2) of \cite{os21}.

For idQM, we restrict parameters:
\begin{align}
  \!\!\!\!\!\text{XMP}:&\ \ell\ \text{even},
  \label{paraMPrest}\\
  \!\!\!\!\!\text{XW}:&\ a_1,a_2\in\mathbb{R},
  \ \ \{a_3^*,a_4^*\}=\{a_3,a_4\} \ (\text{as a set}),
  \ \ 0<a_j<\text{Re}\,a_k\ \ (j=1,2;k=3,4),\!\!
  \label{paraWrest}\\
  \!\!\!\!\!\text{XAW}:&\ a_1,a_2\in\mathbb{R},
  \ \ \{a_3^*,a_4^*\}=\{a_3,a_4\} \ (\text{as a set}),
  \ \ 1>a_j>|a_k|\ \ (j=1,2;k=3,4).\!\!
  \label{paraAWrest}
\end{align}
The deforming polynomials for XMP, XW and XAW are:
\begin{equation}
  \xi_{\ell}(\eta;\bm{\lambda})\eqdef
  P_{\ell}\bigl(\eta;\mathfrak{t}
  \bigl(\bm{\lambda}+(\ell-1)\bm{\delta}\bigr)\bigr),\quad
  \mathfrak{t}(\bm{\lambda})\eqdef\left\{
  \begin{array}{ll}
  -\lambda_1&:\text{XMP}\\
  (-\lambda_1,-\lambda_2,\lambda_3,\lambda_4)&:\text{XW, XAW}
  \end{array}\right..
\end{equation}
The two identities relating $\xi_{\ell}(\eta(x);\bm{\lambda})$ and
$\xi_{\ell}(\eta(x);\bm{\lambda}+\bm{\delta})$ are \cite{os20}:
\begin{align}
  &\tfrac{i}{\varphi(x)}\bigl(
  v_1^*(x;\bm{\lambda}+\ell\bm{\delta})e^{\frac{\gamma}{2}p}
  -v_1(x;\bm{\lambda}+\ell\bm{\delta})e^{-\frac{\gamma}{2}p}\bigr)
  \xi_{\ell}(\eta(x);\bm{\lambda})
  =\hat{f}_{\ell,0}(\bm{\lambda})\xi_{\ell}(\eta(x);\bm{\lambda}+\bm{\delta}),
  \label{xil(l+d)idQM}\\
  &\tfrac{-i}{\varphi(x)}\bigl(
  v_2(x;\bm{\lambda}+(\ell-1)\bm{\delta})e^{\frac{\gamma}{2}p}
  -v_2^*(x;\bm{\lambda}+(\ell-1)\bm{\delta})e^{-\frac{\gamma}{2}p}\bigr)
  \xi_{\ell}(\eta(x);\bm{\lambda}+\bm{\delta})\n
  &\qquad\quad
  =\hat{b}_{\ell,0}(\bm{\lambda})\xi_{\ell}(\eta(x);\bm{\lambda}),
  \label{xil(l)idQM}
\end{align}
where $v_1(x;\bm{\lambda})$, $v_2(x;\bm{\lambda})$ are the factors of
the potential function $V(x;\bm{\lambda})$:
\begin{align}
  &\hspace{40mm}V(x;\bm{\lambda})=-\sqrt{\kappa}\,
  \frac{v_1(x;\bm{\lambda})v_2(x;\bm{\lambda})}
  {\varphi(x)\varphi(x-i\frac{\gamma}{2})},
  \label{factorV}\\
  &\!\!\!v_1(x;\bm{\lambda})\eqdef\left\{
  \begin{array}{ll}
  \!\!i(a+ix)&\!\!:\text{XMP}\\
  \!\!\prod_{j=1}^2(a_j+ix)&\!\!:\text{XW}\\
  \!\!e^{-ix}\prod_{j=1}^2(1-a_je^{ix})&\!\!:\text{XAW}
  \end{array}\right.\!\!\!,
  \ \ v_2(x;\bm{\lambda})\eqdef\left\{
  \begin{array}{ll}
  \!\!i&\!\!:\text{XMP}\\
  \!\!\prod_{j=3}^4(a_j+ix)&\!\!:\text{XW}\\
  \!\!e^{-ix}\prod_{j=3}^4(1-a_je^{ix})&\!\!:\text{XAW}
  \end{array}\right.\!\!\!.\!\!\!
  \label{v1v2}
\end{align}
The constants $\hat{f}_{\ell,n}(\bm{\lambda})$ and
$\hat{b}_{\ell,n}(\bm{\lambda})$ are given by
\begin{equation}
  \hat{f}_{\ell,n}(\bm{\lambda})\eqdef\left\{
  \begin{array}{ll}
  \!\!2a+n&\!\!\!:\text{XMP}\\
  \!\!a_1+a_2+n&\!\!\!:\text{XW}\\
  \!\!-q^{-\frac{n-\ell}{2}}(1-a_1a_2q^n)&\!\!\!:\text{XAW}
  \end{array}\right.\!\!\!,
  \ \ \hat{b}_{\ell,n}(\bm{\lambda})\eqdef\left\{
  \begin{array}{ll}
  \!\!2&\!\!\!:\text{XMP}\\
  \!\!a_3+a_4+n+2\ell-1&\!\!\!:\text{XW}\\
  \!\!-q^{-\frac{n+\ell}{2}}(1-a_3a_4q^{n+2\ell-1})&\!\!\!:\text{XAW}
  \end{array}\right.\!\!\!.\!\!\!
  \label{hatfhatb}
\end{equation}
These two identities \eqref{xil(l+d)idQM} and \eqref{xil(l)idQM}
imply the difference equation for the deforming polynomial,
\begin{align}
  &\Bigl(V\bigl(x;\mathfrak{t}(\bm{\lambda}+(\ell-1)\bm{\delta})\bigr)
  (e^{\gamma p}-1)
  +V^*\bigl(x;\mathfrak{t}(\bm{\lambda}+(\ell-1)\bm{\delta})\bigr)
  (e^{-\gamma p}-1)\Bigr)\xi_{\ell}(\eta(x);\bm{\lambda})\n
  &\qquad\quad
  =\mathcal{E}(\ell;\mathfrak{t}(\bm{\lambda}))
  \xi_{\ell}(\eta(x);\bm{\lambda}),
  \label{xildiffeqidQM}
\end{align}
which should be compared with the difference equation for the eigenpolynomial
\begin{equation}
  \bigl(V(x;\bm{\lambda})(e^{\gamma p}-1)
  +V^*(x;\bm{\lambda})(e^{-\gamma p}-1)\bigr)P_n(\eta(x);\bm{\lambda})
  =\mathcal{E}(n;\bm{\lambda})P_n(\eta(x);\bm{\lambda}).
  \label{PndiffeqidQM}
\end{equation}
Note that $\mathcal{E}(\ell;\mathfrak{t}(\bm{\lambda}))
=\mathcal{E}(\ell;\mathfrak{t}(\bm{\lambda}+(\ell-1)\bm{\delta}))$.
For the appropriate parameter ranges, the deforming polynomial
$\xi_{\ell}(\eta(x);\bm{\lambda})$ has no zero in the rectangular domain
$x_1\leq\text{Re}\,x\leq x_2$, $|\text{Im}\,x|\leq|\gamma|$,
which is necessary for the hermiticity of the Hamiltonian.

For rdQM, the deforming polynomials for XM, XR and X$q$R are:
\begin{align}
  &\check{\xi}_{\ell}(x;\bm{\lambda})\eqdef
  \xi_{\ell}\bigl(\eta(x;\bm{\lambda}+(\ell-1)\bm{\delta});\bm{\lambda}\bigr),
  \quad
  \check{P}_n(x;\bm{\lambda})
  \eqdef P_n\bigl(\eta(x;\bm{\lambda});\bm{\lambda}\bigr),\\
  &\check{\xi}_{\ell}(x;\bm{\lambda})\eqdef\left\{
  \begin{array}{ll}
  c^{\ell}\check{P}_{\ell}\bigl(-(x+\beta+\ell-1);
  \bm{\lambda}+(\ell-1)\bm{\delta}\bigr)
  &\!\!\!:\text{XM}\\
  \check{P}_{\ell}\bigl(x;\mathfrak{t}
  \bigl(\bm{\lambda}+(\ell-1)\bm{\delta}\bigr)\bigr),
  \ \ \mathfrak{t}(\bm{\lambda})\eqdef
  (\lambda_4-\lambda_1,\lambda_4-\lambda_2,\lambda_3,\lambda_4)
  &\!\!\!:\text{XR, X$q$R}
  \end{array}\right.\!\!\!,\!
\end{align}
which satisfies the normalisation
\begin{equation}
  \xi_{\ell}(0;\bm{\lambda})=1.
\end{equation}
The two identities relating $\check{\xi}_{\ell}(x;\bm{\lambda})$ and
$\check{\xi}_{\ell}(x;\bm{\lambda}+\bm{\delta})$ are \cite{os23}:
\begin{align}
  &\frac{1}{\varphi(x;\bm{\lambda}+\ell\bm{\delta}+\tilde{\bm{\delta}})}
  \Bigl(
  v_1^B(x;\bm{\lambda}+\ell\bm{\delta})
  -v_1^D(x;\bm{\lambda}+\ell\bm{\delta})e^{\partial}\Bigr)
  \check{\xi}_{\ell}(x;\bm{\lambda})
  =\hat{f}_{\ell,0}(\bm{\lambda})
  \check{\xi}_{\ell}(x;\bm{\lambda}+\bm{\delta}),
  \label{xil(l+d)rdQM}\\
  &\frac{1}{\varphi(x;\bm{\lambda}+(\ell-1)\bm{\delta}+\tilde{\bm{\delta}})}
  \Bigl(
  v_2^B(x;\bm{\lambda}+(\ell-1)\bm{\delta})
  -v_2^D(x;\bm{\lambda}+(\ell-1)\bm{\delta})e^{-\partial}\Bigr)
  \check{\xi}_{\ell}(x;\bm{\lambda}+\bm{\delta})\n
  &\qquad\quad
  =\hat{b}_{\ell,0}(\bm{\lambda})\check{\xi}_{\ell}(x;\bm{\lambda}).
  \label{xil(l)rdQM}
\end{align}
Here $v_1^B(x;\bm{\lambda})$, $v_2^B(x;\bm{\lambda})$,
$v_1^D(x;\bm{\lambda})$, $v_2^D(x;\bm{\lambda})$ are the factors of
the potential functions $B(x;\bm{\lambda})$ and $D(x;\bm{\lambda})$:
\begin{align}
  B(x;\bm{\lambda})&=-\sqrt{\kappa}\,
  \frac{v_1^B(x;\bm{\lambda})v_2^B(x;\bm{\lambda})}
  {\varphi(x;\bm{\lambda}+\tilde{\bm{\delta}})
  \varphi(x+\frac12;\bm{\lambda}+\tilde{\bm{\delta}})},
  \label{factorB}\\
  D(x;\bm{\lambda})&=-\sqrt{\kappa}\,
  \frac{v_1^D(x;\bm{\lambda})v_2^D(x;\bm{\lambda})}
  {\varphi(x;\bm{\lambda}+\tilde{\bm{\delta}})
   \varphi(x-\frac12;\bm{\lambda}+\tilde{\bm{\delta}})},
  \label{factorD}\\
  v_1^B(x;\bm{\lambda})&\eqdef
  \left\{
  \begin{array}{ll}
  -c^{\frac12}&:\text{XM}\\
  d^{-1}(x+a)(x+b)&:\text{XR}\\
  {\displaystyle
  \frac{q^{-x}}{1-d}(1-aq^x)(1-bq^x)
  }&:\text{X$q$R}
  \end{array}\right.\!,
  \label{v1B}\\
  v_2^B(x;\bm{\lambda})&\eqdef
  \left\{
  \begin{array}{ll}
  c^{-\frac12}B(x;\bm{\lambda})&:\text{XM}\\
  d^{-1}(x+c)(x+d)&:\text{XR}\\
  {\displaystyle
  \frac{q^{-x}}{1-d}(1-cq^x)(1-dq^x)
  }&:\text{X$q$R}
  \end{array}\right.\!,
  \label{v2B}\\
  v_1^D(x;\bm{\lambda})&\eqdef
  \left\{
  \begin{array}{ll}
  -c^{-\frac12}&:\text{XM}\\
  d^{-1}(x+d-a)(x+d-b)&:\text{XR}\\
  {\displaystyle
  \frac{q^{-x}}{1-d}abd^{-1}(1-a^{-1}dq^x)(1-b^{-1}dq^x)
  }&:\text{X$q$R}
  \end{array}\right.\!,
  \label{v1D}\\
  v_2^D(x;\bm{\lambda})&\eqdef
  \left\{
  \begin{array}{ll}
  c^{\frac12}D(x;\bm{\lambda})&:\text{XM}\\
  d^{-1}(x+d-c)x&:\text{XR}\\
  {\displaystyle
  \frac{q^{-x}}{1-d}c(1-c^{-1}dq^x)(1-q^x)
  }&:\text{X$q$R}
  \end{array}\right.\!,
  \label{v2D}
\end{align}
where $\tilde{\bm{\delta}}$ is
\begin{equation}
  \tilde{\bm{\delta}}\eqdef\left\{
  \begin{array}{ll}
  (-1,0)&:\text{XM}\\
  (0,0,-1,-1)&:\text{XR,X$q$R}
  \end{array}\right..
\end{equation}
The constants $\hat{f}_{\ell,n}(\bm{\lambda})$ and
$\hat{b}_{\ell,n}(\bm{\lambda})$ are given by
\begin{align}
  \hat{f}_{\ell,n}(\bm{\lambda})&\eqdef
  \left\{
  \begin{array}{ll}
  {\displaystyle
  \frac{1-c}{\sqrt{c}}\frac{\beta+n+2\ell-1}{\beta+\ell-1}
  }&:\text{XM}\\
  {\displaystyle
  (a+b-d+n)\frac{c+2\ell+n-1}{c+\ell-1}
  }&:\text{XR}\\
  {\displaystyle
  q^{-n}(1-abd^{-1}q^n)\frac{1-cq^{2\ell+n-1}}{1-cq^{\ell-1}}
  }&:\text{X$q$R}
  \end{array}\right.,\\
  \hat{b}_{\ell,n}(\bm{\lambda})&\eqdef
  \left\{
  \begin{array}{ll}
  {\displaystyle
  \frac{\sqrt{c}}{1-c}(\beta+\ell-1)
  }&:\text{XM}\\
  c+\ell-1&:\text{XR}\\
  1-cq^{\ell-1}&:\text{X$q$R}
  \end{array}\right.\!.
  \label{fhlnbhln}
\end{align}
Again these two identities \eqref{xil(l+d)rdQM} and \eqref{xil(l)rdQM}
imply the difference equation for the deforming polynomial,
\begin{align}
  &\bigl(\widetilde{B}(x;\bm{\lambda})(1-e^{\partial})
  +\widetilde{D}(x;\bm{\lambda})(1-e^{-\partial})\bigr)
  \check{\xi}_{\ell}(x;\bm{\lambda})
  =\widetilde{\mathcal{E}}(\ell;\bm{\lambda})
  \check{\xi}_{\ell}(x;\bm{\lambda}),
  \label{xildiffeqrdQM}\\
  &\quad\widetilde{B}(x;\bm{\lambda})\eqdef\left\{
  \begin{array}{ll}
  -D\bigl(-(x+\beta+\ell-1);\bm{\lambda}+(\ell-1)\bm{\delta}\bigr)
  &:\text{XM}\\
  B\bigl(x;\mathfrak{t}(\bm{\lambda}+(\ell-1)\bm{\delta})\bigr)
  &:\text{XR, X$q$R}
  \end{array}\right.,\\
  &\quad\widetilde{D}(x;\bm{\lambda})\eqdef\left\{
  \begin{array}{ll}
  -B\bigl(-(x+\beta+\ell-1);\bm{\lambda}+(\ell-1)\bm{\delta}\bigr)
  &:\text{XM}\\
  D\bigl(x;\mathfrak{t}(\bm{\lambda}+(\ell-1)\bm{\delta})\bigr)
  &:\text{XR, X$q$R}
  \end{array}\right.,\\
  &\quad\widetilde{\mathcal{E}}(\ell;\bm{\lambda})\eqdef\left\{
  \begin{array}{ll}
  -\mathcal{E}(\ell;\bm{\lambda})
  &:\text{XM}\\
  \mathcal{E}(\ell;\mathfrak{t}(\bm{\lambda}))
  &:\text{XR, X$q$R}
  \end{array}\right.,
\end{align}
to be compared with the difference equation for the eigenpolynomial
\begin{equation}
  \bigl(B(x;\bm{\lambda})(1-e^{\partial})
  +D(x;\bm{\lambda})(1-e^{-\partial})\bigr)\check{P}_n(x;\bm{\lambda})
  =\mathcal{E}(n;\bm{\lambda})\check{P}_n(x;\bm{\lambda}).
  \label{PndiffeqrdQM}
\end{equation}
Note that $\mathcal{E}(\ell;\mathfrak{t}(\bm{\lambda}))
=\mathcal{E}(\ell;\mathfrak{t}(\bm{\lambda}+(\ell-1)\bm{\delta}))$.
For the appropriate parameter ranges, the deforming polynomial
$\check{\xi}_{\ell}(x;\bm{\lambda})$ is positive at integer points
$x=0,1,\ldots,x_{\text{max}}^{\ell}+1$
($x_{\text{max}}^{\ell}\eqdef x_{\text{max}}-\ell$).

Note that the range of parameters can be enlarged.
For XJ, see \cite{gomez201101}.

%%%%%%%%%%%%%%%%%%%%%%%%%%%%%%%%%%%%%%%%%%%
%                                         %
% 4.2 New Orthogonal Polynomials          %
%                                         %
%%%%%%%%%%%%%%%%%%%%%%%%%%%%%%%%%%%%%%%%%%%
%\subsection{Explicit Forms of the New Orthogonal Polynomials}
\subsection{New Orthogonal Polynomials}
\label{subsec:New}

In this subsection we present the explicit forms of the exceptional
orthogonal polynomials and the orthogonality relations.
They are bilinear in the deforming
polynomial $\xi_{\ell}$ (with $\bm{\lambda}$ and $\bm{\lambda}+\bm{\delta}$)
and the eigenpolynomial $P_n$
(with $\bm{\lambda}+\ell\bm{\delta}+\tilde{\bm{\delta}}$).

For oQM, the exceptional polynomials are:
\begin{align}
  &P_{\ell,n}(\eta;\bm{\lambda})\eqdef
  \frac{2}{\hat{f}_{\ell,n}(\bm{\lambda})}
  \Bigl(d_2(\eta)\xi_{\ell}(\eta;\bm{\lambda})\partial_{\eta}
  P_n(\eta;\bm{\lambda}+\ell\bm{\delta}+\bm{\tilde{\delta}})\n[-2pt]
  &\phantom{P_{\ell,n}(\eta;\bm{\lambda})
  =\frac{2}{\hat{f}_{\ell,n}(\bm{\lambda})}}
  -d_1(\bm{\lambda})\xi_{\ell}(\eta;\bm{\lambda}+\bm{\delta})
  P_n(\eta;\bm{\lambda}+\ell\bm{\delta}+\bm{\tilde{\delta}})\Bigr)
  \times\left\{
  \begin{array}{ll}
  \pm 1&:\text{XL1/XL2}\\
  \pm 1&:\text{XJ1/XJ2}
  \end{array}\right.\!\!,
  \label{PlnoQM}
\end{align}
where $\bm{\tilde{\delta}}$ is
\begin{equation}
  \bm{\tilde{\delta}}\eqdef\left\{
  \begin{array}{ll}
  \mp 1&:\text{XL1/XL2}\\
  \mp(1,-1)&:\text{XJ1/XJ2}
  \end{array}\right.,
  \label{deltatoQM}
\end{equation}
and $\hat{f}_{\ell,n}(\bm{\lambda})$ will be given in \eqref{fhlnbhlnoQM}.
They satisfy the orthogonality relation:
\begin{equation}
  \int_{x_1}^{x_2}\!\!\psi_{\ell}(x;\bm{\lambda})^2\,
  P_{\ell,n}(\eta(x);\bm{\lambda})P_{\ell,m}(\eta(x);\bm{\lambda})dx
  =h_{\ell,n}(\bm{\lambda})\delta_{nm},
  \label{orthooQM}
\end{equation}
where the weight function $\psi_{\ell}(x;\bm{\lambda})^2$ will be given
in \eqref{psiloQM}.
The normalisation constant $h_{\ell,n}(\bm{\lambda})$ is related to
$h_n(\bm{\lambda})$ \eqref{hnoQM} as
\begin{equation}
  h_{\ell,n}(\bm{\lambda})
  =\frac{\hat{b}_{\ell,n}(\bm{\lambda})}{\hat{f}_{\ell,n}(\bm{\lambda})}
  h_n(\bm{\lambda}+\ell\bm{\delta}+\bm{\tilde{\delta}})
  =\frac{\hat{b}_{\ell,n}(\bm{\lambda})}{\hat{f}_{\ell,n}(\bm{\lambda})}
  \frac{\hat{f}_{0,n}(\bm{\lambda}+\ell\bm{\delta})}
  {\hat{b}_{0,n}(\bm{\lambda}+\ell\bm{\delta})}
  h_n(\bm{\lambda}+\ell\bm{\delta}),
  \label{hlnoQM}
\end{equation}
where $\hat{f}_{\ell,n}(\bm{\lambda})$ and $\hat{b}_{\ell,n}(\bm{\lambda})$
are given by
\begin{align}
  &\hat{f}_{\ell,n}(\bm{\lambda})\eqdef 2\times\left\{
  \begin{array}{ll}
  -1&:\text{XL1}\\
  n+g+\frac12&:\text{XL2}\\
  -(n+h+\frac12)&:\text{XJ1}\\
  n+g+\frac12&:\text{XJ2}
  \end{array}\right.\!\!,
  \quad
  \hat{b}_{\ell,n}(\bm{\lambda})\eqdef 2\times\left\{
  \begin{array}{ll}
  -(n+g+2\ell-\frac12)&:\text{XL1}\\
  1&:\text{XL2}\\
  -(n+g+2\ell-\frac12)&:\text{XJ1}\\
  n+h+2\ell-\frac12&:\text{XJ2}
  \end{array}\right.\!\!.
  \label{fhlnbhlnoQM}
\end{align}
The deforming polynomial $\xi_{\ell}(\eta;\bm{\lambda})$ has the same
sign in the orthogonality domain, that is $(0,\infty)$ for XL and
$(-1,1)$ for XJ.

The two types of the XJ polynomials are the mirror images of each other
\begin{align}
  &P_{\ell,n}^{\text{XJ2}}(x;g,h)
  =(-1)^{\ell+n}P_{\ell,n}^{\text{XJ1}}(-x;h,g),
  \label{plnJ2=plnJ1}
\end{align}
due to the parity property of the Jacobi polynomial
$P_n^{(\alpha,\,\beta)}(-x)=(-1)^nP_n^{(\beta,\,\alpha)}(x)$,
see (50)--(54) of \cite{os19}.
The Laguerre polynomials are known to be obtained from the Jacobi
polynomials in a certain limit:
\begin{equation}
  \lim_{\beta\to\infty}P_n^{(\alpha,\,\pm\beta)}
  \bigl(1-2x\beta^{-1}\bigr)
  =L_n^{(\alpha)}(\pm x).
  \label{JtoL}
\end{equation}
This connects the XJ polynomials to the XL polynomials as shown in (43),
(58) of \cite{os19}.

For idQM, the exceptional polynomials are \cite{os20}:
\begin{align}
  &P_{\ell,n}(\eta(x);\bm{\lambda})
  \eqdef\frac{-i}{\hat{f}_{\ell,n}(\bm{\lambda})\varphi(x)}\Bigl(
  v_1(x;\bm{\lambda}+\ell\bm{\delta})
  \xi_{\ell}(\eta(x+i\tfrac{\gamma}{2});\bm{\lambda})
  P_n(\eta(x-i\tfrac{\gamma}{2});\bm{\lambda}+\ell\bm{\delta}
  +\bm{\tilde{\delta}})\n
  &\qquad\qquad\qquad\qquad
  -v_1^*(x;\bm{\lambda}+\ell\bm{\delta})
  \xi_{\ell}(\eta(x-i\tfrac{\gamma}{2});\bm{\lambda})
  P_n(\eta(x+i\tfrac{\gamma}{2});\bm{\lambda}+\ell\bm{\delta}
  +\bm{\tilde{\delta}})\Bigr),
  \label{PlnidQM}
\end{align}
where $\bm{\tilde{\delta}}$ is
\begin{equation}
  \bm{\tilde{\delta}}\eqdef\left\{
  \begin{array}{ll}
  \tfrac12&:\text{XMP}\\
  (\tfrac12,\tfrac12,-\tfrac12,-\tfrac12)&:\text{XW, XAW}
  \end{array}\right..
  \label{deltatidQM}
\end{equation}
They satisfy the orthogonality relation:
\begin{equation}
  \int_{x_1}^{x_2}\!\!\psi_{\ell}(x;\bm{\lambda})^2\,
  P_{\ell,n}(\eta(x);\bm{\lambda})P_{\ell,m}(\eta(x);\bm{\lambda})dx
  =h_{\ell,n}(\bm{\lambda})\delta_{nm},
  \label{orthoidQM}
\end{equation}
where the weight function $\psi_{\ell}(x;\bm{\lambda})^2$ will be given
in \eqref{psilidQM}.
The normalisation constant $h_{\ell,n}(\bm{\lambda})$ is related to
$h_n(\bm{\lambda})$ \eqref{hnidQM} as
\begin{equation}
  h_{\ell,n}(\bm{\lambda})
  =\frac{\hat{b}_{\ell,n}(\bm{\lambda})}{\hat{f}_{\ell,n}(\bm{\lambda})}
  h_n(\bm{\lambda}+\ell\bm{\delta}+\bm{\tilde{\delta}})
  =\frac{\hat{b}_{\ell,n}(\bm{\lambda})}{\hat{f}_{\ell,n}(\bm{\lambda})}
  \frac{\hat{f}_{0,n}(\bm{\lambda}+\ell\bm{\delta})}
  {\hat{b}_{0,n}(\bm{\lambda}+\ell\bm{\delta})}
  h_n(\bm{\lambda}+\ell\bm{\delta}).
  \label{hlnidQM}
\end{equation}
We have verified by numerical calculation for the parameter ranges in
\eqref{paraMPrest}--\eqref{paraAWrest} that the deforming polynomial
$\xi_{\ell}(\eta(x);\bm{\lambda})$ has the same sign in the orthogonality
domain, that is $x\in(-\infty,\infty)$ for XM, $x\in(0,\infty)$ for XW
and $x\in(0,\pi)$ for XAW.

For rdQM, the exceptional polynomials are \cite{os23}:
\begin{align}
  &\check{P}_{\ell,n}(x;\bm{\lambda})
  \eqdef P_{\ell,n}(\eta(x;\bm{\lambda}+\ell\bm{\delta});\bm{\lambda})\n
  &\eqdef\frac{1}{\hat{f}_{\ell,n}(\bm{\lambda})}
  \frac{1}{\varphi(x;\bm{\lambda}+\ell\bm{\delta}+\tilde{\bm{\delta}})}
  \Bigl(v_1^B(x;\bm{\lambda}+\ell\bm{\delta})
  \check{\xi}_{\ell}(x;\bm{\lambda})
  \check{P}_{n}(x+1;\bm{\lambda}+\ell\bm{\delta}+\tilde{\bm{\delta}})\n
  &\qquad\qquad\qquad\qquad\qquad\qquad
  -v_1^D(x;\bm{\lambda}+\ell\bm{\delta})\check{\xi}_{\ell}(x+1;\bm{\lambda})
  \check{P}_{n}(x;\bm{\lambda}+\ell\bm{\delta}+\tilde{\bm{\delta}})\Bigr),
  \label{PlnrdQM}
\end{align}
which satisfy the normalisation
\begin{equation}
  P_{\ell,n}(0;\bm{\lambda})=1.
\end{equation}
Here $\bm{\tilde{\delta}}$ is
\begin{equation}
  \tilde{\bm{\delta}}\eqdef\left\{
  \begin{array}{ll}
  (-1,0)&:\text{XM}\\
  (0,0,-1,-1)&:\text{XR, X$q$R}
  \end{array}\right..
  \label{deltatrdQM}
\end{equation}
The orthogonality relation reads
($x_{\text{max}}^{\ell}\eqdef x_{\text{max}}-\ell$,
$n_{\text{max}}^{\ell}\eqdef x_{\text{max}}-\ell$):
\begin{equation}
  \sum_{x=0}^{x_{\text{max}}^{\ell}}
  \frac{\psi_{\ell}(x;\bm{\lambda})^2}{\check{\xi}_{\ell}(1;\bm{\lambda})}\,
  \check{P}_{\ell,n}(x;\bm{\lambda})\check{P}_{\ell,m}(x;\bm{\lambda})
  =\frac{\delta_{nm}}{d_{\ell,n}(\bm{\lambda})^2}\quad
  (n,m=0,1,\ldots,n_{\text{max}}^{\ell}),
  \label{orthordQM}
\end{equation}
where the weight function $\psi_{\ell}(x;\bm{\lambda})^2$ will be given
in \eqref{psilrdQM}.
The normalisation constant $d_{\ell,n}(\bm{\lambda})^2$ is related to
$d_n(\bm{\lambda})^2$ \eqref{dnrdQM} as
\begin{equation}
  d_{\ell,n}(\bm{\lambda})^2
  =d_n(\bm{\lambda}+\ell\bm{\delta}+\tilde{\bm{\delta}})^2\,
  \frac{\hat{f}_{\ell,n}(\bm{\lambda})}{\hat{b}_{\ell,n}(\bm{\lambda})}
  \frac{1}{s_{\ell}(\bm{\lambda})}
  =d_n(\bm{\lambda}+\ell\bm{\delta})^2\,
  \frac{\hat{f}_{\ell,n}(\bm{\lambda})}{\hat{b}_{\ell,n}(\bm{\lambda})}
  \frac{\hat{b}_{0,n}(\bm{\lambda}+\ell\bm{\delta})}
  {\hat{f}_{0,n}(\bm{\lambda}+\ell\bm{\delta})}
  \frac{s_0(\bm{\lambda}+\ell\bm{\delta})}{s_{\ell}(\bm{\lambda})},
  \label{dln2}
\end{equation}
where $s_{\ell}(\bm{\lambda})$ is
\begin{equation}
  s_{\ell}(\bm{\lambda})\eqdef\left\{
  \begin{array}{ll}
  {\displaystyle
  \frac{1-c}{c}\frac{1}{\beta+\ell-1}
  }&:\text{XM}\\  
  {\displaystyle
  -\frac{(d-a)(d-b)}{(c+\ell-1)(d+\ell)}
  }&:\text{XR}\\
  {\displaystyle
  -abd^{-1}q^{\ell}
  \frac{(1-a^{-1}d)(1-b^{-1}d)}{(1-cq^{\ell-1})(1-dq^{\ell})}
  }&:\text{X$q$R}
  \end{array}\right..
  \label{sl}
\end{equation}
We have verified by numerical calculation for the parameter ranges in
\eqref{ex7}--\eqref{ex9} that the deforming polynomial
$\check{\xi}_{\ell}(x;\bm{\lambda})$ has the same sign in the
orthogonality domain, that is $x=0,1,\ldots,x_{\text{max}}^{\ell}$.

The exceptional polynomial $P_{\ell,n}(\eta;\bm{\lambda})$ is a degree
$\ell+n$ polynomial in $\eta$ but has only $n$ zeros in $(0,\infty)$ for
XL, $(-1,1)$ for XJ, $(\eta(x_1),\eta(x_2))$ for idQM and
$(0,\eta(x_{\text{max}}))$ for rdQM.
For $\ell=0$, the expressions of $P_{\ell,n}(\eta;\bm{\lambda})$ reduce
to those of the original polynomials.

These new orthogonal polynomials do not satisfy the ordinary three term
recurrence relations \eqref{threeterm}. For the XL and XJ polynomials,
a {\em new type of bi-spectrality} is demonstrated in \S4 of \cite{stz}.
The bi-spectrality of the other new orthogonal polynomials, XW, XAW, XR
and X$q$R are to be explored.
A different type of three term recurrence relations for the XL and XJ
polynomials is reported in \S10 of \cite{hos}.
The generating functions $\sum_{n=0}^\infty t^nP_{\ell,n}(\eta;\bm{\lambda})$
for the XL and XJ polynomials are reported in \S9 of \cite{hos}.
For the XL1 and XL2 polynomials, the closed form expression of the
{\em double generating functions}
\[
 G(s,t,\eta;\bm{\lambda})\eqdef \sum_{\ell=0}^\infty\sum_{n=0}^\infty
 s^{\ell}t^nP_{\ell,n}(\eta;\bm{\lambda})
\]
are given there, too.
The zeros of orthogonal polynomials have always attracted the interest of
researchers. In \cite{hs4} the behaviours of the zeros of the XL1, XL2
and XJ2 polynomials $P_{\ell,n}(\eta;\bm{\lambda})$ are explored as the
parameters change. The structure of the extra zeros of the other new
orthogonal polynomials, XW, XAW, XR and X$q$R are also an interesting problem.

%%%%%%%%%%%%%%%%%%%%%%%%%%%%%%%%%%%%%%%%%%%
%                                         %
% 4.3 Deformed Hamiltonians               %
%                                         %
%%%%%%%%%%%%%%%%%%%%%%%%%%%%%%%%%%%%%%%%%%%
\subsection{Deformed Hamiltonians}
\label{sec:defhams}

The new orthogonal polynomials are discovered as the main part of the
eigenfunctions of quantum mechanical systems in which the original
Hamiltonians are deformed in terms of the deforming polynomial $\xi_{\ell}$.
The parameters are so chosen that the deformed systems retain shape
invariance \cite{os16,os17,os18,os19,os20,stz,os23}.

The deformed Hamiltonian and their eigenfunctions have the following forms:
\begin{align}
  &\mathcal{H}_{\ell}(\bm{\lambda})\eqdef
  \mathcal{A}_{\ell}(\bm{\lambda})^{\dagger}\mathcal{A}_{\ell}(\bm{\lambda}),
  \label{Hldef}\\
  &\mathcal{H}_{\ell}(\bm{\lambda})\phi_{\ell,n}(x;\bm{\lambda})
  =\mathcal{E}_{\ell,n}(\bm{\lambda})\phi_{\ell,n}(x;\bm{\lambda}),\quad
  \mathcal{E}_{\ell,n}(\bm{\lambda})
  =\mathcal{E}(n;\bm{\lambda}+\ell\bm{\delta}),\\
  &\phi_{\ell,n}(x;\bm{\lambda})=\psi_{\ell}(x;\bm{\lambda})
  P_{\ell,n}(\eta(x;\bm{\lambda}+\ell\bm{\delta});\bm{\lambda}).
  \label{philn}
\end{align}
The deformation is achieved additively at the level of the prepotential
for oQM and multiplicatively at the level of the potential functions $V(x)$,
$V^*(x)$, $B(x)$ and $D(x)$ for dQM.
For oQM, operators $\mathcal{A}_{\ell}(\bm{\lambda})$,
$\mathcal{A}_{\ell}(\bm{\lambda})^{\dagger}$ and
$\psi_{\ell}(x;\bm{\lambda})$ are \cite{os16,os19}:
\begin{align}
  &\mathcal{A}_{\ell}(\bm{\lambda})\eqdef
  \frac{d}{dx}-\partial_xw_{\ell}(x;\bm{\lambda}),\quad
  \mathcal{A}_{\ell}(\bm{\lambda})^{\dagger}=
  -\frac{d}{dx}-\partial_xw_{\ell}(x;\bm{\lambda}),
  \label{AloQM}\\
  &w_{\ell}(x;\bm{\lambda})\eqdef
  w(x;\bm{\lambda}+\ell\bm{\delta})
  +\log\frac{\xi_{\ell}(\eta(x);\bm{\lambda}+\bm{\delta})}
  {\xi_{\ell}(\eta(x);\bm{\lambda})},
  \label{wl}\\
  &  \psi_{\ell}(x;\bm{\lambda})\eqdef
  \frac{\phi_0(x;\bm{\lambda}+\ell\bm{\delta})}
  {\xi_{\ell}(\eta(x);\bm{\lambda})},
  \quad
  \phi_{\ell,0}(x;\bm{\lambda})=e^{w_{\ell}(x;\bm{\lambda})}
  =\psi_{\ell}(x;\bm{\lambda})\xi_{\ell}(\eta(x);\bm{\lambda}+\bm{\delta}).
  \label{psiloQM}
\end{align}
For idQM, they are \cite{os17}:
\begin{align}
  &\mathcal{A}_{\ell}(\bm{\lambda})\eqdef
   i\bigl(e^{\frac{\gamma}{2}p}\sqrt{V_{\ell}^*(x;\bm{\lambda})}
  -e^{-\frac{\gamma}{2}p}\sqrt{V_{\ell}(x;\bm{\lambda})}\,\bigr),\n
  &\mathcal{A}_{\ell}(\bm{\lambda})^{\dagger}=
  -i\bigl(\sqrt{V_{\ell}(x;\bm{\lambda})}\,e^{\frac{\gamma}{2}p}
  -\sqrt{V_{\ell}^*(x;\bm{\lambda})}\,e^{-\frac{\gamma}{2}p}\bigr),
  \label{AlidQM}\\
  &V_{\ell}(x;\bm{\lambda})\eqdef
  V(x;\bm{\lambda}+\ell\bm{\delta})\,
  \frac{\xi_{\ell}(\eta(x+i\frac{\gamma}{2});\bm{\lambda})}
  {\xi_{\ell}(\eta(x-i\frac{\gamma}{2});\bm{\lambda})}
  \frac{\xi_{\ell}(\eta(x-i\gamma);\bm{\lambda}+\bm{\delta})}
  {\xi_{\ell}(\eta(x);\bm{\lambda}+\bm{\delta})},
  \label{Vl}\\
  &V_{\ell}^*(x;\bm{\lambda})=
  V^*(x;\bm{\lambda}+\ell\bm{\delta})\,
  \frac{\xi_{\ell}(\eta(x-i\frac{\gamma}{2});\bm{\lambda})}
  {\xi_{\ell}(\eta(x+i\frac{\gamma}{2});\bm{\lambda})}
  \frac{\xi_{\ell}(\eta(x+i\gamma);\bm{\lambda}+\bm{\delta})}
  {\xi_{\ell}(\eta(x);\bm{\lambda}+\bm{\delta})},
  \label{Vl*}\\
  &\psi_{\ell}(x;\bm{\lambda})
  \eqdef\frac{\phi_0(x;\bm{\lambda}+\ell\bm{\delta})}
  {\sqrt{\xi_{\ell}(\eta(x+i\frac{\gamma}{2});\bm{\lambda})
  \xi_{\ell}(\eta(x-i\frac{\gamma}{2});\bm{\lambda})}}.
  \label{psilidQM}
\end{align}
For rdQM, they are \cite{os23}:
\begin{align}
  &\mathcal{A}_{\ell}(\bm{\lambda})
  \eqdef\sqrt{B_{\ell}(x;\bm{\lambda})}
  -e^{\partial}\sqrt{D_{\ell}(x;\bm{\lambda})},\quad
  \mathcal{A}_{\ell}(\bm{\lambda})^{\dagger}
  =\sqrt{B_{\ell}(x;\bm{\lambda})}
  -\sqrt{D_{\ell}(x;\bm{\lambda})}\,e^{-\partial},
  \label{AlrdQM}\\
  &B_{\ell}(x;\bm{\lambda})\eqdef B(x;\bm{\lambda}+\ell\bm{\delta})
  \frac{\check{\xi}_{\ell}(x;\bm{\lambda})}
  {\check{\xi}_{\ell}(x+1;\bm{\lambda})}
  \frac{\check{\xi}_{\ell}(x+1;\bm{\lambda}+\bm{\delta})}
  {\check{\xi}_{\ell}(x;\bm{\lambda}+\bm{\delta})},
  \label{Bl}\\
  &D_{\ell}(x;\bm{\lambda})\eqdef D(x;\bm{\lambda}+\ell\bm{\delta})
  \frac{\check{\xi}_{\ell}(x+1;\bm{\lambda})}
  {\check{\xi}_{\ell}(x;\bm{\lambda})}
  \frac{\check{\xi}_{\ell}(x-1;\bm{\lambda}+\bm{\delta})}
  {\check{\xi}_{\ell}(x;\bm{\lambda}+\bm{\delta})},
  \label{Dl}\\
  &\psi_{\ell}(x;\bm{\lambda})\eqdef\phi_0(x;\bm{\lambda}+\ell\bm{\delta})
  \sqrt{\frac{\check{\xi}_{\ell}(1;\bm{\lambda})}
  {\check{\xi}_{\ell}(x;\bm{\lambda})\check{\xi}_{\ell}(x+1;\bm{\lambda})}}.
  \label{psilrdQM}
\end{align}
These potential functions satisfy the boundary conditions
$D_{\ell}(0;\bm{\lambda})=0$ and
$B_{\ell}(N-\ell;\bm{\lambda})=0$, and
the functions $\phi_{\ell,0}(x;\bm{\lambda})$ and
$\psi_{\ell}(x;\bm{\lambda})$ satisfy the normalisation conditions
$\phi_{\ell,0}(0;\bm{\lambda})=1$ and $\psi_{\ell}(0;\bm{\lambda})=1$.

These deformed systems are shape invariant:
\begin{equation}
  \mathcal{A}_{\ell}(\bm{\lambda})\mathcal{A}_{\ell}(\bm{\lambda})^{\dagger}
  =\kappa\mathcal{A}_{\ell}(\bm{\lambda+\bm{\delta}})^{\dagger}
  \mathcal{A}_{\ell}(\bm{\lambda}+\bm{\delta})
  +\mathcal{E}_{\ell,1}(\bm{\lambda}),
\end{equation}
or equivalently \eqref{shapeinvoQM}--\eqref{shapeinvrdQM2} with
replacements $(w,V,B,D)\to(w_{\ell},V_{\ell},B_{\ell},D_{\ell})$.
The shape invariance of these deformed systems is shown analytically
by making use of the two identities of the deforming polynomials
\eqref{xil(l+d)oQM}--\eqref{xil(l)oQM},
\eqref{xil(l+d)idQM}--\eqref{xil(l)idQM} and
\eqref{xil(l+d)rdQM}--\eqref{xil(l)rdQM} \cite{os18,os20,os23}.

The action of the operators $\mathcal{A}_{\ell}(\bm{\lambda})$ and
$\mathcal{A}_{\ell}(\bm{\lambda})^{\dagger}$ on the eigenfunctions is
\begin{align}
  \mathcal{A}_{\ell}(\bm{\lambda})\phi_{\ell,n}(x;\bm{\lambda})
  &=f_{\ell,n}(\bm{\lambda})
  \phi_{\ell,n-1}(x;\bm{\lambda}+\bm{\delta})
  \times\left\{
  \begin{array}{ll}
  1&:\text{oQM, idQM}\\
  \frac{1}{\sqrt{B_{\ell}(0;\bm{\lambda})}}&:\text{rdQM}
  \end{array}\right.,
  \label{Alphiln=flnphiln}\\
  \mathcal{A}_{\ell}(\bm{\lambda})^{\dagger}
  \phi_{\ell,n-1}\bigl(x;\bm{\lambda}+\bm{\delta}\bigr)
  &=b_{\ell,n-1}(\bm{\lambda})\phi_{\ell,n}(x;\bm{\lambda})
  \times\left\{
  \begin{array}{ll}
  1&:\text{oQM, idQM}\\
  \sqrt{B_{\ell}(0;\bm{\lambda})}&:\text{rdQM}
  \end{array}\right.,
  \label{Aldphiln=blnphiln}
\end{align}
where $f_{\ell,n}(\bm{\lambda})$ and $b_{\ell,n-1}(\bm{\lambda})$ are
\begin{align}
  f_{\ell,n}(\bm{\lambda})&=f_n(\bm{\lambda}+\ell\bm{\delta})
  \times\left\{
  \begin{array}{ll}
  \frac{c}{1-c}(\beta+\ell)&:\text{XM}\\
  1&:\text{other examples}
  \end{array}\right.,
  \label{fln}\\
  b_{\ell,n-1}(\bm{\lambda})&=b_{n-1}(\bm{\lambda}+\ell\bm{\delta})
  \times\left\{
  \begin{array}{ll}
  \bigl(\frac{c}{1-c}(\beta+\ell)\bigr)^{-1}&:\text{XM}\\
  1&:\text{other examples}
  \end{array}\right..
  \label{bln}
\end{align}
The forward shift operator $\mathcal{F}_{\ell}(\bm{\lambda})$ and the
backward shift operator $\mathcal{B}_{\ell}(\bm{\lambda})$ are defined by
\begin{align}
  \mathcal{F}_{\ell}(\bm{\lambda})&\eqdef
  \psi_{\ell}(x;\bm{\lambda}+\bm{\delta})^{-1}\circ
  \mathcal{A}_{\ell}(\bm{\lambda})\circ\psi_{\ell}(x;\bm{\lambda})
  \times\left\{
  \begin{array}{ll}
  1&:\text{oQM, idQM}\\
  \sqrt{B_{\ell}(0;\bm{\lambda})}&:\text{rdQM}
  \end{array}\right.,
  \label{Fldef}\\
  \mathcal{B}_{\ell}(\bm{\lambda})&\eqdef
  \psi_{\ell}(x;\bm{\lambda})^{-1}\circ
  \mathcal{A}_{\ell}(\bm{\lambda})^{\dagger}
  \circ\psi_{\ell}\bigl(x;\bm{\lambda}+\bm{\delta}\bigr)
  \times\left\{
  \begin{array}{ll}
  1&:\text{oQM, idQM}\\
  \frac{1}{\sqrt{B_{\ell}(0;\bm{\lambda})}}&:\text{rdQM}
  \end{array}\right.,
  \label{Bldef}
\end{align}
and their action on the polynomials is
($\check{P}_{\ell,n}(x;\bm{\lambda})\eqdef P_{\ell,n}(\eta(x);\bm{\lambda})$
for oQM and idQM)
\begin{align}
  \mathcal{F}_{\ell}(\bm{\lambda})\check{P}_{\ell,n}(x;\bm{\lambda})
  &=f_{\ell,n}(\bm{\lambda})
  \check{P}_{\ell,n-1}(x;\bm{\lambda}+\bm{\delta}),
  \label{FlPln=flnPln}\\
  \mathcal{B}_{\ell}(\bm{\lambda})
  \check{P}_{\ell,n-1}(x;\bm{\lambda}+\bm{\delta})
  &=b_{\ell,n-1}(\bm{\lambda})\check{P}_{\ell,n}(x;\bm{\lambda}).
  \label{BlPln=blnPln}
\end{align}
The explicit forms of $\mathcal{F}_{\ell}(\bm{\lambda})$ and
$\mathcal{B}_{\ell}(\bm{\lambda})$ are the following:\\
oQM:
\begin{align}
  \mathcal{F}_{\ell}(\bm{\lambda})
  &=\cF\frac{\xi_{\ell}(\eta;\bm{\lambda}+\bm{\delta})}
  {\xi_{\ell}(\eta;\bm{\lambda})}\Bigl(\frac{d}{d\eta}
  -\partial_{\eta}\log\xi_{\ell}(\eta;\bm{\lambda}+\bm{\delta})\Bigr),
  \label{FloQM}\\
  \mathcal{B}_{\ell}(\bm{\lambda})
  &=-4\cF^{-1}c_2(\eta)\frac{\xi_{\ell}(\eta;\bm{\lambda})}
  {\xi_{\ell}(\eta;\bm{\lambda}+\bm{\delta})}
  \Bigl(\frac{d}{d\eta}
  +\frac{c_1(\eta,\bm{\lambda}+\ell\bm{\delta})}{c_2(\eta)}
  -\partial_{\eta}\log\xi_{\ell}(\eta;\bm{\lambda})\Bigr),
  \label{BloQM}
\end{align}
idQM:
\begin{align}
  \mathcal{F}_{\ell}(\bm{\lambda})
  &=\frac{i}{\varphi(x)\xi_{\ell}(\eta(x);\bm{\lambda})}
  \Bigl(\xi_{\ell}\bigl(\eta(x+i\tfrac{\gamma}{2});\bm{\lambda}
  +\bm{\delta}\bigr)e^{\frac{\gamma}{2}p}
  -\xi_{\ell}\bigl(\eta(x-i\tfrac{\gamma}{2});\bm{\lambda}
  +\bm{\delta}\bigr)e^{-\frac{\gamma}{2}p}\Bigr),
  \label{FlidQM}\\
  \mathcal{B}_{\ell}(\bm{\lambda})
  &=\frac{-i}{\xi_{\ell}(\eta(x);\bm{\lambda}+\bm{\delta})}
  \Bigl(V(x;\bm{\lambda}+\ell\bm{\delta})
  \xi_{\ell}\bigl(\eta(x+i\tfrac{\gamma}{2});\bm{\lambda}\bigr)
  e^{\frac{\gamma}{2}p}\n
  &\qquad\qquad\qquad\qquad
  -V^*(x;\bm{\lambda}+\ell\bm{\delta})
  \xi_{\ell}\bigl(\eta(x-i\tfrac{\gamma}{2});\bm{\lambda}\bigr)
  e^{-\frac{\gamma}{2}p}\Bigr)\varphi(x),
  \label{BlidQM}
\end{align}
rdQM:
\begin{align}
  \mathcal{F}_{\ell}(\bm{\lambda})
  &=\frac{B(0,\bm{\lambda}+\ell\bm{\delta})}
  {\varphi(x;\bm{\lambda}+\ell\bm{\delta})
  \check{\xi}_{\ell}(x+1;\bm{\lambda})}
  \Bigl(\check{\xi}_{\ell}(x+1;\bm{\lambda}+\bm{\delta})
  -\check{\xi}_{\ell}(x;\bm{\lambda}+\bm{\delta})e^{\partial}\Bigr),
  \label{FlrdQM}\\
  \mathcal{B}_{\ell}(\bm{\lambda})
  &=\frac{1}{B(0;\bm{\lambda}+\ell\bm{\delta})}
  \frac{1}{\check{\xi}_{\ell}(x;\bm{\lambda}+\bm{\delta})}\n
  &\quad\times
  \Bigl(B(x;\bm{\lambda}+\ell\bm{\delta})\check{\xi}_{\ell}(x;\bm{\lambda})
  -D(x;\bm{\lambda}+\ell\bm{\delta})\check{\xi}_{\ell}(x+1;\bm{\lambda})
  e^{-\partial}\Bigl)
  \varphi(x;\bm{\lambda}+\ell\bm{\delta}).
  \label{BlrdQM}
\end{align}

%%%%%%%%%%%%%%%%%%%%%%%%%%%%%%%%%%%%%%%%%%%
%                                         %
% 4.4 Second Order Equations for the ...  %
%                                         %
%%%%%%%%%%%%%%%%%%%%%%%%%%%%%%%%%%%%%%%%%%%
\subsection{Second Order Equations for the New Polynomials}
\label{sec:secondeq}

The new orthogonal polynomials $P_{\ell,n}(\eta;\bm{\lambda})$ satisfy
second order differential or difference equations with
{\em rational coefficients}. Here we list the explicit forms.

The similarity transformed Hamiltonian
$\widetilde{\mathcal{H}}_{\ell}(\bm{\lambda})$ is defined by
\begin{equation}
  \widetilde{\mathcal{H}}_{\ell}(\bm{\lambda})\eqdef
  \psi_{\ell}(x;\bm{\lambda})^{-1}\circ
  \mathcal{H}_{\ell}(\bm{\lambda})\circ\psi_{\ell}(x;\bm{\lambda})
  =\mathcal{B}_{\ell}(\bm{\lambda})\mathcal{F}_{\ell}(\bm{\lambda}),
  \label{Htl}
\end{equation}
and its action on the polynomials is
\begin{equation}
  \widetilde{\mathcal{H}}_{\ell}(\bm{\lambda})
  \check{P}_{\ell,n}(x;\bm{\lambda})
  =\mathcal{E}_{\ell,n}(\bm{\lambda})\check{P}_{\ell,n}(x;\bm{\lambda}).
\end{equation}
{}From \eqref{FlPln=flnPln}--\eqref{BlPln=blnPln} and
\eqref{fln}--\eqref{bln}, we have
\begin{equation}
  \mathcal{E}_{\ell,n}(\bm{\lambda})
  =f_{\ell,n}(\bm{\lambda})b_{\ell,n-1}(\bm{\lambda})
  =f_n(\bm{\lambda}+\ell\bm{\delta})b_{n-1}(\bm{\lambda}+\ell\bm{\delta})
  =\mathcal{E}(n;\bm{\lambda}+\ell\bm{\delta}).
\end{equation}
The explicit forms of $\widetilde{\mathcal{H}}_{\ell}(\bm{\lambda})$ are
the following:\\
oQM:
\begin{align}
  \widetilde{\mathcal{H}}_{\ell}(\bm{\lambda})
  &=-4\Bigl(c_2(\eta)\frac{d^2}{d\eta^2}
  +\bigl(c_1(\eta,\bm{\lambda}+\ell\bm{\delta})-2c_2(\eta)
  \partial_{\eta}\log\xi_{\ell}(\eta;\bm{\lambda})\bigr)
  \frac{d}{d\eta}\n
  &\phantom{=-4\Bigl(}
  -2d_2(\eta)d_3(\bm{\lambda},\ell)
  \partial_{\eta}\log\xi_{\ell}(\eta;\bm{\lambda})
  -\tfrac14\,\widetilde{\mathcal{E}}_{\ell}(\bm{\lambda})\Bigr),
  \label{HtloQM}
\end{align}
idQM:
\begin{align}
  \widetilde{\mathcal{H}}_{\ell}(\bm{\lambda})
  &=V(x;\bm{\lambda}+\ell\bm{\delta})
  \frac{ \xi_{\ell}\bigl(\eta(x+i\tfrac{\gamma}{2});\bm{\lambda}\bigr)}
  {\xi_{\ell}\bigl(\eta(x-i\tfrac{\gamma}{2});\bm{\lambda}\bigr)}
  \biggl(e^{\gamma p}
  -\frac{\xi_{\ell}\bigl(\eta(x-i\gamma);\bm{\lambda}+\bm{\delta}\bigr)}
  {\xi_{\ell}\bigl(\eta(x);\bm{\lambda}+\bm{\delta}\bigr)}\biggr)\n
  &\quad+V^*(x;\bm{\lambda}+\ell\bm{\delta})
  \frac{ \xi_{\ell}\bigl(\eta(x-i\tfrac{\gamma}{2});\bm{\lambda}\bigr)}
  {\xi_{\ell}\bigl(\eta(x+i\tfrac{\gamma}{2});\bm{\lambda}\bigr)}
  \biggl(e^{-\gamma p}
  -\frac{\xi_{\ell}\bigl(\eta(x+i\gamma);\bm{\lambda}+\bm{\delta}\bigr)}
  {\xi_{\ell}\bigl(\eta(x);\bm{\lambda}+\bm{\delta}\bigr)}\biggr),
  \label{HtlidQM}
\end{align}
rdQM:
\begin{align}
  \widetilde{\mathcal{H}}_{\ell}(\bm{\lambda})
  &=B(x;\bm{\lambda}+\ell\bm{\delta})
  \frac{\check{\xi}_{\ell}(x;\bm{\lambda})}
  {\check{\xi}_{\ell}(x+1;\bm{\lambda})}
  \Bigl(\frac{\check{\xi}_{\ell}(x+1;\bm{\lambda}+\bm{\delta})}
  {\check{\xi}_{\ell}(x;\bm{\lambda}+\bm{\delta})}-e^{\partial}\Bigr)\n
  &\quad+D(x;\bm{\lambda}+\ell\bm{\delta})
  \frac{\check{\xi}_{\ell}(x+1;\bm{\lambda})}
  {\check{\xi}_{\ell}(x;\bm{\lambda})}
  \Bigl(\frac{\check{\xi}_{\ell}(x-1;\bm{\lambda}+\bm{\delta})}
  {\check{\xi}_{\ell}(x;\bm{\lambda}+\bm{\delta})}-e^{-\partial}\Bigr).
  \label{HtlrdQM}
\end{align}

For the XL and XJ polynomials in oQM we write concretely \cite{hos}:
\begin{align}
  &\text{XL1}:
  \eta\partial_{\eta}^2P_{\ell,n}(\eta;\bm{\lambda})
  +\Bigl(g+\ell+\tfrac12-\eta
  -2\frac{\eta\,\partial_{\eta}\xi_{\ell}(\eta;\bm{\lambda})}
  {\xi_{\ell}(\eta;\bm{\lambda})}\Bigr)
  \partial_{\eta}P_{\ell,n}(\eta;\bm{\lambda})\n[2pt]
  &\qquad\qquad\qquad\qquad\qquad
  +\Bigl(2\frac{\eta\,\partial_{\eta}
  \xi_{\ell}(\eta;\bm{\lambda}+\bm{\delta})}
  {\xi_{\ell}(\eta;\bm{\lambda})}+n-\ell\Bigr)
  P_{\ell,n}(\eta;\bm{\lambda})=0,
  \label{L1eq}\\
  &\text{XL2}:
  \eta\partial_{\eta}^2P_{\ell,n}(\eta;\bm{\lambda})
  +\Bigl(g+\ell+\tfrac12-\eta
  -2\frac{\eta\,\partial_{\eta}\xi_{\ell}(\eta;\bm{\lambda})}
  {\xi_{\ell}(\eta;\bm{\lambda})}\Bigr)
  \partial_{\eta}P_{\ell,n}(\eta;\bm{\lambda})\n[2pt]
  &\qquad\qquad\qquad\qquad\qquad
  +\Bigl(-2\frac{(g+\tfrac12)\partial_{\eta}
  \xi_{\ell}(\eta;\bm{\lambda}+\bm{\delta})}
  {\xi_{\ell}(\eta;\bm{\lambda})}+n+\ell\Bigr)
  P_{\ell,n}(\eta;\bm{\lambda})=0,
  \label{L2eq}\\
  &\text{XJ1}:
  (1-\eta^2)\partial_{\eta}^2P_{\ell,n}(\eta;\bm{\lambda})\n
  &\qquad
  +\Bigl(h-g-(g+h+2\ell+1)\eta
  -2\frac{(1-\eta^2)\partial_{\eta}\xi_{\ell}(\eta;\bm{\lambda})}
  {\xi_{\ell}(\eta;\bm{\lambda})}\Bigr)
  \partial_{\eta}P_{\ell,n}(\eta;\bm{\lambda})\n[2pt]
  &\qquad
  +\Bigl(-\frac{2(h+\tfrac12)(1-\eta)\partial_{\eta}
  \xi_{\ell}(\eta;\bm{\lambda}+\bm{\delta})}
  {\xi_{\ell}(\eta;\bm{\lambda})}\n
  &\qquad\qquad
  +\ell(\ell+g-h-1)+n(n+g+h+2\ell)\Bigr)
  P_{\ell,n}(\eta;\bm{\lambda})=0,
  \label{J1eq}\\
  &\text{XJ2}:
  (1-\eta^2)\partial_{\eta}^2P_{\ell,n}(\eta;\bm{\lambda})\n
  &\qquad
  +\Bigl(h-g-(g+h+2\ell+1)\eta-2\frac{(1-\eta^2)\partial_{\eta}
  \xi_{\ell}(\eta;\bm{\lambda})}
  {\xi_{\ell}(\eta;\bm{\lambda})}\Bigr)
  \partial_{\eta}P_{\ell,n}(\eta;\bm{\lambda})\n[2pt]
  &\qquad
  +\Bigl(\frac{2(g+\tfrac12)(1+\eta)\partial_{\eta}
  \xi_{\ell}(\eta;\bm{\lambda}+\bm{\delta})}
  {\xi_{\ell}(\eta;\bm{\lambda})}\n
  &\qquad\qquad
  +\ell(\ell+h-g-1)+n(n+g+h+2\ell)\Bigr)
  P_{\ell,n}(\eta;\bm{\lambda})=0.
  \label{J2eq}
\end{align}
The zeros of the deforming polynomial $\xi_{\ell}(\eta_j;\bm{\lambda})=0$
($j=1,\ldots,\ell$) are the extra regular singularities on top of those
for the original Laguerre and Jacobi polynomials. In all the four cases,
the characteristic exponents at these extra $\ell$ zeros are the same:
\begin{equation}
  \rho=0,\ 3.
\end{equation}
The new polynomial solutions have $\rho=0$ at all the extra singularities.
The XJ polynomials provide an infinite family ($n=0,1,2,\ldots,$) of
global solutions of the Fuchsian differential equations
\eqref{J1eq}-\eqref{J2eq} with $3+\ell$ regular singularities.
To the best of our knowledge, global solutions of a Fuchsian differential
equation with more than four singularities had been utterly unknown,
although in this case, the locations of the singularities are very special.

%%%%%%%%%%%%%%%%%%%%%%%%%%%%%%%%%%%%%%%%%%%
%                                         %
% 4.5 Connection between the Original ... %
%                                         %
%%%%%%%%%%%%%%%%%%%%%%%%%%%%%%%%%%%%%%%%%%%
\subsection{Intertwining the Original and the Deformed Systems}
\label{sec:Dar-Cru}

Here we present the transformation intertwining the original and the
deformed Hamiltonian systems. It provides a simple derivation of the
bi-linear expression of the new orthogonal polynomials in the original
and the deforming polynomials. It also yields another proof of the
shape invariance of the deformed systems. This type of transformations
are sometimes called Darboux-Crum transformations.

The general setting of the transformation theory is common to oQM, idQM
and rdQM.
Let us define a pair of Hamiltonians
$\hat{\mathcal{H}}_{\ell}^{(\pm)}(\bm{\lambda})$ in terms of the
operators $\hat{\mathcal{A}}_{\ell}(\bm{\lambda})$ and
$\hat{\mathcal{A}}_{\ell}(\bm{\lambda})^{\dagger}$,
\begin{equation}
  \hat{\mathcal{H}}_{\ell}^{(+)}(\bm{\lambda})\eqdef
  \hat{\mathcal{A}}_{\ell}(\bm{\lambda})^{\dagger}
  \hat{\mathcal{A}}_{\ell}(\bm{\lambda}),\quad
  \hat{\mathcal{H}}_{\ell}^{(-)}(\bm{\lambda})\eqdef
  \hat{\mathcal{A}}_{\ell}(\bm{\lambda})
  \hat{\mathcal{A}}_{\ell}(\bm{\lambda})^{\dagger},
  \label{H+-}
\end{equation}
and consider their Schr\"{o}dinger equations:
\begin{equation}
  \hat{\mathcal{H}}_{\ell}^{(\pm)}(\bm{\lambda})
  \hat{\phi}_{\ell,n}^{(\pm)}(x;\bm{\lambda})
  =\hat{\mathcal{E}}_{\ell,n}^{(\pm)}(\bm{\lambda})
  \hat{\phi}_{\ell,n}^{(\pm)}(x;\bm{\lambda})\quad
  (n=0,1,2,\ldots).
  \label{H+-Scheq}
\end{equation}
By definition, all the eigenfunctions must be square integrable (square
summable). The pair of Hamiltonians are intertwined:
\begin{align}
  &\hat{\mathcal{H}}_{\ell}^{(+)}(\bm{\lambda})
  \hat{\mathcal{A}}_{\ell}(\bm{\lambda})^{\dagger}
  =\hat{\mathcal{A}}_{\ell}(\bm{\lambda})^{\dagger}
  \hat{\mathcal{A}}_{\ell}(\bm{\lambda})
  \hat{\mathcal{A}}_{\ell}(\bm{\lambda})^{\dagger}
  =\hat{\mathcal{A}}_{\ell}(\bm{\lambda})^{\dagger}
  \hat{\mathcal{H}}_{\ell}^{(-)}(\bm{\lambda}),\\
  &\hat{\mathcal{A}}_{\ell}(\bm{\lambda})
  \hat{\mathcal{H}}_{\ell}^{(+)}(\bm{\lambda})
  =\hat{\mathcal{A}}_{\ell}(\bm{\lambda})
  \hat{\mathcal{A}}_{\ell}(\bm{\lambda})^{\dagger}
  \hat{\mathcal{A}}_{\ell}(\bm{\lambda})
  =\hat{\mathcal{H}}_{\ell}^{(-)}(\bm{\lambda})
  \hat{\mathcal{A}}_{\ell}(\bm{\lambda}).
\end{align}
In all the cases to be discussed here, we have
$\hat{\mathcal{A}}_{\ell}(\bm{\lambda})
\hat{\phi}_{\ell,n}^{(+)}(x;\bm{\lambda})\neq 0$ and
$\hat{\mathcal{A}}_{\ell}(\bm{\lambda})^{\dagger}
\hat{\phi}_{\ell,n}^{(-)}(x;\bm{\lambda})\neq 0$.
This situation is called the `broken susy' case in the supersymmetric
quantum mechanics.
This means that the two systems are exactly iso-spectral and there
is one-to-one correspondence between the eigenfunctions:
\begin{align}
  &\hat{\mathcal{E}}_{\ell,n}^{(+)}(\bm{\lambda})
  =\hat{\mathcal{E}}_{\ell,n}^{(-)}(\bm{\lambda}),\\
  &\hat{\phi}_{\ell,n}^{(-)}(x;\bm{\lambda})\propto
  \hat{\mathcal{A}}_{\ell}(\bm{\lambda})
  \hat{\phi}_{\ell,n}^{(+)}(x;\bm{\lambda}),\quad
  \hat{\phi}_{\ell,n}^{(+)}(x;\bm{\lambda})\propto
  \hat{\mathcal{A}}_{\ell}(\bm{\lambda})^{\dagger}
  \hat{\phi}_{\ell,n}^{(-)}(x;\bm{\lambda}).
  \label{phi+-form}
\end{align}
The strategy is to find the operators $\hat{\mathcal{A}}_{\ell}$ and
$\hat{\mathcal{A}}_{\ell}^{\dagger}$ in such a way that
$\hat{\mathcal{H}}_{\ell}^{(+)}$ is the original Hamiltonian (up to
an additive constant) shown in \S\ref{sec:H} and
$\hat{\mathcal{H}}_{\ell}^{(-)}$ becomes the deformed Hamiltonian
(up to an additive constant) demonstrated in \S\ref{sec:defhams}.
Then the above formula \eqref{phi+-form} gives the expression of the
new orthogonal polynomials in terms of the original ones.

We define operators $\hat{\mathcal{A}}_{\ell}(\bm{\lambda})$ and
$\hat{\mathcal{A}}_{\ell}^{\dagger}(\bm{\lambda})$ as follows:\\
oQM:
\begin{align}
  &\hat{\mathcal{A}}_{\ell}(\bm{\lambda})\eqdef
  \frac{d}{dx}-\partial_x\hat{w}_{\ell}(x;\bm{\lambda}),\quad
  \hat{\mathcal{A}}_{\ell}(\bm{\lambda})^{\dagger}=
  -\frac{d}{dx}-\partial_x\hat{w}_{\ell}(x;\bm{\lambda}),
  \label{AhloQM}\\
  &\hat{w}_{\ell}(x;\bm{\lambda})\eqdef
  \log\xi_{\ell}(\eta(x);\bm{\lambda})
  +\left\{\begin{array}{ll}
  \frac12x^2+(g+\ell-1)\log x&:\text{XL1}\\
  w\bigr(x;\mathfrak{t}(\bm{\lambda}+(\ell-1)\bm{\delta})\bigr)
  &:\text{XL2, XJ1, XJ2}
  \end{array}\right..
  \label{hatwl}
\end{align}
(Note that $\frac12x^2+(g+\ell-1)\log x
=w(ix;\bm{\lambda}+(\ell-1)\bm{\delta})+\text{const.}$ for XL1.)\\
idQM:
\begin{align}
  &\hat{\mathcal{A}}_{\ell}(\bm{\lambda})\eqdef
  i\bigl(e^{\frac{\gamma}{2}p}\sqrt{\hat{V}_{\ell}^*(x;\bm{\lambda})}
  -e^{-\frac{\gamma}{2}p}\sqrt{\hat{V}_{\ell}(x;\bm{\lambda})}\,\bigr),\n
  &\hat{\mathcal{A}}_{\ell}(\bm{\lambda})^{\dagger}=
  -i\bigl(\sqrt{\hat{V}_{\ell}(x;\bm{\lambda})}\,e^{\frac{\gamma}{2}p}
  -\sqrt{\hat{V}_{\ell}^*(x;\bm{\lambda})}\,e^{-\frac{\gamma}{2}p}\bigr),
  \label{AhlidQM}\\
  &\hat{V}_{\ell}(x;\bm{\lambda})\eqdef
  V\bigl(x;\mathfrak{t}(\bm{\lambda}+(\ell-1)\bm{\delta})\bigr)
  \frac{\xi_{\ell}(\eta(x-i\gamma);\bm{\lambda})}
  {\xi_{\ell}(\eta(x);\bm{\lambda})},
  \label{Vhl}\\
  &\hat{V}_{\ell}^*(x;\bm{\lambda})=
  V^*\bigl(x;\mathfrak{t}(\bm{\lambda}+(\ell-1)\bm{\delta})\bigr)
  \frac{\xi_{\ell}(\eta(x+i\gamma);\bm{\lambda})}
  {\xi_{\ell}(\eta(x);\bm{\lambda})},
  \label{Vhls}
\end{align}
rdQM:
\begin{align}
  &\hat{\mathcal{A}}_{\ell}(\bm{\lambda})
  \eqdef\sqrt{\hat{B}_{\ell}(x;\bm{\lambda})}
  -e^{\partial}\sqrt{\hat{D}_{\ell}(x;\bm{\lambda})},\quad
  \hat{\mathcal{A}}_{\ell}(\bm{\lambda})^{\dagger}
  =\sqrt{\hat{B}_{\ell}(x;\bm{\lambda})}
  -\sqrt{\hat{D}_{\ell}(x;\bm{\lambda})}\,e^{-\partial},
  \label{AhlrdQM}\\
  &\hat{B}_{\ell}(x;\bm{\lambda})
  \eqdef\frac{\check{\xi}_{\ell}(x+1;\bm{\lambda})}
  {\check{\xi}_{\ell}(x;\bm{\lambda})}
  \times\left\{
  \begin{array}{ll}
  -D\bigl(-(x+\beta+\ell-1);\bm{\lambda}+(\ell-1)\bm{\delta}\bigr)
  &:\text{XM}\\
  B\bigl(x;\mathfrak{t}(\bm{\lambda}+(\ell-1)\bm{\delta})\bigr)
  &:\text{XR, X$q$R}
  \end{array}\right.,
  \label{Bhl}\\
  &\hat{D}_{\ell}(x;\bm{\lambda})
  \eqdef\frac{\check{\xi}_{\ell}(x-1;\bm{\lambda})}
  {\check{\xi}_{\ell}(x;\bm{\lambda})}
  \times\left\{
  \begin{array}{ll}
  -B\bigl(-(x+\beta+\ell-1);\bm{\lambda}+(\ell-1)\bm{\delta}\bigr)
  &:\text{XM}\\
  D\bigl(x;\mathfrak{t}(\bm{\lambda}+(\ell-1)\bm{\delta})\bigr)
  &:\text{XR, X$q$R}
  \end{array}\right..
  \label{Dhl}
\end{align}
Then, by using the two identities of the deforming polynomials
\eqref{xil(l+d)oQM}--\eqref{xil(l)oQM},
\eqref{xil(l+d)idQM}--\eqref{xil(l)idQM} and
\eqref{xil(l+d)rdQM}--\eqref{xil(l)rdQM},
we can show that \cite{stz,os21,os20,os23}
\begin{align}
  \hat{\mathcal{H}}_{\ell}^{(+)}(\bm{\lambda})
  &=\hat{\kappa}_{\ell}(\bm{\lambda})
  \bigl(\mathcal{H}(\bm{\lambda}+\ell\bm{\delta}+\bm{\tilde{\delta}})
  +\hat{f}_{\ell,0}(\bm{\lambda})\hat{b}_{\ell,0}(\bm{\lambda})\bigr),
  \label{Hl+=H}\\
  \hat{\mathcal{H}}_{\ell}^{(-)}(\bm{\lambda})
  &=\hat{\kappa}_{\ell}(\bm{\lambda})
  \bigl(\mathcal{H}_{\ell}(\bm{\lambda})
  +\hat{f}_{\ell,0}(\bm{\lambda})\hat{b}_{\ell,0}(\bm{\lambda})\bigr),
  \label{Hl-=Hl}
\end{align}
where $\hat{\kappa}_{\ell}(\bm{\lambda})$ is
\begin{align}
  \hat{\kappa}_{\ell}(\bm{\lambda})\eqdef 1\ :\text{oQM},
  \quad
  &\hat{\kappa}_{\ell}(\bm{\lambda})\eqdef
  \left\{
  \begin{array}{ll}
  1&:\text{XMP, XW}\\
  (a_1a_2q^{\ell})^{-1}&:\text{XAW}
  \end{array}\right.\!,\n
  &\hat{\kappa}_{\ell}(\bm{\lambda})\eqdef
  \left\{
  \begin{array}{ll}
  1&:\text{XM, XR}\\
  (abd^{-1}q^{\ell})^{-1}&:\text{X$q$R}
  \end{array}\right.\!.
  \label{kappah}
\end{align}
Therefore the original system with the shifted parameters
($\mathcal{H}(\bm{\lambda}+\ell\bm{\delta}+\tilde{\bm{\delta}})$)
and the deformed system ($\mathcal{H}_{\ell}(\bm{\lambda})$) are
exactly isospectral.

Based on the results \eqref{Hl+=H}--\eqref{Hl-=Hl}, we have
\begin{gather}
  \hat{\phi}_{\ell,n}^{(+)}(x;\bm{\lambda})
  =\phi_n(x;\bm{\lambda}+\ell\bm{\delta}+\bm{\tilde{\delta}}),\quad
  \hat{\phi}_{\ell,n}^{(-)}(x;\bm{\lambda})
  =\phi_{\ell,n}(x;\bm{\lambda}),
  \label{phi+-=..}\\
  \!\!\hat{\mathcal{E}}_{\ell,n}^{(\pm)}(\bm{\lambda})
  =\hat{\kappa}_{\ell}(\bm{\lambda})\bigl(
  \mathcal{E}_n(\bm{\lambda}+\ell\bm{\delta}+\bm{\tilde{\delta}})
  +\hat{f}_{\ell,0}(\bm{\lambda})\hat{b}_{\ell,0}(\bm{\lambda})\bigr)
  =\hat{\kappa}_{\ell}(\bm{\lambda})\bigl(
  \mathcal{E}_{\ell,n}(\bm{\lambda})
  +\hat{f}_{\ell,0}(\bm{\lambda})\hat{b}_{\ell,0}(\bm{\lambda})\bigr).\!\!
  \label{E+-=..}
\end{gather}
The correspondence of the pair of eigenfunctions
$\hat{\phi}_{\ell,n}^{(\pm)}(x)$ with their own normalisation specified
in the preceding sections is related by
\begin{align}
  \hat{\phi}_{\ell,n}^{(-)}(x;\bm{\lambda})
  &=\frac{\hat{\mathcal{A}}_{\ell}(\bm{\lambda})
  \hat{\phi}_{\ell,n}^{(+)}(x;\bm{\lambda})}
  {\sqrt{\hat{\kappa}_{\ell}(\bm{\lambda})}\,\hat{f}_{\ell,n}(\bm{\lambda})}
  \times\left\{
  \begin{array}{ll}
  1&:\text{oQM, idQM}\\
  \sqrt{\check{\xi}_{\ell}(1;\bm{\lambda})s_{\ell}(\bm{\lambda})}
  &:\text{rdQM}
  \end{array}\right.,
  \label{phi-<->phi+}\\
  \hat{\phi}_{\ell,n}^{(+)}(x;\bm{\lambda})
  &=\frac{\hat{\mathcal{A}}_{\ell}(\bm{\lambda})^{\dagger}
  \hat{\phi}_{\ell,n}^{(-)}(x;\bm{\lambda})}
  {\sqrt{\hat{\kappa}_{\ell}(\bm{\lambda})}\,\hat{b}_{\ell,n}(\bm{\lambda})}
  \times\left\{
  \begin{array}{ll}
  1&:\text{oQM, idQM}\\
  \frac{1}{\sqrt{\check{\xi}_{\ell}(1;\bm{\lambda})s_{\ell}(\bm{\lambda})}}
  &:\text{rdQM}
  \end{array}\right..
  \label{phi+<->phi-}
\end{align}
By removing the effects of the orthogonality weight functions, we define
the operators $\hat{\mathcal{F}}_{\ell}(\bm{\lambda})$ and
$\hat{\mathcal{B}}_{\ell}(\bm{\lambda})$
\begin{align}
  \hat{\mathcal{F}}_{\ell}(\bm{\lambda})&\eqdef
  \psi_{\ell}(x;\bm{\lambda})^{-1}\circ
  \frac{\hat{\mathcal{A}}_{\ell}(\bm{\lambda})}
  {\sqrt{\hat{\kappa}_{\ell}(\bm{\lambda})}}\circ
  \phi_0(x;\bm{\lambda}+\ell\bm{\delta}+\bm{\tilde{\delta}})
  \times\left\{
  \begin{array}{ll}
  1&:\text{oQM, idQM}\\
  \sqrt{\check{\xi}_{\ell}(1;\bm{\lambda})s_{\ell}(\bm{\lambda})}
  &:\text{rdQM}
  \end{array}\right.\!\!,
  \label{Fhldef}\\
  \hat{\mathcal{B}}_{\ell}(\bm{\lambda})&\eqdef
  \phi_0(x;\bm{\lambda}+\ell\bm{\delta}+\bm{\tilde{\delta}})^{-1}\circ
  \frac{\hat{\mathcal{A}}_{\ell}(\bm{\lambda})^{\dagger}}
  {\sqrt{\hat{\kappa}_{\ell}(\bm{\lambda})}}\circ
  \psi_{\ell}(x;\bm{\lambda})
  \times\left\{
  \begin{array}{ll}
  1&:\text{oQM, idQM}\\
  \frac{1}{\sqrt{\check{\xi}_{\ell}(1;\bm{\lambda})s_{\ell}(\bm{\lambda})}}
  &:\text{rdQM}
  \end{array}\right.\!\!.\!\!
  \label{Bhldef}
\end{align}
Their explicit forms are the following:\\
oQM:
\begin{align}
  \hat{\mathcal{F}}_{\ell}(\bm{\lambda})&=
  2\Bigl(d_2(\eta)\xi_{\ell}(\eta;\bm{\lambda})\frac{d}{d\eta}
  -d_1(\bm{\lambda})\xi_{\ell}(\eta;\bm{\lambda}+\bm{\delta})\Bigr)
  \times\left\{
  \begin{array}{ll}
  \pm 1&:\text{XL1/XL2}\\
  \pm 1&:\text{XJ1/XJ2}
  \end{array}\right.,
  \label{FhloQM}\\
  \hat{\mathcal{B}}_{\ell}(\bm{\lambda})&=
  \frac{-2}{\xi_{\ell}(\eta;\bm{\lambda})}\Bigl(
  \frac{c_2(\eta)}{d_2(\eta)}\frac{d}{d\eta}+d_3(\bm{\lambda},\ell)\Bigr)
  \times\left\{
  \begin{array}{ll}
  \pm 1&:\text{XL1/XL2}\\
  \pm 1&:\text{XJ1/XJ2}
  \end{array}\right.,
  \label{BhloQM}
\end{align}
idQM:
\begin{align}
  \!\!\!\!\!\hat{\mathcal{F}}_{\ell}(\bm{\lambda})&=
  \frac{-i}{\varphi(x)}\Bigl(
  v_1(x;\bm{\lambda}+\ell\bm{\delta})
  \xi_{\ell}(\eta(x+i\tfrac{\gamma}{2});\bm{\lambda})e^{\frac{\gamma}{2}p}
  -v_1^*(x;\bm{\lambda}+\ell\bm{\delta})
  \xi_{\ell}(\eta(x-i\tfrac{\gamma}{2});\bm{\lambda})e^{-\frac{\gamma}{2}p}
  \Bigr),\!\!\!
  \label{FhlidQM}\\
  \!\!\!\!\!\hat{\mathcal{B}}_{\ell}(\bm{\lambda})&=
  \frac{1}{\xi_{\ell}(\eta(x);\bm{\lambda})}\frac{-i}{\varphi(x)}\Bigl(
  v_2(x;\bm{\lambda}+(\ell-1)\bm{\delta})e^{\frac{\gamma}{2}p}
  -v_2^*(x;\bm{\lambda}+(\ell-1)\bm{\delta})e^{-\frac{\gamma}{2}p}\Bigr),
  \!\!\!
  \label{BhlidQM}
\end{align}
rdQM:
\begin{align}
  \hat{\mathcal{F}}_{\ell}(\bm{\lambda})&=
  \frac{1}{\varphi(x;\bm{\lambda}+\ell\bm{\delta}+\tilde{\bm{\delta}})}
  \Bigl(v_1^B(x;\bm{\lambda}+\ell\bm{\delta})
  \check{\xi}_{\ell}(x;\bm{\lambda})e^{\partial}
  -v_1^D(x;\bm{\lambda}+\ell\bm{\delta})
  \check{\xi}_{\ell}(x+1;\bm{\lambda})\Bigr),
  \label{FhlrdQM}\\
  \hat{\mathcal{B}}_{\ell}(\bm{\lambda})&=
  \frac{1}{\check{\xi}_{\ell}(x;\bm{\lambda})}
  \frac{1}{\varphi(x;\bm{\lambda}+(\ell-1)\bm{\lambda}+\tilde{\bm{\delta}})}
  \Bigl(v_2^B(x;\bm{\lambda}+(\ell-1)\bm{\delta})
  -v_2^D(x;\bm{\lambda}+(\ell-1)\bm{\delta})e^{-\partial}\Bigr).
  \label{BhlrdQM}
\end{align}
The operators $\hat{\mathcal{F}}_{\ell}(\bm{\lambda})$ and
$\hat{\mathcal{B}}_{\ell}(\bm{\lambda})$ act as the forward and backward
shift operators connecting the original polynomials $P_n$ and
the exceptional polynomials $P_{\ell,n}$:
\begin{align}
  \hat{\mathcal{F}}_{\ell}(\bm{\lambda})
  \check{P}_n(x;\bm{\lambda}+\ell\bm{\delta}+\bm{\tilde{\delta}})
  &=\hat{f}_{\ell,n}(\bm{\lambda})\check{P}_{\ell,n}(x;\bm{\lambda}),
  \label{FhatPn=Pln}\\
  \hat{\mathcal{B}}_{\ell}(\bm{\lambda})\check{P}_{\ell,n}(x;\bm{\lambda})
  &=\hat{b}_{\ell,n}(\bm{\lambda})
  \check{P}_n(x;\bm{\lambda}+\ell\bm{\delta}+\bm{\tilde{\delta}}).
  \label{BhatPln=Pn}
\end{align}
The former relation \eqref{FhatPn=Pln} with the explicit forms of
$\hat{\mathcal{F}}_{\ell}(\bm{\lambda})$ \eqref{FhloQM}, \eqref{FhlidQM}
and \eqref{FhlrdQM} provides the explicit expressions \eqref{PlnoQM},
\eqref{PlnidQM} and \eqref{PlnrdQM} of the exceptional orthogonal polynomials.
In terms of $\hat{\mathcal{F}}_{\ell}(\bm{\lambda})$ and
$\hat{\mathcal{B}}_{\ell}(\bm{\lambda})$,
the relations \eqref{Hl+=H}--\eqref{Hl-=Hl} become
\begin{align}
  &\hat{\mathcal{B}}_{\ell}(\bm{\lambda})
  \hat{\mathcal{F}}_{\ell}(\bm{\lambda})
  =\widetilde{\mathcal{H}}(\bm{\lambda}+\ell\bm{\delta}+\tilde{\bm{\delta}})
  +\hat{f}_{\ell,0}(\bm{\lambda})\hat{b}_{\ell,0}(\bm{\lambda}),\\
  &\hat{\mathcal{F}}_{\ell}(\bm{\lambda})
  \hat{\mathcal{B}}_{\ell}(\bm{\lambda})
  =\widetilde{\mathcal{H}}_{\ell}(\bm{\lambda})
  +\hat{f}_{\ell,0}(\bm{\lambda})\hat{b}_{\ell,0}(\bm{\lambda}).
\end{align}
The other simple consequences of these relations are
\begin{equation}
  \hat{\mathcal{E}}_{\ell,n}^{(\pm)}(\bm{\lambda})
  =\hat{\kappa}_{\ell}(\bm{\lambda})\hat{f}_{\ell,n}(\bm{\lambda})
  \hat{b}_{\ell,n}(\bm{\lambda}),\quad
   \mathcal{E}_n(\bm{\lambda}+\ell\bm{\delta})
  =\hat{f}_{\ell,n}(\bm{\lambda})\hat{b}_{\ell,n}(\bm{\lambda})
  -\hat{f}_{\ell,0}(\bm{\lambda})\hat{b}_{\ell,0}(\bm{\lambda}).
  \label{Eln+-}
\end{equation}
Orthogonality relations \eqref{orthooQM}--\eqref{hlnoQM},
\eqref{orthoidQM}--\eqref{hlnidQM} and \eqref{orthordQM}--\eqref{dln2}
can be shown by using these intertwining relations \cite{os21,os20,os23}.

It is interesting to note that the operator
$\hat{\mathcal{A}}_{\ell}(\bm{\lambda})$ intertwines those of the original
and deformed systems $\mathcal{A}(\bm{\lambda})$ and
$\mathcal{A}_{\ell}(\bm{\lambda})$:
\begin{align}
  &\hat{\mathcal{A}}_{\ell}(\bm{\lambda}+\bm{\delta})
  \mathcal{A}(\bm{\lambda}+\ell\bm{\delta}+\bm{\tilde{\delta}})
  =\mathcal{A}_{\ell}(\bm{\lambda})
  \hat{\mathcal{A}}_{\ell}(\bm{\lambda}),
  \label{AhA=AlAh}\\
  &\hat{\mathcal{A}}_{\ell}(\bm{\lambda})
  \mathcal{A}(\bm{\lambda}+\ell\bm{\delta}+\bm{\tilde{\delta}})^{\dagger}
  =\mathcal{A}_{\ell}(\bm{\lambda})^{\dagger}
  \hat{\mathcal{A}}_{\ell}(\bm{\lambda}+\bm{\delta}).
  \label{AhAd=AldAh}
\end{align}
In terms of the definitions of the forward shift operators
$\mathcal{F}(\bm{\lambda})$ \eqref{Fdef},
$\mathcal{F}_{\ell}(\bm{\lambda})$ \eqref{Fldef},
$\hat{\mathcal{F}}_{\ell}(\bm{\lambda})$ \eqref{Fhldef}, and
$\mathcal{B}(\bm{\lambda})$ \eqref{Bdef},
$\mathcal{B}_{\ell}(\bm{\lambda})$ \eqref{Bldef},
the above relations are rewritten as:
\begin{align}
  &\hat{s}_{\ell}(\bm{\lambda}+\bm{\delta})
  \hat{\mathcal{F}}_{\ell}(\bm{\lambda}+\bm{\delta})
  \mathcal{F}(\bm{\lambda}+\ell\bm{\delta}+\bm{\tilde{\delta}})
  =\hat{s}_{\ell}(\bm{\lambda})
  \mathcal{F}_{\ell}(\bm{\lambda})
  \hat{\mathcal{F}}_{\ell}(\bm{\lambda}),
  \label{FlhF=FlFlh}\\
  &\hat{s}_{\ell}(\bm{\lambda})
  \hat{\mathcal{F}}_{\ell}(\bm{\lambda})
  \mathcal{B}(\bm{\lambda}+\ell\bm{\delta}+\bm{\tilde{\delta}})
  =\hat{s}_{\ell}(\bm{\lambda}+\bm{\delta})
  \mathcal{B}_{\ell}(\bm{\lambda})
  \hat{\mathcal{F}}_{\ell}(\bm{\lambda}+\bm{\delta}),
  \label{FlhB=BlFlh}
\end{align}
where $\hat{s}_{\ell}(\bm{\lambda})$ is
\begin{equation}
  \hat{s}_{\ell}(\bm{\lambda})\eqdef 1\ :\text{oQM},
  \quad
  \hat{s}_{\ell}(\bm{\lambda})
  \eqdef\sqrt{\hat{\kappa}_{\ell}(\bm{\lambda})}
  \ :\text{idQM},
  \quad
  \hat{s}_{\ell}(\bm{\lambda})
  \eqdef\hat{\kappa}_{\ell}(\bm{\lambda})\times\left\{
  \begin{array}{ll}
  \beta+\ell-1&:\text{XM}\\
  c+\ell-1&:\text{XR}\\
  1-cq^{\ell-1}&:\text{X$q$R}
  \end{array}\right.\!.
\end{equation}
These relations can be proven by explicit calculation with the help of
the two identities of the deforming polynomial
\eqref{xil(l+d)oQM}--\eqref{xil(l)oQM},
\eqref{xil(l+d)idQM}--\eqref{xil(l)idQM} and
\eqref{xil(l+d)rdQM}--\eqref{xil(l)rdQM},
and provide a proof of \eqref{Alphiln=flnphiln}--\eqref{Aldphiln=blnphiln}
and \eqref{FlPln=flnPln}--\eqref{BlPln=blnPln}
without recourse to the shape invariance of the deformed system,
see \cite{os21,os20,os23}.

\bigskip

The intertwining relation offers another proof of the shape invariance
of the $\ell$-th new orthogonal polynomials through the established shape
invariance of the original polynomials as depicted in the following
commutative diagram:
\begin{align*}
\begin{CD}
  \fbox{\begin{tabular}{@{}c@{}}
  $\hat{\mathcal{H}}_{\ell}^{(+)}(\bm{\lambda}+\bm{\delta})$\\
  $\propto\mathcal{H}(\bm{\lambda}+(\ell+1)\bm{\delta}+\bm{\tilde{\delta}})
  +c(\bm{\lambda}+\bm{\delta})$
  \end{tabular}}
  @>{\textstyle\ \hat{\mathcal{A}}_{\ell}(\bm{\lambda}+\bm{\delta})\ }>>
  \fbox{\begin{tabular}{@{}c@{}}
  $\hat{\mathcal{H}}_{\ell}^{(-)}(\bm{\lambda}+\bm{\delta})$\\
  $\propto\mathcal{H}_{\ell}(\bm{\lambda}+\bm{\delta})
  +c(\bm{\lambda}+\bm{\delta})$
  \end{tabular}}\\[4pt]
  @A\begin{tabular}{@{}c@{}}\text{established}\\\text{shape invariance}
  \end{tabular}AA
  @AA\begin{tabular}{@{}c@{}}\text{shape}\\\text{invariance}
  \end{tabular}A\\[4pt]
  \fbox{\begin{tabular}{@{}c@{}}
  $\hat{\mathcal{H}}_{\ell}^{(+)}(\bm{\lambda})$\\
  $\propto\mathcal{H}(\bm{\lambda}+\ell\bm{\delta}+\bm{\tilde{\delta}})
  +c(\bm{\lambda})$
  \end{tabular}}
  @>{\textstyle{\qquad \hat{\mathcal{A}}_{\ell}(\bm{\lambda})\qquad}}>>
  \fbox{\begin{tabular}{@{}c@{}}
  $\hat{\mathcal{H}}_{\ell}^{(-)}(\bm{\lambda})$\\
  $\propto\mathcal{H}_{\ell}(\bm{\lambda})+c(\bm{\lambda})$
  \end{tabular}}\\[4pt]
  \text{original polynomial}
  @.
  \text{new polynomial}
\end{CD}\\[6pt]
  \text{Two ways of proving shape invariance of the new orthogonal
  polynomials system.}\hspace{5mm}
\end{align*}

Historically the first members of the $X_1$ Laguerre polynomials were 
discussed in the framework of `conditionally exactly solvable problems'
\cite{junkroy} in 1997.
About ten years later the first members of the $X_1$ Laguerre and Jacobi
polynomials were constructed by G\'{o}mez-Ullate et al \cite{gomez} in 2008
in the framework of the Sturm-Liouville theory. They were rederived as
the main part of the eigenfunctions of shape invariant quantum mechanical
Hamiltonians by Quesne and collaborators \cite{quesne}.
In 2009 the present authors derived the infinitely many $X_{\ell}$ Laguerre
and Jacobi polynomials by deforming the Hamiltonian systems of the radial
oscillator and the P\"{o}schl-Teller potential in terms of the
eigenpolynomials of degree $\ell$ \cite{os16}.
The main idea was very simple. The $X_1$ Jacobi Hamiltonian of Quesne
\cite{quesne} in our notation read
\begin{align}
  \mathcal{H}_1(g,h)=&-\frac{d^2}{dx^2}+\frac{g(g+1)}{\sin^2 x}
  +\frac{h(h+1)}{\cos^2 x}-(2+g+h)^2\n
  &+\frac{8(g+h+1)}{1\!+\!g\!+\!h\!+\!(g-\!h)\cos2x}
  -\frac{8(2g+1)(2h+1)}{(1\!+\!g\!+\!h\!+\!(g-\!h)\cos2x)^2}.
\end{align}
Its groundstate wavefunction was
\begin{align}
  \phi_{1,0}(x;g,h)&=(\sin x)^{g+1}(\cos x)^{h+1}\frac{3+g+h+(g-h)\cos2x}
  {1+g+h+(g-h)\cos2x}\n
  &=(\sin x)^{g+1}(\cos x)^{h+1}
  \frac{P_1^{(g+2-\frac32,-h-2-\frac12)}(\cos2x)}
  {P_1^{(g+1-\frac32,-h-1-\frac12)}(\cos2x)}.
\end{align}
{}From this it did not take long to guess the general $\ell$ form:
\begin{equation}
  \phi_{\ell,0}(x;g,h)=(\sin x)^{g+\ell}(\cos x)^{h+\ell}
  \frac{P_{\ell}^{(g+\ell+1-\frac32,-h-\ell-1-\frac12)}(\cos2x)}
  {P_{\ell}^{(g+\ell-\frac32,-h-\ell-\frac12)}(\cos2x)},
\end{equation}
which was $e^{w_{\ell}(x,\bm{\lambda})}$ in \eqref{psiloQM} for XJ1.
After the discovery of the second lowest members ($\ell=2$) of the XL2
family \cite{quesne2}, the entire XJ2 and XL2 families were constructed
in \cite{os19}.
Then the construction of the exceptional Wilson and Askey-Wilson
polynomials \cite{os17} and the exceptional Racah and $q$-Racah
polynomials \cite{os23} followed by taking shape invariance as a
guiding principle.
The Fuchsian properties and many other aspects of XL and XJ polynomials
were demonstrated in \cite{hos}.
The intertwining relations were developed in \cite{stz,os20,os23}.
The general knowledge of the solution spaces of exactly solvable
(discrete) quantum mechanical systems governed by Crum's theorem
\cite{crum} and its modifications \cite{adler,os15,gos,os22} had been
very helpful for the discovery of various exceptional orthogonal
polynomials. Slightly different formulation of the Darboux-Crum
transformation for the XL polynomials were reported in \cite{duttaroy,gomez2}.

Naively one expects that most exactly solvable QM Hamiltonian systems
would admit similar shape invariant deformations. Since the `twisting'
of the parameters is essential, the systems corresponding to the Hermite
and the Gegenbauer ($g=h$ case of the Jacobi) polynomials do not admit
such a deformation.
So far, in addition to the examples presented in this review,
the exceptional polynomials for the continuous Hahn \cite{os20},
dual ($q$)-Hahn and little $q$-Jacobi \cite{os23} are explicitly known.
Various exceptional orthogonal polynomials can be obtained from
XAW and X$q$R polynomials by taking appropriate limits.
It is a big challenge to construct various new orthogonal polynomials,
for example, those corresponding to the Morse and Scarf potentials and
the ($q$)-Hahn polynomials, and to study them in detail.

%%%%%%%%%%%%%%%%%%%%%%%%%%%%%%%%%%%%%%%%%%%%%%%%%%%%%%%%%%%%%%%
%                                                             %
% 5. Summary and Comments                                     %
%                                                             %
%%%%%%%%%%%%%%%%%%%%%%%%%%%%%%%%%%%%%%%%%%%%%%%%%%%%%%%%%%%%%%%
\section{Summary and Comments}
\label{sec:Summ}
\setcounter{equation}{0}

Discrete quantum mechanics, as formulated and developed by the present
authors during the last decade, is briefly reviewed following the logical
structure rather than the historical developments. The parallelism and
contrast among the ordinary quantum mechanics (oQM) and the discrete
quantum mechanics (dQM) are emphasised.
As far as possible, one universal formula valid for oQM and dQM with the
pure imaginary shifts (idQM) and the real shifts (rdQM) is presented first
and various ramifications follow. In one aspect, this is a reformulation
of the theory of bi-spectral orthogonal polynomials in the language and
logic of quantum mechanics and every quantity is given explicitly without
recourse to the moment problem as exemplified in \S\ref{dQM}.
While it offers structural understanding of various specific properties
of the bi-spectral polynomials, a new unified theory is established to
generate all the known bi-spectral polynomials from the first principle,
as reviewed in \S\ref{sec:Uni}.
As evidenced by the discovery of various kinds of infinite families of
new orthogonal polynomials reviewed in \S\ref{sec:Exce}, this
reformulation has been extremely successful.

Here are a few words on how our project of discrete quantum mechanics
started. The classical and quantum integrability were very closely
related in the multi-particle dynamics of Calogero-Moser systems
\cite{Cal-Sut}. Reflecting the quantised eigenvalues etc, the
corresponding classical quantities showed Diophantine properties at
various levels \cite{cs2}. In particular, the classical equilibrium
points of the Calogero-Moser systems were described by the zeros of
the classical polynomials, the Hermite, Laguerre and Jacobi
\cite{stiel,calmat,calpere}, which in turn constituted the
eigenpolynomials of single particle oQM. The classical equilibrium point,
namely the minimum of the classical potential, could be better rephrased
by the {\em maximum of the groundstate eigenfunction} $\phi_0(x)=e^{w(x)}$,
or the maximum of the prepotential $w(x)$. Stieltjes investigated the
equilibrium problems of multi-particle systems having
`logarithmic potential' which was nothing but the corresponding prepotential.
Investigation of the discrete counterparts of the Calogero-Moser systems,
that is, the Ruijsenaars-Schneider-van-Diejen \cite{RSvD} systems
revealed similar Diophantine properties \cite{rags}.
Later it was discovered that the classical equilibrium points of the
Ruijsenaars-Schneider-van-Diejen systems were described by the zeros of
the Wilson and Askey-Wilson polynomials \cite{os2,vDiej}.
Then it was quite natural to seek for a single particle quantum
mechanical formulation in which the Wilson and Askey-Wilson polynomials
were the eigenpolynomials.
Thus the first discrete quantum mechanics was born \cite{os4}.
The factorised Hamiltonian was the guiding principle.

The discrete quantum mechanics is still in its infancy.
Although the single particle dynamics has been understood rather well
by now, the multi-particle dynamics is virtually untouched.
A lot of interesting problems, for example, the construction of
multi-particle version of various results in this review, are waiting
to be clarified.

Due to the length limit, many interesting and important topics could
not be covered in this review, for example, the limits from dQM to oQM
and those of the corresponding polynomials, continuous $\ell$ version
of XL and XJ, etc.
We refer to our papers \cite{os16}--\cite{os21} and \cite{os2} and others.

%%%%%%%%%%%%%%%%%%%%%%%%%%%%%%%%%%%%%%%%%%%%%%%%%%%%%%%%%%%%%%%
%                                                             %
%  Acknowledgments                                            %
%                                                             %
%%%%%%%%%%%%%%%%%%%%%%%%%%%%%%%%%%%%%%%%%%%%%%%%%%%%%%%%%%%%%%%
\section*{Acknowledgements}
R.\,S. is supported in part by Grant-in-Aid for Scientific Research from
the Ministry of Education, Culture, Sports, Science and Technology,
No.22540186. R.\,S. thanks Orlando Ragnisco and Dept. Phys. Univ. Roma
Tre for the hospitality, where part of this work was done.

%%%%%%%%%%%%%%%%%%%%%%%%%%%%%%%%%%%%%%%%%%%%%%%%%%%%%%%%%%%%%%%
%                                                             %
%  Appendix                                                   %
%                                                             %
%%%%%%%%%%%%%%%%%%%%%%%%%%%%%%%%%%%%%%%%%%%%%%%%%%%%%%%%%%%%%%%
\section*{Appendix: Symbols and Definitions}
\label{append}
\setcounter{equation}{0}
\renewcommand{\theequation}{A.\arabic{equation}}

Here are several symbols and definitions related to
the ($q$)-hypergeometric functions.

\noindent
%%%%%%%%%%%%%%%%%%%
$\circ$ shifted factorial (Pochhammer symbol) $(a)_n$ :
\begin{equation}
   (a)_n\eqdef\prod_{k=1}^n(a+k-1)=a(a+1)\cdots(a+n-1)
   =\frac{\Gamma(a+n)}{\Gamma(a)}.
   \label{defPoch}
\end{equation}
%%%%%%%%%%%%%%%%%%%
$\circ$ $q$-shifted factorial ($q$-Pochhammer symbol) $(a\,;q)_n$ :
\begin{equation}
   (a\,;q)_n\eqdef\prod_{k=1}^n(1-aq^{k-1})=(1-a)(1-aq)\cdots(1-aq^{n-1}).
   \label{defqPoch}
\end{equation}
%%%%%%%%%%%%%%%%%%%
$\circ$ Hypergeometric series ${}_rF_s$ :
\begin{equation}
   {}_rF_s\Bigl(\genfrac{}{}{0pt}{}{a_1,\,\cdots,a_r}{b_1,\,\cdots,b_s}
   \Bigm|z\Bigr)
   \eqdef\sum_{n=0}^{\infty}
   \frac{(a_1,\,\cdots,a_r)_n}{(b_1,\,\cdots,b_s)_n}\frac{z^n}{n!}\,,
   \label{defhypergeom}
\end{equation}
where $(a_1,\,\cdots,a_r)_n\eqdef\prod_{j=1}^r(a_j)_n
=(a_1)_n\cdots(a_r)_n$.\\
%%%%%%%%%%%%%%%%%%%
$\circ$ $q$-Hypergeometric series (the basic hypergeometric series)
${}_r\phi_s$ :
\begin{equation}
   {}_r\phi_s\Bigl(
   \genfrac{}{}{0pt}{}{a_1,\,\cdots,a_r}{b_1,\,\cdots,b_s}
   \Bigm|q\,;z\Bigr)
   \eqdef\sum_{n=0}^{\infty}
   \frac{(a_1,\,\cdots,a_r\,;q)_n}{(b_1,\,\cdots,b_s\,;q)_n}
   (-1)^{(1+s-r)n}q^{(1+s-r)n(n-1)/2}\frac{z^n}{(q\,;q)_n}\,,
   \label{defqhypergeom}
\end{equation}
where $(a_1,\,\cdots,a_r\,;q)_n\eqdef\prod_{j=1}^r(a_j\,;q)_n
=(a_1\,;q)_n\cdots(a_r\,;q)_n$.

%%%%%%%%%%%%%%%%%%%%%%%%%%%%%%%%%%%%%%%%%%%%%%%%%%%%%%%%%%%%%%%
%                                                             %
%  References                                                 %
%                                                             %
%%%%%%%%%%%%%%%%%%%%%%%%%%%%%%%%%%%%%%%%%%%%%%%%%%%%%%%%%%%%%%%

\end{document}